\documentclass[smallextended]{svjour3}       
\smartqed  
\usepackage{graphicx}
\usepackage{natbib}
\begin{document}

\title{Radio AGN in the local universe: unification, triggering and evolution
}

\titlerunning{Radio AGN in the local universe}        

\author{Clive Tadhunter
}

\authorrunning{Radio AGN in the local universe} 

\institute{C. Tadhunter \at
              University of Sheffield \\
              Tel.: +44-114-2224300\\
              Fax: +123-45-678910\\
              \email{c.tadhunter@sheffield.ac.uk}
}

\date{Received: date / Accepted: date}

\maketitle

\begin{abstract}
Associated with one of the most important forms of active galactic nucleus (AGN) feedback, and showing a strong preference for giant elliptical host galaxies, radio AGN ($L_{1.4GHz} > 10^{24}$~W Hz$^{-1}$) are a key sub-class of the overall AGN population. Recently their study has benefitted dramatically from the availability of high-quality data covering the X-ray to far-IR wavelength range obtained with the current generation of  ground- and space-based telescope facilities. 
Reflecting this progress, here I review 
our current state of understanding of the population of radio AGN at low and intermediate redshifts
($z < 0.7$), concentrating on their nuclear AGN and host galaxy properties, and covering three interlocking themes: the classification of radio AGN and its interpretation; the triggering and fuelling of the jet and AGN activity; and the evolution of the host galaxies. I show that much of the observed diversity in the AGN  properties of radio AGN can be explained in terms of a combination of orientation/anisotropy, mass accretion rate, and variability effects. The detailed morphologies of the host galaxies are consistent with the triggering of strong line radio galaxies (SLRG) in galaxy mergers. However, the star formation properties and cool ISM contents suggest that the triggering mergers are relatively minor in terms of their gas masses in most cases, and would not lead to major growth of the supermassive black holes and stellar bulges; therefore, apart from a minority ($<$20\%) that show evidence for higher star formation rates and more massive cool ISM reservoirs, the SLRG represent late-time re-triggering of activity in mature giant elliptical galaxies. In contrast, the host and environmental properties of weak line radio galaxies (WLRG) with Fanaroff-Riley class I (FRI) radio morphologies are consistent with more gradual fuelling of the activity via gas accretion at low rates onto the supermassive black holes.

\end{abstract}

\section{Introduction}
\label{intro}

Radio AGN have always held an important place in our understanding of AGN and their link to galaxy evolution. Among the first active galaxies to be discovered following the opening of the radio window in first half of the 20th century
\citep{bolton49}, it was soon recognised that they are associated with giant elliptical 
galaxies \citep{matthews64}. Their extraordinary luminosities at radio wavelengths also mean that they stand out at high redshifts. In this way, they act as important signposts to the early Universe, in particular probing the evolution of the most massive galaxies in some of the highest density environments \citep{mccarthy93,miley08}. Most recently it has been acknowledged that the mechanical energy imparted by the expanding jets and lobes  is one of the most important forms of AGN-induced feedback. This is because it prevents the hot X-ray emitting gas of the host galaxies and clusters from cooling to form stars, and thereby influences the shape of the high luminosity end of the galaxy luminosity function \citep[e.g.][]{benson03,mcnamara07}. In addition, the jets drive massive outflows 
of warm, neutral and molecular gas that can potentially influence the star formation histories of the central bulge regions of galaxies
\citep{holt03,morganti05,holt08,morganti13}. 

The physics of the exquisite jets and lobes of relativistic particles that produce the radio emission, and how these components interact with their gaseous environments, has been a major focus for many detailed studies at radio wavelengths. However, much of our current understanding of the fuelling and triggering of the activity in radio AGN has been derived from observations at other wavelengths. These include: optical and infrared observations of the host galaxies; X-ray, optical and infrared observations of the nuclei; and X-ray and optical observations of the large-scale environments. 

\begin{table}
\begin{tabular}{llllllll}
\hline
 & &\multicolumn{6}{c}{\bf References} \\
{\bf Sample}               &{\bf Selection}  &{\bf Mast.} &{\bf Radio} &{\bf Opt.} &{\bf NIR} &{\bf MFIR} &{\bf Xray} \\
\hline \\
{\bf 2Jy}            &$S_{2.7GHz} > 2$Jy &1,9, &2,3, &6,7 &10 &5,11, &14,15, \\
Dicken   &$0.05 < z < 0.7$    &40 &4,5, &8,9 & &8,12,  &38 \\
et al. (2009)                        &$\delta < +10^{\circ}$ & &39 & & &13 & \\
                        &$\alpha^{4.8}_{2.7} > +0.5$  & & & & & & \\
			&($F_{\nu} \propto \nu^{-\alpha}$) & & & & & & \\
\hline \\ 
{\bf 3CR}              &$S_{178MHz} > 9$Jy &16,17, &19,20, &23,24, &26,27, &30,31,    &33,34,\\
Buttiglione             &$z < 0.3$           &18     &21,22 &25     &28,29 &32,11,     &35,36, \\
et al. (2009)                    &$\delta > -5^{\circ}$ &     &   &       &      &12,34     &37,38\\
\hline \\
\end{tabular}
\caption{Selection criteria and references for the two main samples of radio AGN considered in this review. Note that both
of these samples were selected from larger master samples with less restrictive selection criteria, the
references for which are given in column 3. Columns 4 to 8 give the references for the optical, radio,
near-IR, mid- to far-IR (MFIR), and X-ray data for the samples. Reference key: 1. \citet{wall85}; 2. \citet[][and references therein]{morganti93};
3. \citet{morganti97a}; 4. \citet{morganti99}; 5. \citet{dicken08}; 6. \citet[][and references therein]{tadhunter93}; 7. \citet{shaw95}, 8. \citet{morganti97b}; 9. \citet{tadhunter98}; 10. \citet{inskip10};  11. \citet{dicken12}; 12.\citet{dicken14}; 13. \citet{dicken16}; 14. \citet{siebert96}; 15. \citet{mingo14}; 16. \citet{bennett62a}; 17. \citet{bennett62b}; 18. \citet{spinrad85}; 19.  http:$//$www.jb.man.ac.uk$/$atlas 
; 20. \citet{black92}; 21. \citet{leahy97}; 22. \citet[][and references therein]{hardcastle97}; 23. \citet{buttiglione09}; 24. \citet{buttiglione10};25. \citet{buttiglione11}; 26. \citet{lilly84}; 27. \citet{madrid06}; 28. \citet{donzelli07}; 29. \citet{baldi10}; 30. \citet{haas04}; 31. \citet{ogle06}; 32. \citet{dicken10}; 33. \citet{hardcastle06}; 34. \citet{hardcastle09}; 35. \citet{massaro10}; 36. \citet{massaro12}; 37. \citet{massaro15}; 38. \citet{ineson15}; 39. \citet{tzioumis02}; 40. \citet{diserego94}.}
\label{samples}
\end{table}

Aside from their high radio luminosities, perhaps the most important feature that the majority of radio AGN have in common is that they are associated with massive early-type galaxies. This feature holds considerable advantages for investigating the triggering of the AGN activity, since it allows relatively ``clean'' searches to be made for the morphological, star formation, and gas content signatures of the triggering events. This is in contrast to Seyfert galaxies, for example, which are more commonly associated with spiral galaxy hosts \citep[e.g.][]{adams77}; the morphological complexity of such hosts, coupled with the large gas reservoirs of their quiescent disks and associated star formation activity, make it challenging to disentangle the triggering events from the normal, non-AGN-related 
evolution of their host galaxies.

Two of the key outstanding questions for radio AGN are as follows.
\begin{itemize}
\item What do their radio morphological and optical spectroscopic classifications imply about the fuelling and evolution of their activity?
\item How are radio AGN triggered, and how does the triggering tie in with the evolution of their host galaxies?
\end{itemize}

Answering these questions is crucial, for example, if we want to properly incorporate radio AGN into
models of galaxy evolution. It has become clear that addressing them requires a multi-wavelength approach that encompasses not only deep radio and optical observations, but also observations at X-ray and infrared wavelengths. Therefore this is a field that has particularly benefitted from the availability of large space observatories such as Chandra, XMM, Spitzer and Herschel over the last 20 years. At the same time, deeper and higher resolution optical observations with 8m-class telescopes and the Hubble Space Telescope (HST) respectively have allowed the host galaxy morphologies to be examined in unprecedented depth. Concerted efforts have also been made to improve the completeness of optical spectroscopic classifications for samples of radio AGN. 

In this article I review the considerable progress that has been made in the study of radio AGN and their host galaxies based on deep multi-wavelength observations with ground- and space-based observatories over the last 20 years. I take the approach of concentrating on modest-sized samples of radio AGN selected from radio surveys with relatively bright radio flux limits, in particular the southern 2Jy sample of \citet{dicken09}, and the northern 3CR\footnote{It is important to distinguish between the 3CR sample \citep[][see Table \ref{samples}]{bennett62a,bennett62b} and the 3CRR sample of \citet{laing83}. The latter has has more restrictive selection criteria: flux densities $S_{178MHz} > 10.9$~Jy, declinations $\delta > 10^{\circ}$, and Galactic latitudes $|b| > 10^{\circ}$; the 3CRR sample selection is also based on higher quality radio data.} sample
of \citet{buttiglione09}; full details of the selection criteria and references for these samples are given in Table \ref{samples}. I also concentrate on objects at low- to intermediate-redshifts ($z < 0.7$) because this ensures a high degree of completeness in optical spectroscopic classifications and in the detection of individual objects at X-ray and infrared wavelengths. I aim to fill an important gap between highly detailed studies of individual iconic radio AGN in the local universe such as Centaurus A and Cygnus A, and the more statistical, but less detailed, studies of much larger samples of radio AGN selected using a combination of deep wide field optical spectroscopic and radio
surveys.

No single review can encompass all aspects of radio AGN. In particular, this review will not consider the detailed physics of the synchrotron-emitting jets and lobes, radio galaxies at high redshifts ($z > 0.7$), or the feedback effect of the expanding radio components. These aspects are covered by excellent reviews elsewhere:  \citet{miley80} and \citet{worrall09} review the detailed jet/lobe physics from radio and X-ray perspectives; \citet{miley08}  review high redshift radio galaxies; \citet{mcnamara07} review  the impact of the radio sources on the hot ISM of the host galaxies and galaxy clusters; and 
\citet{fabian12} presents a broad overview of the AGN feedback effect. Readers might
also find the detailed reviews by \citet{israel98} and \citet{carilli96} on, respectively, the archetypal FRI and FRII sources Centaurus A and Cygnus A useful. Finally, \citet{heckman14} present a comprehensive overview of the key results on nearby AGN ($z <  0.2$) from the Sloan Digital Sky Survey (SDSS), which helps to
place radio AGN in the broader context of other AGN populations.

Throughout this review I assume a cosmology with $H_0 = 71$ km s$^{-1}$ Mpc$^{-1}$, $\Omega_m = 0.73$ and $\Omega_{\lambda} = 0.27$, and a spectral index of $\alpha = 0.7$ (for $F_{\nu} \propto \nu^{-\alpha}$) when converting radio fluxes between different frequencies. 

\section{AGN and jet properties}
\label{sec:1}

Before considering their detailed properties, it is important to start by defining
radio AGN as a class. One approach is to use the 
shapes of the AGN spectral energy distributions (SEDs), for example defining radio AGN to have a minimum ratio of radio to optical luminosity \citep[e.g. $R = L_{5GHz}/L_B > 10$:][]{kellermann89}. However, the disadvantage of this method is that the optical AGN luminosity estimates may be affected by dust obscuration --- a particular problem for narrow-line AGN --- and contamination
by the direct starlight of the host galaxies. There is also evidence that the 
radio-to-optical luminosity ratio that defines the boundary between
radio-loud and radio-quiet AGN increases substantially towards low AGN luminosities and Eddington ratios \citep{sikora07,chiaberge11}. Use of a mid-IR-to-radio flux ratio \citep[e.g. $q_{24} = L_{24\mu m}/L_{1.4 GHz}$:][]{appleton04,ibar08} would avoid the extinction problem; however, in this case there is the potential for strong contamination of the mid-IR luminosities by star formation in the host galaxies.

An alternative to the SED-based approach is to use a single radio power cut above which an AGN can be considered as
radio-loud. This is the approach adopted for this review: I define a radio AGN to have a monochromatic radio luminosity measured at 1.4~GHz of $L_{1.4GHz} > 10^{24}$~W Hz$^{-1}$. Although this cut may seem arbitrary given
that the general population of AGN shows a continuous range of radio power, it does have some physical basis in the sense that the populations of objects in the local universe that are generally considered radio-quiet in terms of AGN properties, such as starburst galaxies and classical, UV-selected Seyfert galaxies, have steeply declining luminosity functions above $L_{1.4GHz} = 10^{23}$~W Hz$^{-1}$; such objects are rare above
$L_{1.4GHz} = 10^{24}$~W Hz$^{-1}$ \citep{meurs84,condon89,sadler02}.

In what follows I will also sometimes draw a distinction between luminous, quasar-like AGN and their less luminous counterparts.
Originally quasars were defined by their star-like appearance in optical images. However, with the detection and characterisation of the 
underlying host galaxies, it has become clear that quasars represent the higher luminosity end of a general AGN
population that shows a continuous range of luminosities. Here I define  quasars to have bolometric luminosity $L_{bol} > 10^{38}$~W. Roughly, this corresponds
to 2 -- 10 keV X-ray luminosities $L_{X-ray} > 10^{37}$~W, optical absolute magnitudes $M_B < -23$ and
optical [OIII]$\lambda$5007 emission line luminosities $L_{[OIII]} > 10^{35}$~W, depending on the precise SEDs and
bolometric correction factors assumed. 
However, I will not distinguish between broad-line radio galaxies (BLRG: see below for description) and radio-loud quasars, because there is a significant overlap in the properties of these two groups, and some BLRG show evidence for relatively high levels of dust extinction \citep[e.g.][]{osterbrock76} consistent with them being partially obscured quasars; I will refer to these objects collectively as BLRG/Q sources, but they also sometimes
labelled broad-line objects (BLO).

\begin{table}
\begin{center}
\begin{tabular}{lll}
\hline \\
{\bf Abbr.} &{\bf Meaning} &{\bf Ref} \\
\hline \\
NLRG &Narrow-line radio galaxy &1 \\
BLRG &Broad-line radio galaxy &2 \\
WLRG &Weak-line radio galaxy &3 \\
SLRG &Strong-line radio galaxy &4 \\
Quasar &Quasi-stellar radio source &5 \\
LEG &Low-excitation galaxy &6 \\
HEG &High-excitation galaxy &6 \\
ELEG &Extreme low-excitation galaxy &6 \\
BLRQ/Q &Broad-line radio galaxy or quasar &7 \\
BLO &Broad-line object &6 \\
OVV &Optically violently variable (quasar) &8 \\
\hline 
FRI &Fanaroff-Riley class I source &9 \\
FRII &Fanaroff-Riley class II source &9 \\
FR0 &Fanaroff-Riley class 0 source &10 \\
FSRQ &Flat-spectrum radio-loud quasar &11 \\
SSRQ &Steep-spectrum radio-loud quasar &11 \\
CSS &Compact steep spectrum radio source &12 \\
GPS &Gigahertz-peaked radio source &13 \\
FD &Fat-double radio source &14 \\
RD &Relaxed-double radio source &15 \\
\hline \\
\end{tabular}
\caption{Summary of the main abbreviations of the labels used to classify radio AGN. The top half of the
table relates to optical classifications, while the lower half relates to radio classifications. The final column gives references to some of the first uses of the labels. Reference key: 1. \citet{costero77}; 2. \citet{osterbrock76}; 3.\citet{tadhunter98}; 4. \citet{dicken14}; 5. \citet{schmidt63}; 5. \citet{buttiglione10}; 7. \citet{dicken09}; 8. \citet{penston71}; 9. \citet{fanaroff74}; 10. \citet{ghisellini11}; 11. \citet{urry95}; 12. \citet{fanti90}; 13. \citet{odea91}; 14. \citet{owen89}; 15. \citep{leahy93}.
Note that LEGs and HEGs are sometimes labelled LERGs (low excitation radio galaxies) and HERGs (high excitation radio galaxies)
in the literature.}
\label{abbr}
\end{center}
\end{table}

\subsection{Classification}
\label{sec:1.1}

In many sciences, including astronomy, the classification of a phenomenon often precedes its
interpretation in terms of the underlying mechanisms. Faced with a diverse phenomenon, the process of classification
is an attempt to sort things out, a first step in elucidating underlying patterns of behaviour. While the classifications
themselves remain relatively fixed, their interpretation in terms of the underlying physical processes, structures, geometries etc.
may not be unique and may change with time. For these reasons it is important to keep the classifications separate from
the interpretations. In this sub-section I  summarise the radio and optical classifications of
radio AGN; possible interpretations are then discussed in the following sub-sections. For ease of reference, the  labels used to classify radio AGN are summarised in Table \ref{abbr}.

\begin{figure}
\includegraphics[width=12.0cm]{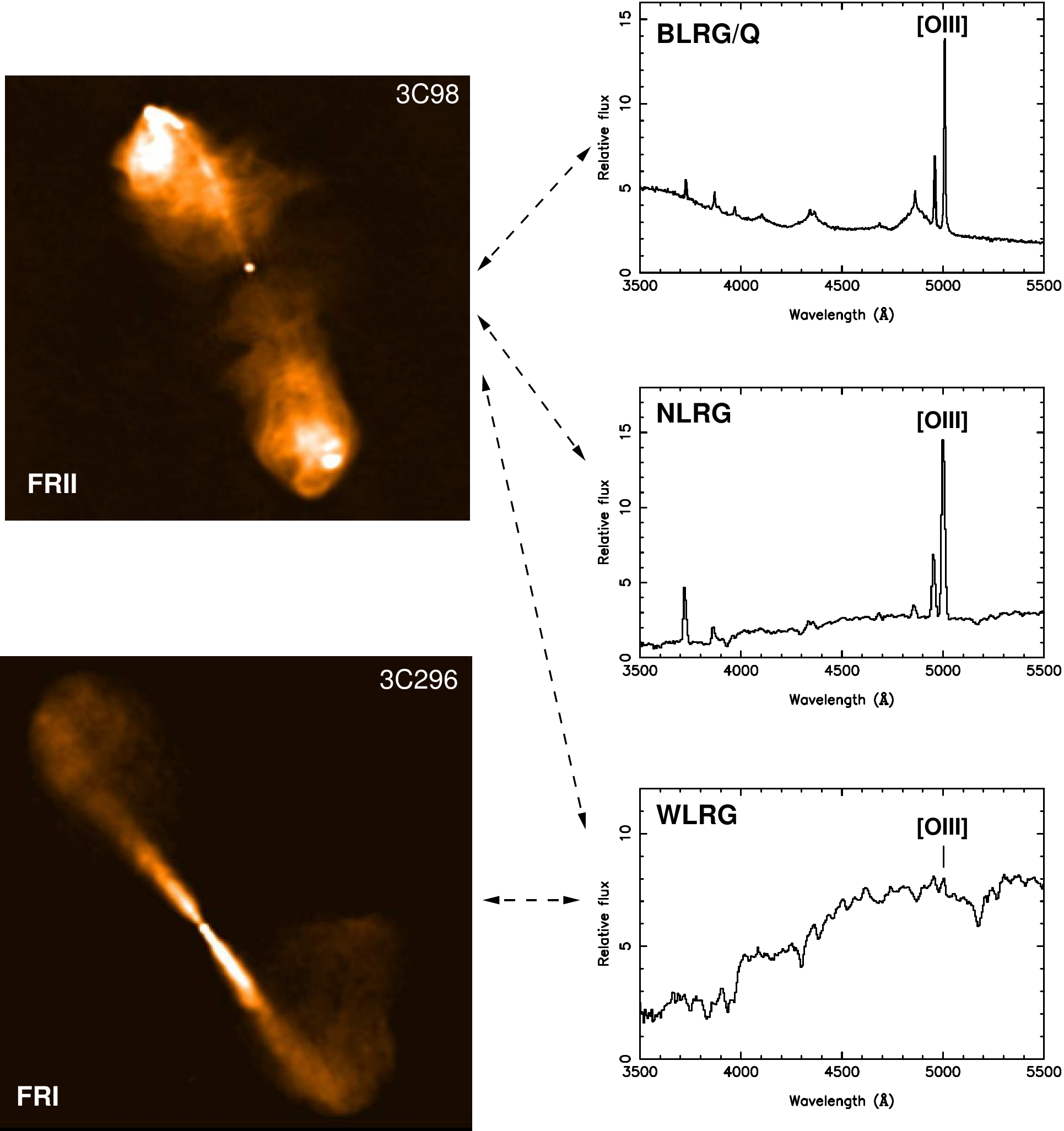}
\caption{The main radio and optical classifications of radio AGN. Left: radio morphological classifications. Right: optical spectroscopic classifications. The dashed lines indicate
links between the radio and optical classifications. Credits: the radio images were taken from
http://www.jb.man.ac.uk/atlas.}
\label{class}       
\end{figure}

An interesting feature of classification in astronomy is that a particular object may me classified in different
ways at different wavelengths. For radio AGN some of the greatest insights have been obtained when attempting to understand the
relationships between their classifications at the different wavelengths.
The main radio and optical classifications of radio AGN are illustrated in
Figure \ref{class}.

The radio classification of radio AGN followed the development of large radio interferometers in the 1970s that were able to map the extended radio-emitting structures in some detail. It was found that extragalactic radio sources are generally double in character, with synchrotron-emitting lobes that sometimes contain
compact higher surface brightness concentrations called hot spots, connected to radio cores sources
in the nuclei of the galaxies by jets \citep[e.g.][]{miley80}. The radio structures often extend on scales that are larger than the host galaxies ($\sim$50~kpc to 1~Mpc). 

One of the
main radio morphological classifications is that of \citet{fanaroff74}, who
divided objects according to whether the distance between the highest surface brightness extended radio features 
on either side of the nucleus was more or less than 50\% of the total diameter of the radio source, corresponding
to the FRII and FRI classifications respectively (see Figure \ref{class} for examples). Fanaroff \& Riley showed that the FRII sources have higher radio powers on average than FRI sources, with the division between the two types occurring at a radio power of $L_{178MHz} \sim 10^{26}$ W Hz$^{-1}$ (or $L_{1.4GHz} \sim 3\times10^{25}$ W Hz$^{-1}$). However, it was already recognised by Fanaroff \& Riley
that a minority of sources ($\sim$17\% in their sample) could not be unambiguously assigned to either the FRI or the
FRII class. Certainly, with the higher resolution radio maps available since the early 1980s it has become clear that 
some sources have an apparently hybrid FRI/FRII character (see the example of Hercules A in Figure \ref{hera}). Also, the radio power
division is not  sharp: some sources with an FRII morphology have radio powers well below
$L_{178MHz} = 10^{26}$ W Hz$^{-1}$, and vice versa. Some ambiguity in this classification may be caused by the fact that,
with the availability of higher resolution radio maps,
the original quantitative criterion of Fanaroff \& Riley has often been replaced by a subjective assessment of whether the morphology of a source appears ``edge brightened'' or ``edge darkened''.

Although the FRI/FRII classification is the most common, other radio morphological classifications are also used\footnote{See http:$//$www.jb.man.ac.uk$/$atlas 
for a more detailed discussion of radio 
classifications schemes.}. For example, some radio sources resemble double-lobed, edge-brightened FRII sources, with similar steep-spectrum
radio synchrotron emission, but have much smaller linear diameters. The class of compact steep-spectrum (CSS) sources has been defined \citep[see][]{fanti90,odea98} to include steep spectrum radio sources with total diameters $D \le 20$~kpc.  Further diversity in the radio classification is added by objects that, rather than showing the classical double-lobed
morphology, exhibit a core-jet or core-halo structure \citep[e.g.][]{antonucci85a}. Finally,
some extended double-lobed radio sources have been classified as ``fat doubles'' (FD) based on the fact that their lobes are unusually extended in the direction
perpendicular to the axis defined by the radio jets \citep{owen89}. Such sources are also sometimes labelled
``relaxed doubles'' (RD) on the basis that their radio lobes lack bright hotspots and jets \citep{leahy93}.  However, both the fat and the relaxed classifications lack clear, quantitative 
definitions.

In addition to the morphological classifications, radio sources are also classified on the
basis of their long wavelength SED shapes, often quantified in terms of the spectral index ($\alpha$) of a power law fitted to their radio spectra when expressed in frequency units ($F_\nu \propto \nu^{-\alpha}$). Flat- and steep-spectrum radio
sources are defined to have spectral indices smaller or greater than a particular limiting value (typically $\alpha_{lim} \sim 0.5$,
but definitions vary). It is notable that, whereas steep-spectrum radio sources are commonly associated with FRI, FRII
or CSS radio morphologies, flat-spectrum sources are generally core dominated, appearing as core-jet or core-halo
sources in high resolution very long baseline interferometry (VLBI) observations \citep[e.g.][]{antonucci85a}. However, some
radio sources have more complex radio spectra that cannot be approximated using a simple power-law fit. In particular, the
GHz-peaked sources (GPS) have radio spectra that peak at GHz frequencies and decline at higher and lower frequencies \citep{bolton63}; such spectra are often attributed to synchrotron self-absorption effects \citep{kellermann66}.
Morphologically,  GPS sources often show double-lobed radio structures that are even more compact
($D \le 1$~kpc) than those of CSS sources\footnote{ Since the GPS and related CSS sources show strong morphological similarities with their more extended counterparts \citep{odea98,tzioumis02}, and are thought to represent radio sources in a young evolutionary phase \citep{fanti95,owsianik98,polatidis03}, in most of this review I will not distinguish them from the extended sources. However, note that the detection of GSP/CSS sources in flux-limited samples may be affected by 
selection effects related to the strong interactions between the compact radio sources and dense ISM in the central regions of the galaxies \citep{tadhunter11,morganti11,dicken12}.}.

Most  optical  classifications of radio AGN rest on their emission line properties. 
In the 1970s Osterbrock and colleagues identified a host of 
narrow emission lines in several radio galaxies that they used to measure key properties of the narrow line region (NLR) such
as reddening, density and temperature \citep[e.g.][]{osterbrock75}. Paralleling the classification of Seyfert galaxies
into Seyfert type 1 and Seyfert type 2, based on respectively the presence or absence of broad ($FWHM > 2000$~km s$^{-1}$) permitted lines in the spectra \citep{khachikian71}, they also divided the radio galaxies into two types:
broad-line radio galaxies \citep[BLRG][]{osterbrock76} and narrow-line radio galaxies \citep[NLRG:][]{costero77}, with BLRG showing spectral similarities to
quasars, but having less luminous AGN. Collectively the NLRG and BLRG/Q spectroscopic classes can be labelled as strong-line radio galaxies (SLRG).

By the late 1970s, however, it became clear that the NLRG/BLRG/Q classification does not capture the full spectral diversity of the radio AGN population: in a study of 3CR radio galaxies \citet{hine79} recognised a new class of objects (labelled ``Class B'') that have ``only the absorption spectra typical of giant elliptical galaxies or else very weak [OII]$\lambda$3727''.
Understanding these objects, and how they relate to the SLRG, has been a major focus of
research on radio AGN in the last 20 years; over this period improved spectroscopic data has allowed 
more quantitative classification schemes to be developed. Such objects are now identified in samples of radio AGN on the
basis of their low [OIII]$\lambda$5007 equivalent widths (e.g. the weak-line radio
galaxy [WLRG] classification of Tadhunter et al. 1998) or their low excitation/ionization emission line spectra (e.g. the low excitation galaxy [LEG] class of Buttiglione et al. 2010), or one or other of [OIII] equivalent width and excitation/ionization criteria \citep{laing94,jackson97,best12}. 

The different methods used to classify the Class B/WLRG/LEG objects have the potential to lead to  ambiguity. Fortunately, most objects classified as WLRG on the basis of their low equivalent width [OIII] emission would be also be classified as LEG on the basis of the low excitation/ionization of their emission line spectra, as quantified using emission line ratios and diagnostic diagrams. Indeed, the [OIII] equivalent width is strongly correlated with excitation/ionization state in the radio AGN population (see Figure 1 of Best et al. 2012). However,
the LEG and WLRG classifications are not exactly the same, and a minority of objects classified as LEG are not classified as WLRG. For example,
in the study of \citet{buttiglione10} a few  of the objects classified as LEGs on the basis of their low excitation/ionization emission line spectra have relatively strong, high equivalent width [OIII] emission lines (e.g. 3C84\footnote{N.B. This object would also be classified as a BLRG
on the basis of its broad Balmer emission lines.}, 3C153, 3C196.1, 3C349). 

The data in Buttiglione et al. (2009,2011) can be used to make a more quantitative comparison between the classification schemes. Taking the 99 3CR radio AGN with redshifts $z < 0.3$ in the Buttiglione et al. (2010,2011) study with secure spectroscopic classifications, 2\% are classified as star forming objects, 46\% as LEGs\footnote{This includes the
3 objects classified by \citet{capetti11} as extreme low excitation radio galaxies (ELEGs).}, 33\% as
high excitation galaxies (HEGs), and 18\% as broad line objects (BLOs). In comparison, according to the WLRG/NLRG/BLRG/Q classification scheme of \citet{tadhunter98}, 32\% of the same objects would be classified as WLRG, 48\% as NLRG and 19\% as BLRG/Q objects. Overall, 65\% of the 43 objects classified as LEGs by Buttiglione et al. would be classified as WLRG based on their [OIII] equivalent withs. 

\begin{figure}
\includegraphics[width=12.0cm]{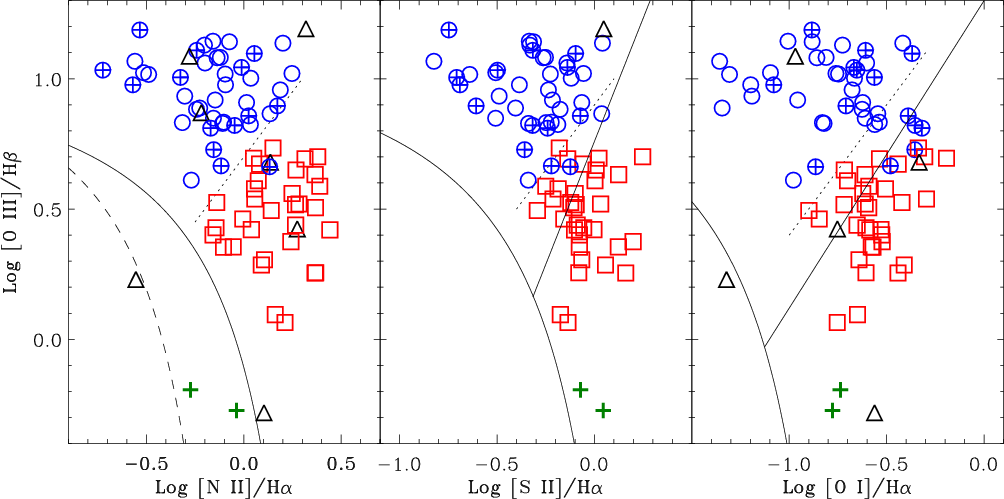}
\caption{Emisssion line ratio diagnostic diagrams for 3CR radio galaxies from \citet{buttiglione10}. In all of
the diagrams the blue circles indicate HEGs, the red squares LEGs, the green crosses objects
in which the ionization of the NLR gas has a significant contribution from stellar photoionization, and the black triangles indicate objects that lack measurements of one of the diagnostic ratios. The dotted line indicates the division between LEGs and HEGs, and the curved black line shows the division between AGN and star forming galaxies. In the $[NII]/H\alpha$ vs $[OIII]/H\beta$ diagram the area between the curved solid and dashed lines indicates the region containg objects with composite AGN/stellar photoionization spectra, while the inclined solid black lines in the other two diagrams indicate the division between Seyfert galaxies and LINERs \citep[from][]{kewley06}. Credits: this figure was originally published as Figure 7 in \citet{buttiglione10}. 
}
\label{diagnostics}       
\end{figure}

An advantage of the \citet{buttiglione10} scheme is that it is based on a clear, quantitative criterion that is motivated by the positions of the points on emission line diagnostic diagrams (see Figure \ref{diagnostics}).
In this scheme LEGs and HEGs are classified according to their excitation indices ($EI$):
\begin{eqnarray}
EI & =  & log([OIII]/H\beta) - 1/3(log([NII]/H\alpha)+ \\
   &    & log([SII]/H\alpha)+log([OI]/H\alpha)) \nonumber
\end{eqnarray}
where $[OIII]/H\beta$ represents the ratio of the flux of the [OIII]$\lambda$5007 line to that of the $H\beta$ line, while $[NII]/H\alpha$, $[SII]/H\alpha$ and $[OI]/H\alpha$ represent respectively the ratios of the fluxes of [NII]$\lambda$6583, [SII]$\lambda\lambda$6717,6731 and [OI]$\lambda$6300 lines to that of H$\alpha$; LEGs are defined to have $EI < 0.95$\footnote{\citet{buttiglione10} and \citet{capetti13} also distinguish a class of radio AGN
showing extremely low excitation emission line spectra that they label extreme low excitation galaxies (ELEGs). However,
in what follows I make no distinction between LEGs and ELEGs and label them collectively as LEGs.},  while HEGs are defined to have $EI > 0.95$.
Importantly, there is evidence that the distribution of excitation class in the sample is bimodal, with one peak representing the HEGs/BLO and the other the LEGs. However, a disadvantage of this method is that it requires accurate  measurements of all the
lines involved in the excitation index. For objects at the lower end of the [OIII] equivalent width (EW) distribution this information is difficult to obtain, since several (or all) of the emission lines are  undetected or have flux measurements with large uncertainties. For this reason, 14 (12\%) of the 113 3CR objects with redshifts $z < 0.3$ in the full sample of Buttiglione et al. (2010,2011), lack spectroscopic classifications. The latter objects would be classified as WLRG in the Tadhunter et al. (1998) scheme, leading to 43\% of the full sample of Buttiglione et al. (2009,2010,2011) being classified as WLRG. 

\begin{table}
\begin{center}
\begin{tabular}{lllll}
Type    &Sample (N)            & \%WLRG         & \%NLRG         & \%BLRG/QSO  \\
\hline \\  
FRI &3CR  (22)            &100                 &0               &0  \\
    &2Jy  (15)         &100                 &0               &0  \\
FRII &3CR (78)            &24              &54                 &21   \\
     &2Jy (39)         &23                 &41              &36 \\
CSS/GPS &3CR (4)       &25                  &75                 &0   \\
        &2Jy (7)       &0                  &71               &29 \\
FRI/FRII &3CR (5)      &60                  &40                 &0   \\
          &2Jy (4)     &100                  &0                 &0   \\
Other    &3CR (4)      &25                   &25        &50 \\
         &2Jy (2)      &0                    &0                &100 \\          
\end{tabular}
\caption{Comparisons between the radio and optical classifications 
for objects in the $z < 0.3$ 3CR sample (Buttiglione et al. 2009, 2010, 2011) and the $z < 0.7$ 2Jy sample \citep{tadhunter98}. In the second column the number in brackets gives the size of the sub-sample. Note that
the ``Other'' classification in the final two rows includes objects with core/halo, core/jet or uncertain, often irregular, radio morphologies that defy clear classification, and objects without the high resolution data required
to make the morphological classification. Note that two of the NLRG objects have narrow line ratios that lead to their
classification as star forming (SF) objects in \citet{buttiglione10}.}
\label{class_comparison}
\end{center}
\end{table}

Therefore a major advantage of the alternative technique of using an upper limiting [OIII] EW is that it captures the objects with low EW or undetected [OIII] that cannot be classified on the basis of excitation class. On the other hand, using a particular [OIII] equivalent width limit ($EW_{[OIII]} < 10$~\AA\, in the case of Tadhunter et al. 1998) is  somewhat arbitrary because, while there is some 
evidence for bi-modality in the [OIII] emission line luminosities and equivalent withs for the high radio power end of the 2Jy sample, at low radio powers the distribution [OIII] equivalent widths appears continuous. Also, the equivalent width of [OIII] depends not only on the [OIII] emission line flux but also on the flux of the underlying stellar continuum. The latter may vary from
object to object (for a given [OIII] luminosity), and be subject to aperture and reddening effects.

An important general point about the LEG and WLRG classes is that, while their emission line ratios are similar to the those of the LINER class of radio-quiet AGN \citep{heckman80}, there are differences in terms of emission line luminosity. As pointed out by \citet{capetti13}, many of the LEGs in the 3CR sample have emission line luminosities that are comparable with those of Seyfert galaxies, but much higher than those of LINERs; the same is true of many of the 2Jy radio AGN classified as WLRG by \citet{tadhunter98}.


One further classification method is based on optical variability. A subset of objects --- labelled blazars --- exhibit extreme variability at optical wavelengths. This class includes BL Lac objects and optically violently variable (OVV) quasars, with the two groups separated on the basis that BL Lac objects show emission lines that are weak relative to the underlying featureless continuum (e.g. $EW < 5$\AA\,: Urry \& Padovani 1995), whereas OVV quasars exhibit broad emission lines as well as a strong non-stellar continuum; a relatively high degree of optical polarisation is also common in both of the latter groups. Moreover, most such objects are
flat-spectrum radio sources with core/halo or core/jet morphologies. 

Finally, it is important to consider the degree to which the various optical and radio classifications are correlated. Table \ref{class_comparison} shows the frequency of the different optical spectroscopic classifications for FRI, FRII, CSS/GPS, and hybrid FRI/FRII or uncertain radio morphological classifications in both the full $z < 0.3$ 3CR sample of Buttiglione et al. (2009,2011: 113 objects) and the full $z < 0.7$ 2Jy sample of Tadhunter et al. (1998: 66 objects)\footnote{Note that this sample extends to lower redshifts than the $0.05 < z < 0.7$ 2Jy sample of \citet{dicken09} described in Table 1}. The most striking feature of this comparison is that {\it all}  the objects unambiguously classified as  FRI radio sources are  WLRG; none are classified as BLRG/Q\footnote{One object deserves particular mention here: 3C84 (also known as NGC1275). This object was originally classified as an FRI by \citet{fanaroff74}, but shows broad Balmer emission that lead to a BLRG optical classification. Also, as expected for a BLRG, it has luminous narrow lines, albeit of low ionization (hence the LEG classification of 3C84 by Buttilgione et al. 2010). On the basis of these properties, 3C84 could be considered as the {\it only} FRI object in the combined 3CR and 2Jy
sample classified as a SLRG. However, its radio structure is highly peculiar, with a strong, highly variable flat spectrum core, an inner steep spectrum double structure, and an outer halo \citep{pedlar90}. Therefore, its radio morphological classification must be regarded as uncertain; it is certainly not a typical FRI source.}. In addition, all SLRG have FRII or CSS/GPS radio morphologies, apart from a small minority that have hybrid FRI/FRII or ambiguous morphologies. It is also clear that, while the majority of FRII sources are SLRG, a significant minority ($\sim$24\%) are WLRG.

\subsection{Orientation-based unification I: SLRG}
\label{sec:1.2}

The orientation-based unified schemes were developed following the recognition in the late 1970s and early 1980s that the continuum emission from AGN is highly anisotropic. In the case of radio AGN, there are two types of anisotropy to consider. First, the {\it beaming} of the non-thermal synchrotron emission due to the bulk relativistic motions of the jets in the core regions of the sources. Direct evidence for the bulk relativistic motions and beaming is provided by the detection of apparently superluminal velocities in VLBI observations of the radio cores of blazar-like objects \citep{cohen77}, and the detection of a polarisation asymmetries in the radio lobes of radio-loud quasars that correlate with sidedness of the inner radio jets \citep{garrington88,laing88}. Second, the {\it blocking} effect of dust and gas in the circum-nuclear dust structures that are often characterised as tori. The presence of this form of anisotropy is supported by detection of significant optical polarisation that is aligned perpendicular to the radio axes in nearby radio galaxies \citep{antonucci82,antonucci84}, the detection of broad emission lines in the polarised intensity spectra of nearby Seyfert galaxies  and  NLRG \citep{antonucci84,antonucci85b,ogle97,cohen99}, and
the detection of ``ionisation cones'' in narrow-band emission line images of Seyfert galaxies \citep[e.g.][]{pogge88,tadhunter89b} and radio galaxies \citep{jackson98}. The effect of such blocking is to heavily attenuate both
the continuum and broad line region (BLR) emission of the AGN at UV and optical wavelengths in objects for which the radio
jet/torus axis is pointing at a large angle to the line of sight.

\begin{figure}
\includegraphics[width=12.0cm]{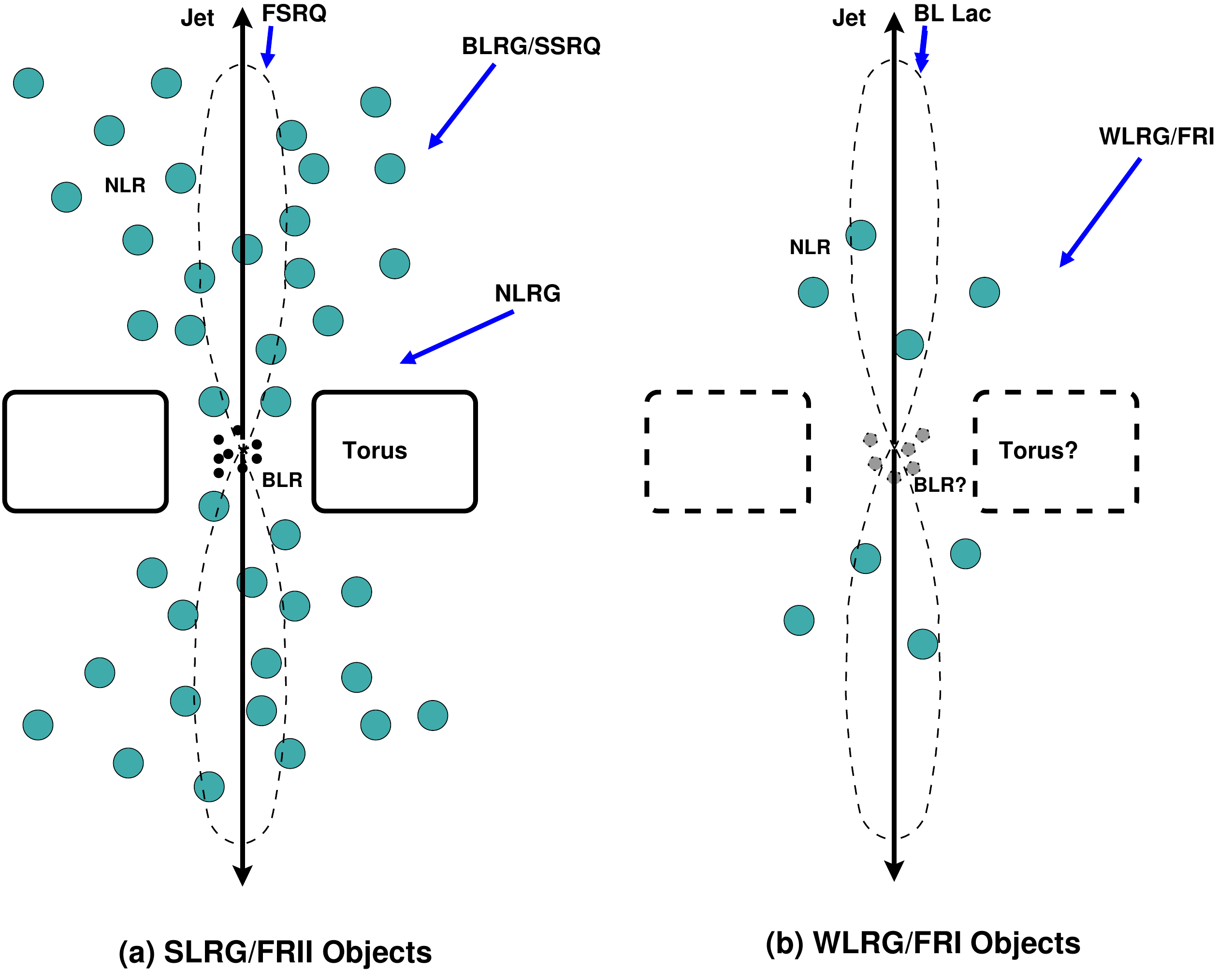}
\caption{Schematic showing the main elements of the orientation-based unified schmes for (a) SLRG/FRII objects, and (b) WLRG/FRI objects. The blue solid arrows indicate the directions from which different classes of objects might be observed, while curved dashed lines indicate the polar diagram of the beamed radio jet emission; 
narrow-line region (NLR) clouds are indicated by turquoise circles.}
\label{unification}       
\end{figure}

The anisotropic AGN continuum emission can help to explain some of the observed diversity in the properties of the radio AGN. Perhaps the most ambitious anisotropy-based unified scheme for radio AGN involves both forms of anisotropy, and attempts to explain the relationship between steep-spectrum radio-loud quasars (SSRQ), flat-spectrum radio-loud quasars  (FSRQ), and radio galaxies,  \citep{barthel89}. This scheme is shown schematically in Figure \ref{unification}(a). The basic idea is that, for a given radio power, the different classes of radio sources are all drawn from the same parent population with similar
central AGN properties, and the differences between their radio and optical properties are explained in terms of anisotropy and orientation. In this scheme, the relative proportions of FSRQ, SSRQ/BLRG and NLRG in a particular radio AGN sample is set by the opening
half-angles of both the torus ($\theta_{tor}$) and the beaming cones of the relativistic jets  ($\theta_{beam} \sim \Gamma^{-1}$, where $\Gamma$ is the Lorentz factor of the bulk relativistic motions in the jet). Recent analysis of the radio and optical properties of the HEG and BLO with FRII morphologies
in the $z < 0.3$ 3CR sample is consistent with $\theta_{tor} = 50\pm5$ degrees and $\Gamma \sim$3 --- 5
\citep{baldi13}.

One way to test such schemes is to compare the statistical properties of the different classes of radio AGN in flux-limited samples selected in well-defined redshift/radio luminosity ranges. For example, \citet{barthel89} compared
the distributions  of the measured (i.e. projected) linear sizes of the radio sources of the radio galaxies and the quasars in his sample of intermediate- to high-redshift ($0.5 < z < 1$) radio galaxies, and found that the radio galaxies have larger diameter radio sources on average than the quasars, as expected in the case that we have a foreshortened view of the radio sources of the quasars. However,  attempts to repeat the linear diameter test on other samples of radio sources, albeit samples selected in different redshift ranges and/or with different radio flux limits, have
failed to find significant differences between the linear size distributions of radio galaxies and quasars \citep{singal93,kapahi95}. 
Note that, this lack of difference between the linear size distributions is not necessarily evidence against the orientation-based unified schemes, but could rather reflect selectional biases in the samples, incompleteness
in the optical classifications of the targets, mixing up different populations of radio sources, or the effects of the radio power evolution of the sources coupled with a luminosity dependence in the opening angle of the torus \citep[e.g.][]{gopal96}. 

Another statistical test involves comparing the radio powers of the extended radio lobes
of powerful radio galaxies and quasars. In this case, since the radio lobes are not thought to be undergoing bulk relativistic motions that would cause anisotropy in their radio emission, we would
expect the radio galaxies and quasars to have similar distributions of lobe power. In general, this is borne out by the observations \citep[e.g.][]{urry95}. Clearly, a major advantage of radio AGN for tests of the unified schemes is the isotropy of the emission from their radio lobes: samples selected on the basis of their low frequency radio emission --- assumed to be lobe-dominated --- can be considered orientation-independent. 

The recent dramatic improvement of the optical and infrared data for samples of nearby radio sources has also allowed tests of orientation-based unification based on the optical and mid-IR emission lines  and mid-IR continuum: if BLRG/Q and NLRG objects are drawn from the same parent population, on average the two groups should show similar luminosities in their narrow emission lines and mid-IR continua, assuming that the latter are emitted isotropically (i.e. they do not suffer attenuation by the circum-nuclear dust).

As one of the brightest optical emission lines that
represents the high ionization conditions typical of the NLR, the [OIII]$\lambda$5007 forbidden line was the first to be used in this way. Early results suggested that the quasars are up to an order of magnitude more luminous in [OIII] than NLRG  \citep{jackson90}. In contrast, the lower ionisation [OII]$\lambda$3727 line --- which is likely to be emitted on larger scales than the [OIII] --- showed no significant difference between the two groups \citep{hes93}. Therefore, rather than providing evidence against the unified schemes, it was proposed that all, or part, of the [OIII]-emitting NLR is emitted on a relatively small scale and is subject to attenuation by the circum-nuclear dust in NLRG. This is supported by the observation of variability in the [OIII] emission lines of the BLRG 3C390.3, which provides evidence that much of the [OIII] emission in that object is emitted on a scale $r < 10$~pc \citep{clavel87,zheng95}.
However, some of the early studies that compared the [OIII] emission line luminosities of radio galaxies and quasars failed to distinguish between WLRG and SLRG. If WLRG represent a separate class of
radio AGN with intrinsically lower luminosity AGN (see discussion in section 2.4 below), this could lead to the apparent differences between the [OIII] luminosities of narrow- and broad-line objects being exaggerated \citep{laing94}.

\begin{figure}
\begin{center}
\includegraphics[width=10.0cm]{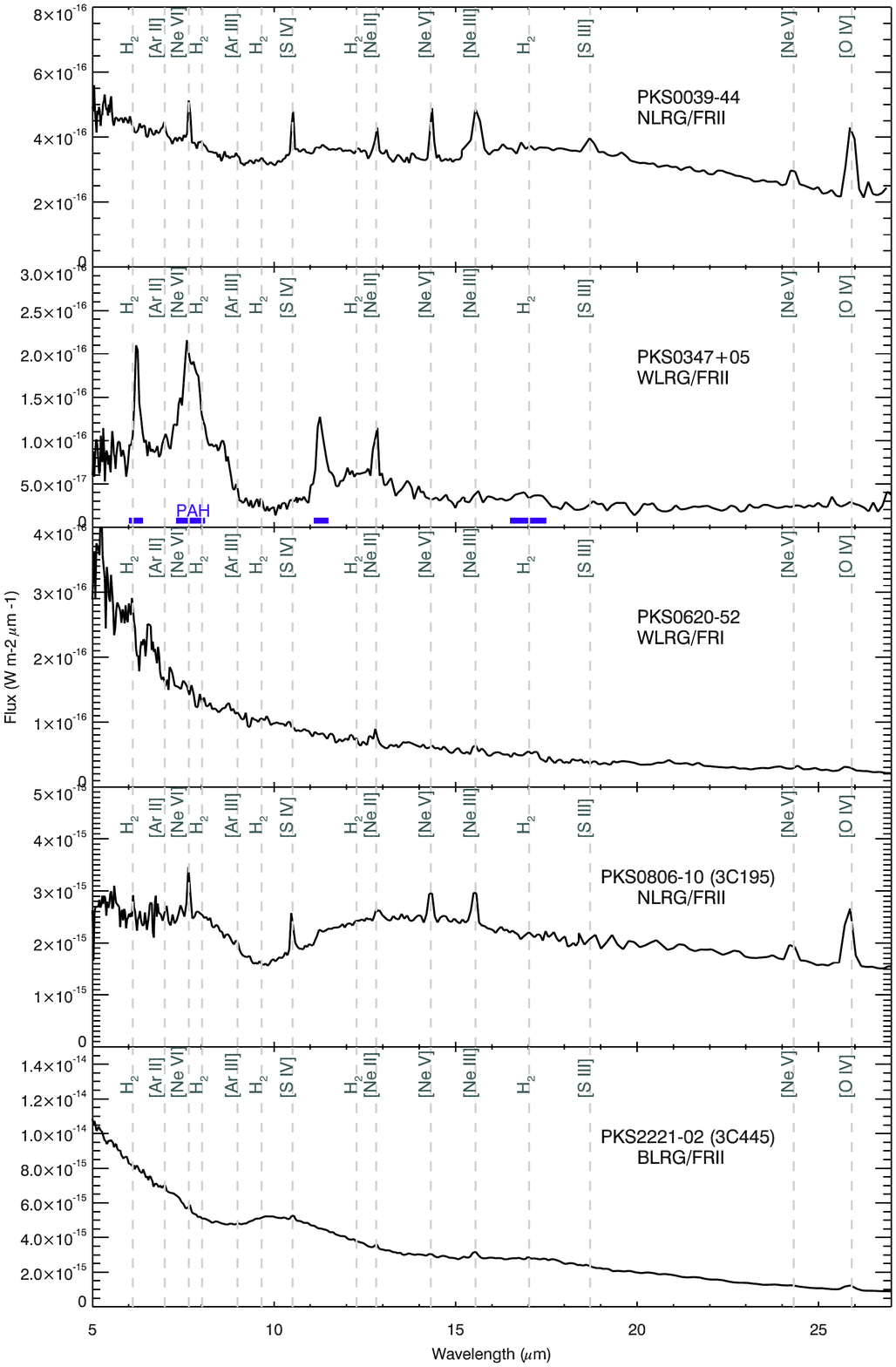}
\caption{Example mid-IR spectra of radio AGN from the 2Jy sample, illustrating the
diversity in the observed spectra. Various mid-IR fine-structure and PAH features are indicated. 
See \citet{dicken12,dicken14} for further details.}
\label{mir_spectra}       
\end{center}
\end{figure}

\begin{figure}
\includegraphics[width=12.0cm]{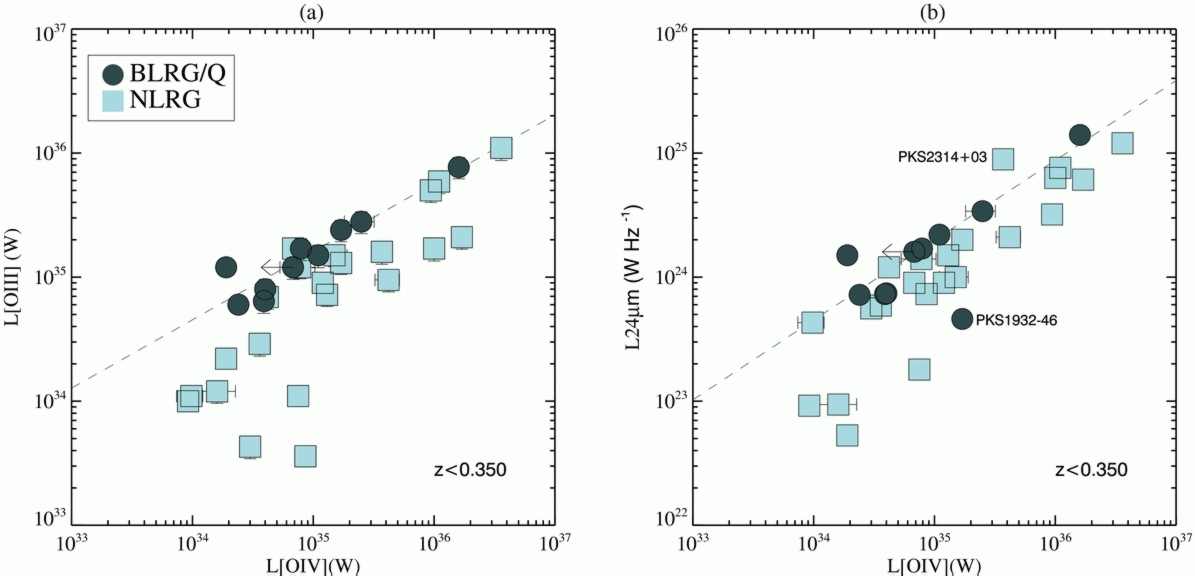}
\caption{Correlation plots showing comparisons between different optical and mid-IR
inidicators of intrinsic AGN power, highlighting the differences between NLRG and BLRG objects the combined 2Jy and 3CRR sample of \citep{dicken14}: (a) $L_{[OIII]}$ vs $L_{[OIV]}$, and (b) $L_{24\mu m}$ vs $L_{[OIV]}$. The dashed
lines show the linear regression fits to the BLRG/Q points. Of the two
objects highlighted in the right-hand plot, the NLRG PKS2314+03 is an object with prodigious starburst activity for which the 24$\mu$m emission is likely to be boosted by starburst heating of the near-nuclear dust, while it is suspected that the BLRG PKS1932-46 is under-luminous in 24$\mu$m emission because its AGN has recently entered a low activity state \citep{inskip07}. 
See \citet{dicken14} for further details.}
\label{mir_corr_slrg}       
\end{figure}

Deep mid-IR spectroscopy observations taken with the IRS spectrograph on the Spitzer satellite have deepened our understanding of the degree of attenuation suffered by the [OIII] narrow line  and the mid-IR continuum emission \citep{haas05,dicken14}. Figure \ref{mir_spectra} shows example Spitzer/IRS spectra for radio AGN from the 2Jy sample and illustrates the diversity in the mid-IR spectra of radio AGN.  Such observations have enabled accurate measurements to be made of the mid-IR [NeIII]$\lambda$15.6$\mu$m and
[OIV]$\lambda$25.9$\mu$m fine structure lines, which are much less likely to suffer dust extinction
than the optical forbidden lines. Figure \ref{mir_corr_slrg} shows the [OIII] emission line and 24$\mu$m continuum luminosities plotted against the [OIV]$\lambda$25.9$\mu$m luminosity for complete sub-samples of 3CRR and 2Jy SLRG, with NLRG and BLRG/Q indicated by different symbols. It is clear that, on average, the NLRG fall below the correlations defined by the BLRG/Q. Under the reasonable assumption that the [OIV] emission is not significantly attenuated, this suggests that both the [OIII] and 24$\mu$m emission suffer a similar mild attenuation by factor of $\sim$2 --- 3 \citep{dicken14}\footnote{ Note that Haas et al. (2005) derive a larger attenuation factor of $\sim$7 based on a smaller, more heterogeneous sample}. This is similar to the level of [OIII] extinction deduced by \citet{baum10} and \citet{lamassa10} for samples of nearby Seyfert 2 galaxies using Spitzer/IRS data.

In fact, it is remarkable that the degree of attenuation is apparently so similar for the [OIII] and 24$\mu$m emission in the radio AGN, given that the equivalent dust extinctions expressed in magnitudes in the V-band, are $A_v \sim 1$ magnitudes and $A_v \sim 20$ magnitudes for the [OIII] emission line and 24$\mu$m continuum respectively. Therefore, if the results are interpreted solely in terms of extinction,
the [OIII] and 24$\mu$m emission must be extinguished by different dust structures, or by different parts
of the same dust structure.  
For example, the 24$\mu$m continuum could be extinguished by the dust in the compact inner parts of the torus, and the [OIII] 
by larger-scale dust, perhaps in the outer parts of the torus, or a kpc-scale 
dust lane in the host galaxy. 
However, it is not possible to entirely rule out the possibility that, rather than being due to extinction, the difference between the 24$\mu$m luminosities of the NLRG and BLRG/Q objects at fixed [OIV] luminosity is due
to contamination by beamed synchrotron emission from the radio cores, given that the degree of synchrotron contamination would be expected to be larger in the case of the BLRG/Q objects, because their jets are pointing closer to the line of sight.

\begin{figure}
\includegraphics[width=12.0cm]{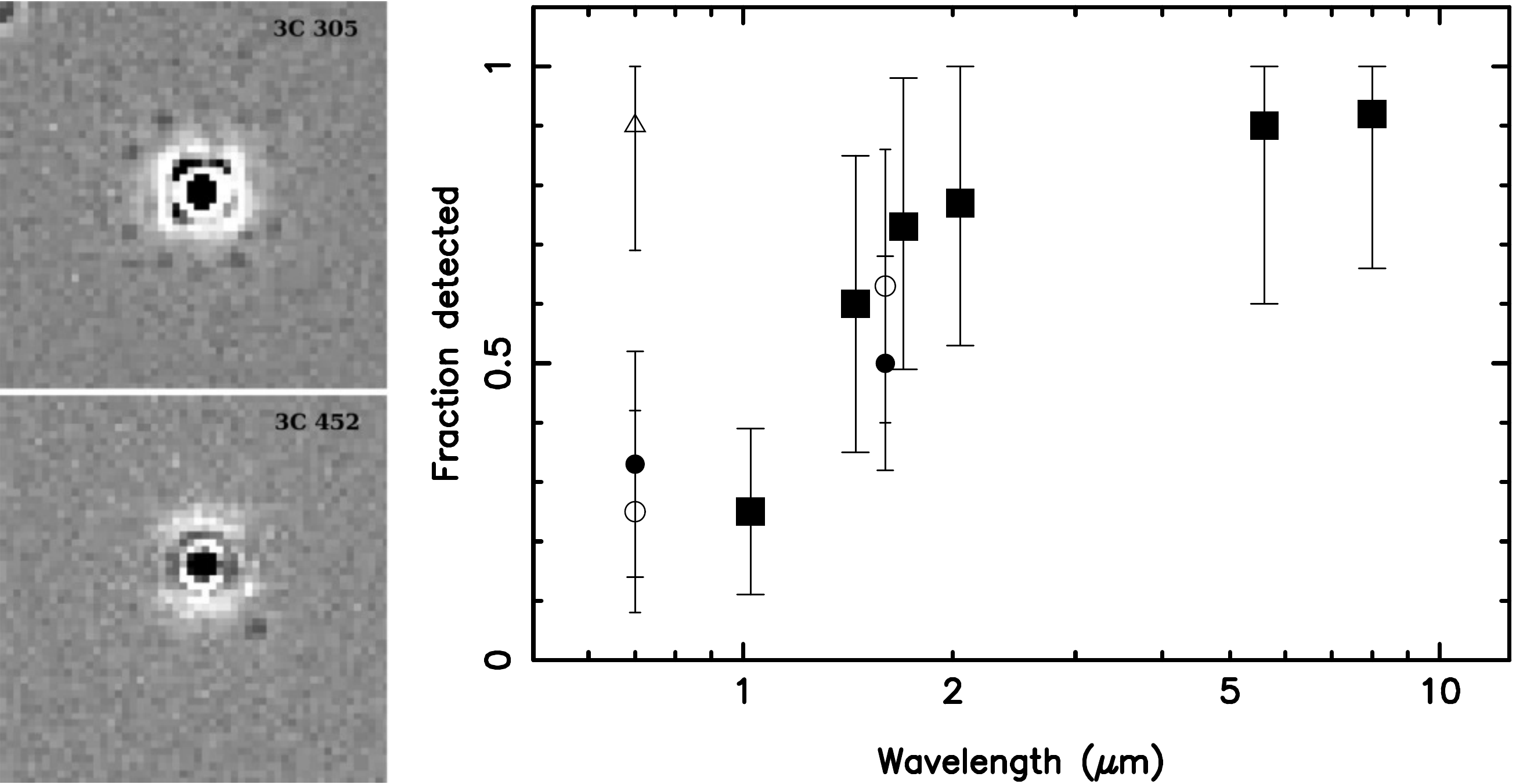}
\caption{The detection of compact cores in NLRG/HEG and WLRG/LEG from the 3CR and 3CRR samples. Left: the  grey-scale images show example 2.05$\mu$m images of 3CRR galaxies from \citep{ramirez14a} for which models of the smooth underlying starlight have been subtracted to highlight the Airy rings surrounding the nuclear point sources. Right: the detection rate of nuclei as a function of wavelength in 3CR and 3CRR HEG and LEG galaxies at $z < 0.11$. Key for the points: black squares represent 3CRR WLRG/FRII sources from \cite{ramirez14a}; black circles represent 3CR NLRG/FRII from 
\citet{baldi10} and \citet{chiaberge02}; open circles represent WLRG/FRII sources from \citet{baldi10} and \citet{chiaberge02}; and the open triangle represents WLRG/FRI sources from \citet{chiaberge99}. Note the increase in detection rate of compact cores with wavelength for the NLRG/FRII and
WLRG/FRII sources, and the fact that WLRG/FRI sources show a much higher detection rate of compact
cores at the shorter, optical wavelengths than the NLRG/FRII and WLRG/FRII sources. See \citet{ramirez14a} for details.}
\label{compact_cores}       
\end{figure}

An alternative technique involves using the high spatial resolution of the HST to investigate the rate of detection of compact core sources at both optical and infrared wavelengths in samples of nearby radio sources. The bright AGN nuclei are expected to be heavily attenuated at optical wavelengths by dust in the torus for objects observed as NLRG. However, due to the decrease in dust extinction towards longer wavelengths, the torus will become more transparent in the near-IR, potentially allowing the direct detection of the compact nuclei. Early attempts to detect the compact near-IR nuclei in radio galaxies using ground-based observations produced some promising results for a few individual radio galaxies \citep{djorgovski91,simpson95}, but were hampered by the difficulty of separating the AGN nuclei from the starlight, given the relatively modest spatial resolution. Therefore the HST, with its order of magnitude better spatial resolution, has revolutionised this area \citep[see][]{baldi10,ramirez14a}. Figure  \ref{compact_cores} shows the detection rate of AGN nuclei in low redshift ($z < 0.11$) NLRG as a function of wavelength from  the optical to the mid-IR\footnote{Note that the mid-IR AGN detection rates are based on photometric measurements with the IRAC instrument on the Spitzer satellite. In this case, rather than using
high spatial resolution to detect the compact nuclei, the AGN are detected as an excess in the 
mid-IR continuum over the flux in the starlight predicted on the basis of an
extrapolation of the near-IR starlight flux, with the latter measured from HST observations using the same aperture as used for the Spitzer/IRAC measurements \citep[see][]{ramirez14a}.}. As expected, the rate of detection
of AGN nuclei in the NLRG sources is low at optical wavelengths ($<$30\%), although 
interestingly not zero. The detection rate for the NLRG then rises with increasing wavelength through the near-IR to the mid-IR bands, reaching 80\% at 2$\mu$m and 95\% at 8$\mu$m \citep{ramirez14a}. This behaviour is entirely as expected on the basis of the standard orientation-based unified schemes. The infra-red fluxes and SED shapes deduced for the nuclei of NLRG are consistent with degrees of dust extinction in the range $10 < A_v < 200$ magnitudes \citep{tadhunter99,ramirez14a}.

At this stage it is important to add a caveat about the nature of the unresolved near-IR cores detected
in NLRG. The near-IR light of type 1 AGN is  dominated by the thermal emission
of dust at close to the sublimation temperature in the inner parts of the obscuring torus, rather than
accretion disk emission. Therefore it is possible that the unresolved nuclei detected in NLRG at near-IR 
wavelengths represent directly transmitted dust emission from the inner parts of the torus. However, alternative possibilities include non-thermal emission from the inner synchrotron jets and light scattered
by dust in the near-nuclear regions (e.g. the far wall of the torus). Unfortunately, the recent detection of high degrees of linear polarisation in the unresolved near- and mid-IR nuclei of some NLRG \citep{tadhunter00,ramirez09,ramirez14b,lopez14} does not entirely resolve this issue, because the synchrotron, scattered AGN  and transmitted AGN mechanisms could all produce significant IR 
polarisation (in the latter case via the dichroic extinction effect of aligned dust grains in the torus: Ramirez et al. 2009,2014b). 
Note that if the near-IR emission of NLRG were dominated by synchrotron or scattered AGN emission, the level of extinction to the AGN would be higher than  estimated based on the assumption that all the near-IR core emission is directly transmitted AGN light.

It is also possible to search for direct AGN continuum emission from NLRG at harder X-ray energies ($>$2 keV), where the level of absorption by gas in the torus is reduced compared to
that at EUV and soft-X-ray wavelengths. Again this is an area that has benefitted from improvements
in technology in the last 20 years, with the launch of the Chandra and XMM satellites providing
improved spatial resolution and sensitivity compared with previous X-ray satellites. One complication of the X-ray studies is that emission from the inner non-thermal jets --- which are likely to suffer less attenuation than that of the AGN themselves ---  may contribute to the X-ray
core fluxes. However, by explicitly accounting for the non-thermal component in the modelling of
the X-ray spectra, it has proved possible to detect the attenuated X-ray emission from AGN in several
NLRG: absorbed X-ray AGN components have been detected in 81\% of the 21 NLRG\footnote{Using updated optical spectral classifications from the data in Buttiglione et al. (2009,2010) and other sources.} at $z < 0.3$
in the sample of 3CRR objects studied by \citet{hardcastle09}, and in 84\% of 19 NLRG in the $0.05 < z < 0.7$
2Jy sample studied by \citet{mingo14}. Thus the X-ray results are consistent with the near- to mid-IR
results in the sense that they show the presence of heavily attenuated AGN is a high proportion of NLRG. However, for objects with both X-ray and near-IR HST observations, the attenuating HI columns are generally higher than predicted from the levels of near-IR dust extinction assuming
the standard Galactic dust-to-gas-ratio \citep{ramirez14a}. This apparent inconsistency
can be explained if AGN have a non-standard (order-of-magnitude higher) gas-to-dust ratio due
to the AGN radiation field destroying dust grains in the inner part of the torus
\citep{maiolino01}.

Despite the success of the X-ray and infrared imaging observations in detecting intrinsically luminous, obscured
AGN in several NLRG, such observations do not allow the detailed spectra of the obscured AGN to be determined --- an important step in demonstrating that the nuclei have quasar-like properties. Therefore, near-IR
spectroscopy and optical spectropolarimetry observations --- which have the potential to detect
broad permitted lines characteristic of quasars in transmitted or scattered light --- provide an important complement to the X-ray and infrared imaging observations.

Attempts have been made to detect the directly transmitted broad Pa$\alpha$ emission using K-band spectroscopy of NLRG. However, despite claims of the detections in a few nearby NLRG \citep{hill96}, the observations have a low S/N, and this technique suffers from the fact that the infrared nuclei tend to be faint relative to the starlight of the cores of the host galaxies in most objects, and the broad Pa$\alpha$ in typical type 1 AGN spectra has a low equivalent width compared with the optical Balmer lines \citep[see discussion in][]{ramirez09}. Together, these factors make the detection of broad Pa$\alpha$ in NLRG challenging, even with 8m-class telescopes. 

Far more successful have been optical spectropolarimetry observations that use the alternative technique of detecting the broad Balmer lines in scattered light: there are now convincing detections of the scattered broad H$\alpha$ lines in five NLRG in the local universe \citep{antonucci84,ogle97,cohen99}. Such observations provide the most direct evidence to support the orientation-based unified schemes for SLRG, because they demonstrate that individual NLRG have nuclei with the spectral characteristics of quasars. 

Although scattered broad lines have so far been detected in a only handful of the dozens of NLRG in the 3CR and 2Jy samples at low redshifts ($z < 0.2$), the relatively low detection rate cannot be taken as strong evidence against the unified schemes. This is because several factors  can confound the detection scattered broad lines, including the strong dilution of the polarised light by the starlight of the host galaxies, geometrical dilution of the polarization, a lack of scattering dust in the NLR, and illuminating AGN that are at the lower end of the intrinsic luminosity range for a given radio power. Moreover, many of the 3CR and 2Jy NLRG still lack deep spectropolarimetry observations.  Unfortunately, it is unlikely that there will be rapid progress in this area in the near future, since making the requisite spectropolarimetry observations of a large sample of NLRG would be prohibitively expensive in observing time with the current generation of 8m telescopes

Taken together, the statistical results from  comparisons between the optical, mid-IR and radio properties of BLRG/Q and NLRG, the detection of highly attenuated AGN nuclei at X-ray, near-IR and mid-IR wavelengths in a high proportion of nearby NLRG, and the detection of polarised broad lines in scattered light in some NLRG, provide compelling evidence that the orientation-based unified schemes for SLRG work to first order: the data are consistent with the idea that all NLRG contain BLRG/Q nuclei that are obscured along our direct line of sight by circum-nuclear dust. 
The unification debate for SLRG now centres on the geometry of the central obscuring region (e.g. smooth torus, clumpy torus, warped disk), and also on whether the properties of the obscuring structures change with luminosity and redshift \citep[e.g.][]{lawrence91,lawrence10,elvis12}. 

\subsection{Orientation-based unification II: WLRG}

\begin{figure}
\includegraphics[width=12.0cm]{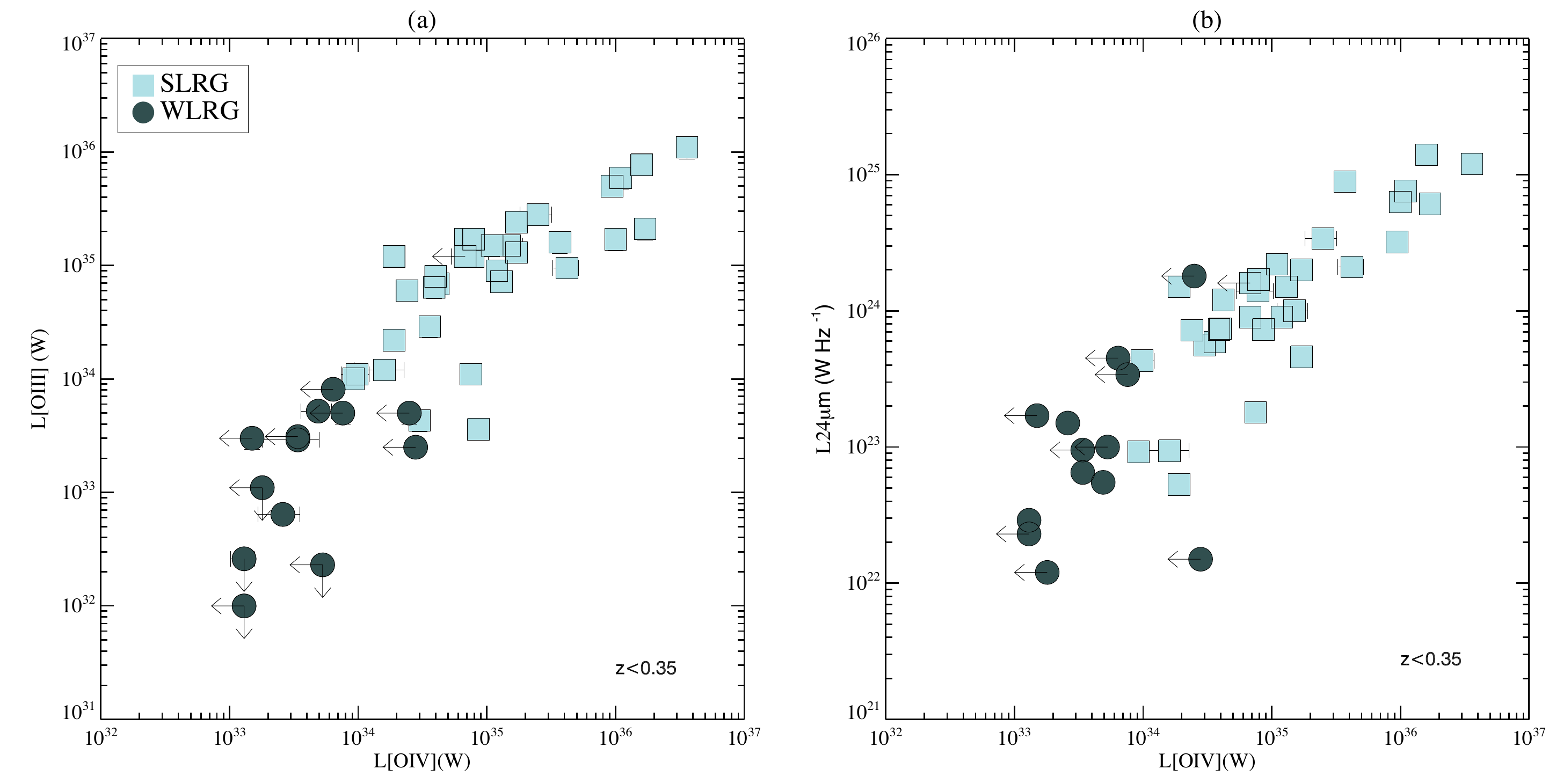}
\caption{Correlation plots showing comparisons between different optical and mid-IR
inidicators of intrinsic AGN power, highlighting the differences between NLRG and WLRG objects in the combined 2Jy and 3CRR sample of \citet{dicken14}: (a) $L_{[OIII]}$ vs $L_{[OIV]}$, and (b) $L_{24\mu m}$ vs $L_{[OIV]}$.  
See \citet{dicken14} for further details of sample selection.}
\label{mir_corr_wlrg}       
\end{figure}

While the orientation-based unified schemes are successful in explaining the relationship between NLRG and BLRG/Q objects amongst the SLRG, they cannot readily explain the relationship between WLRG and SLRG, or between FRI and FRII sources. Although it has been proposed that the WLRG objects might be SLRG radio galaxies in which the NLR is unusually heavily obscured by circumnuclear dust \citep[e.g.][]{cao04}, the Spitzer results presented in Figure \ref{mir_corr_wlrg} demonstrate that this cannot be the case: a heavily obscured SLRG nucleus would be
expected to radiate strongly in the mid-IR continuum and emission lines, whereas the WLRG have weak
mid-IR continuum and [OIV] line emission \citep{hardcastle09,dicken14}; many of the WLRG are also weak at far-IR 
wavelengths \citep{dicken09}\footnote{The exceptions are WLRG that have a strong starburst heated dust component or substantial contamination of their far-IR emission by non-thermal jet emission.}, despite the fact that their continua are often dominated by non-thermal emission \citep{dicken08,leipski09,vanderwolk10}. Similarly, the relationship between FRI and FRII sources cannot be explained in terms of orientation, because it is not possible to ``hide'' the strong radio lobe and hot spot emission typical of FRII sources in the FRI sources.

Noting that FRI sources are almost invariably associated with WLRG optical spectra, and that BL Lac objects also show low equivalent emission lines, an alternative orientation-based unified scheme (see Figure \ref{unification}(b)) has been proposed to explain the link between these two classes of objects \citep{urry91,urry95}. In this case, the required anisotropy in the optical continuum is produced entirely through the beaming effect caused by the bulk relativistic motions of the inner jets; the presence or otherwise of a central obscuring torus in such objects is controversial. The beaming that is such a key part of this scheme explains why BL Lac objects --- viewed with the line of sight close to the direction of the radio axis --- have core-dominated radio morphologies and relatively flat radio spectra; beamed jet emission also explains the relatively strong, point-like,  and  highly polarised optical continuum sources that are observed at optical wavelengths in such objects. 

Interestingly, HST observations also reveal point-like nuclei at optical wavelengths in a high proportion of WLRG/FRI sources \citep[see Figure \ref{compact_cores};][]{chiaberge99,verdoes99,capetti02}. The strong correlations between the luminosities of the optical nuclei and those of the radio cores \citep{chiaberge99}, the  optical to X-ray SED spectral indices \citep{balmaverde06}, and the detection of high degrees of optical polarization \citep{capetti07} are all consistent with the idea that the optical nuclei in the WLRG/FRI objects represent 
synchrotron emission from the inner jets.  At first sight the detection of the optical nuclei in FRI sources might seem to go against the orientation-based unified schemes, since the WLRG/FRI sources are considered to be the unbeamed counterparts of the BL Lac objects. 
However, even in the case that the radiation from the inner jets is strongly beamed, it is expected
that {\it some} radiation will be detected from the jets in objects whose jets are pointing at a large
angle to the line of sight. Moreover, the inner synchrotron jets could extend beyond
any central obscuring tori that would otherwise block the emission of their nuclei at X-ray and optical wavelengths. Finally, as already noted, it is not clear whether the obscuring tori are present in the nuclei of typical WLRG/FRI objects.

The unification of BL Lac objects with FRI radio galaxies in the context of the orientation-based scheme shown in Figure \ref{unification}(b) can be tested in a similar way to unified scheme for SLRG: by statistical comparison of properties that are considered to be isotropic. In this case the results are mixed: while the extended
radio lobe luminosities of BL Lac objects and FRI radio galaxies appear to be similar \citep{urry95}, on average the [OIII] emission line luminosities of the BL Lac objects are significantly higher when the comparison is made for low redshift samples \citep[$z \le 0.2$:][]{urry95,wills04}\footnote{This is consistent with the idea that the relatively low equivalent widths of the optical emission lines in BL Lac objects is due to their strongly beamed optical continuum emission, rather than their low emission line luminosities.}. These results can be
reconciled if there is sufficient circum-nuclear dust to significantly obscure the [OIII] emission in the WLRG/FRI objects. Certainly some WLRG/FRI objects do have kpc-scale dust lanes \citep{dekoff00}. On the other hand, as we have already seen, most WLRG/FRI objects lack evidence for central obscuring tori.

A further relevant result is that broad permitted lines characteristic of type 1 AGN have been detected in some BL Lac objects --- most notably Bl Lac itself \citep{corbett96} and PKS0521-36 \citep{ulrich81} --- whereas, do date, there have been no convincing detections of broad permitted lines in any of the FRI objects in the 3CR and 2Jy samples\footnote{Discounting the 3C84, which, as noted above, has a peculiar radio morphology.}. Even allowing for some obscuring dust in the circum-nuclear regions, by analogy with the SLRG unification scheme, we might expect a significant fraction of the
WLRG/FRI objects --- those observed at intermediate angles (the equivalent of the steep spectrum radio-loud quasars) --- to show broad lines if these objects are truly the unbeamed counterparts of the BL Lac objects, and most BL Lac objects have a BLR in their nucleus. However, the result that FRI sources lack broad lines should be
treated with some caution, because the broad lines might be difficult to detect against the strong stellar emission of the cores of the host elliptical galaxies, especially if the AGN luminosity (including the broad-line emission) scales with the extended radio luminosity \citep{rawlings91}, such that the relatively low radio luminosity FRI sources have correspondingly low luminosity BLR emission.

Taking the results on FRI/BL Lac unification together, it seems unlikely that {\it all} BL Lac objects would
appear as FRI sources if they were observed with their radio jets pointing at a large angle to the line of
sight; however, it is entirely plausible that a {\it subset} of such objects (i.e. those with lower luminosity
[OIII] emission) can be unified with FRI sources in this way.

This leaves the WLRG/FRII sources as true misfits in terms of the orientation-based unified schemes: they cannot be readily unified with WLRG/FRI sources because of the luminosities and morphologies of their extended radio emission, yet they cannot be unified with SLRG/FRII sources because of their low [OIII] emission line luminosities, lack of broad line emission, and low mid-IR continuum and emission line luminosities. Note that the latter mid-IR properties rule out the idea that the WLRG/FRII sources contain luminous AGN that are unusually
heavily obscured, just as they do in the case of the WLRG/FRI sources. 

Overall, it is clear that the orientation-based unified schemes are successful at explaining the relationship between NLRG and BLRG/Q objects in the case of SLRG, and less certainly between FRI radio galaxies and BL Lac objects in the case of WLRG. However, they cannot by themselves explain the full diversity of properties of the radio AGN population. In the following sections I discuss
other factors that are important in determining the observed properties of radio AGN; many of these have come to light in the last 20 years.

\subsection{Accretion rates and modes}
\label{sec:1.3}

Aside from anisotropy/orientation effects, it is natural to consider the possibility that the varied manifestations of the radio AGN phenomenon are related to the mechanisms that produce the AGN radiation and jets in the nuclear regions --- the so-called ``central engines'' of the AGN --- in particular, changes in rate or mode of accretion of material by the central supermassive black holes. 

In this context, it is significant that there is a strong correlation between the radio morphological properties (e.g. Fanaroff \& Riley types I and II) and the optical spectroscopic properties
(e.g. WLRG/SLRG) of radio AGN (see Table \ref{class_comparison}). As we have seen, FRI radio sources are almost invariably associated with WLRG spectra, and SLRG are almost invariably associated with FRII or CSS/GPS radio sources; only the WLRG/FRII objects break the trend. This strongly suggests that radio morphological and optical spectroscopic properties are linked primarily through the natures of the central engines, rather than through external environmental factors. Note, however, that environmental factors may nonetheless affect the radio properties at some level. For example, entrainment of the relatively dense, hot ISM found at the centres of giant elliptical galaxies and clusters of galaxies may be required to explain the detailed properties of FRI jets \citep[e.g.][]{laing14}; and the confinement effect of a hot, dense ISM may act to boost the observed radio luminosity for a given intrinsic mechanical jet power (Barthel \& Arnaud 1996, but see Ramos Almeida et al. 2013).

\begin{figure}
\begin{center}
\includegraphics[width=8.0cm]{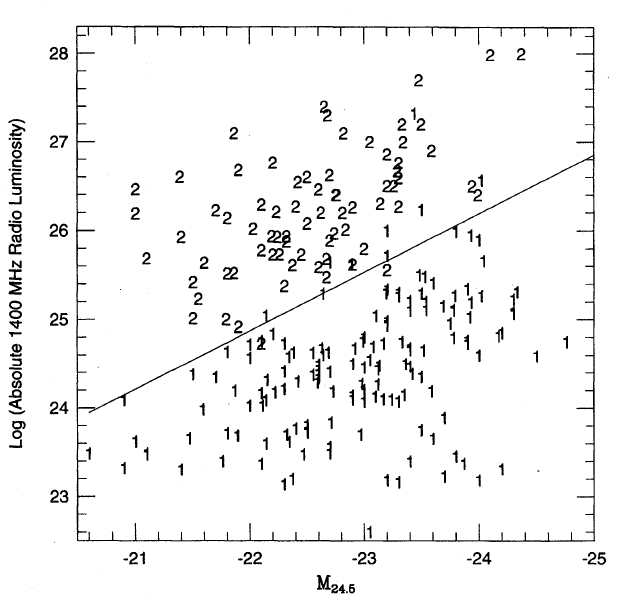}
\caption{A plot of 1.4GHz monochromatic luminosity against  absolute R-Band 
magnitude for a sample of powerful radio sources in galaxy clusters, with
FRI- and FRII-class objects marked with different symbols. Credits: this figure was originally published as Figure 1 in \citet{ledlow96}. 
}
\label{ledlow_owen}       
\end{center}
\end{figure}

Further important clues are provided by examining the links between the levels of AGN activity, the host galaxy absolute magnitudes, and the radio morphological  or optical spectroscopic classifications. Such links were first discussed in detail by \citet{owen93} and \citet{ledlow96} who plotted monochromatic radio power against host galaxy absolute magnitude and identified  FRI and FRII sources with separate symbols (see Figure \ref{ledlow_owen}). Essentially, their plot demonstrates that the radio power that defines the break between FRI and FRII sources increases with increasing host galaxy luminosity: FRI sources hosted by luminous  (and more massive) elliptical galaxies can have higher radio powers than those hosted by less luminous galaxies, without breaching the FRI/FRII radio power limit. 

A possible explanation for the trend shown in Figure \ref{ledlow_owen} in terms of the accretion processes close to the supermassive black holes was suggested by \citet{ghisellini01} who noted that the radio power could be taken as a proxy for the overall level of AGN activity, while the host galaxy absolute magnitude is related to the mass of the central supermassive black hole  \citep[e.g.][]{kormendy13}. Therefore, Figure \ref{ledlow_owen} could be recast as a plot of AGN power against the black hole mass, with the dividing line between FRI and FRII sources representing a fixed ratio of the AGN power to the Eddington luminosity of the black hole. This is the first suggestion that the division between FRI and FRII sources is due to an ``Eddington switch'': a change in the nature of the accretion flow from a radiatively inefficient accretion flow (RIAF) that is geometrically thick but optically thin (FRI objects) to a standard geometrically thin but
optically thick accretion disk (FRII objects) at a critical ratio of the jet power ($Q_{jet}$) to the Eddington luminosity of the central black hole ($L_{edd}$). The critical ratio derived by \citet{ghisellini01} was in the range $10^{-3} < Q_{jet}/L_{edd} <10^{-2}$, although a more recent study by \citet{wold07} that used host galaxy velocity dispersion rather than absolute magnitude as a proxy for black hole mass found $Q_{jet}/L_{edd} \sim 5\times10^{-4}$.

\begin{center}
\begin{figure}
\includegraphics[width=12.0cm]{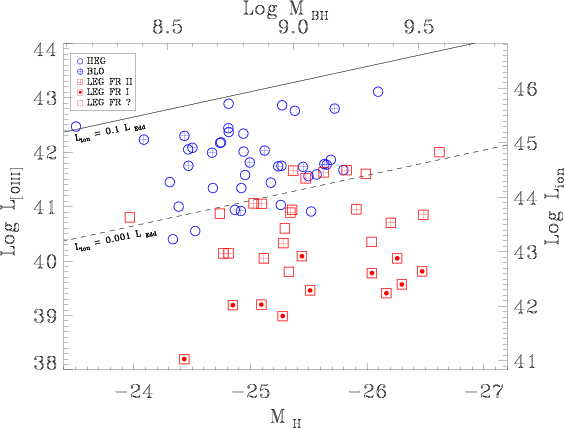}
\caption{[OIII] emission line luminosity plotted against the absolute H-band magnitude for
3CR sources in the sample of Buttiglione et al. (2010). The lines represent different
fractions of the ionizing luminosity relative to the Eddington luminosity of the supermassive black hole. Credits: this figure was originally published as Figure 11 in \citet{buttiglione10}.
}
\label{oiii_mh}       
\end{figure}
\end{center}

Figure \ref{ledlow_owen} can also be redrawn as a plot of emission line luminosity against host galaxy absolute magnitude, this time identifying objects by their optical spectroscopic classifications rather than their radio morphologies \citep{buttiglione10}. The results for the 3CR sample are shown in Figure \ref{oiii_mh}. Just as there is a clear dividing line between FRI and FRII radio galaxies in Figure \ref{ledlow_owen}, there is a clear
demarcation between HEG and LEG objects in Figure \ref{oiii_mh}, with some evidence that the emission line luminosity that represents the
boundary between the two classifications increases with host galaxy luminosity. Again, since the [OIII] luminosity is a good indicator of overall AGN power \citep[e.g.][]{heckman05}, this trend can be interpreted in terms of an Eddington switch: LEGs are associated with radiatively inefficient accretion, and the HEG/BLOs with a standard thin accretion disk, with the transition between the two types occurring at $L_{ion}/L_{edd} \sim 10^{-3}$, where $L_{ion}$ is the ionising luminosity of the AGN. This is also consistent with the recent results of both  \citet{best12} for SDSS-selected radio AGN and \citet{mingo14} for the 2Jy and 3CR samples, who find a clear division between the Eddington ratios  of low- and high-excitation objects at $(L_{bol}+Q_{jet})/L_{edd} \sim 10^{-2}$, where $L_{bol}$ is the bolometric (radiative) luminosity of the AGN. 

Interestingly, an Eddington ratio of $\sim$10$^{-2}$ also represents the division between different accretion states in X-ray binary systems: from the low/hard to the high/soft states \citep{maccarone03}. Therefore, in terms of interpreting diagrams like Figs \ref{ledlow_owen} and \ref{oiii_mh}, analogies have often been drawn with X-ray binary systems \citep[e.g.][]{falcke04,kording06}, and it has been proposed that X-ray binaries and AGN together form part of a main sequence in black hole activity \citep{merloni03}. A potential problem with this picture is that the high/soft state in X-ray binary systems is not commonly associated with the formation of non-thermal jets. However, \citet{nipoti05} have linked radio AGN with a transitory jet flaring stage of activity as sources pass between the low/hard and high/soft states \citep[see also][]{kording06}, and radio-quiet AGN with the more usual non-flaring high/soft state. In this case, the transitory nature of jet flaring stage could be 
consistent with the fact that radio-loud quasars represent $<$10\% of the full quasar population: perhaps objects cycle through radio-loud and radio-quiet phases as part a particular quasar triggering event \citep[but see][]{bessiere12}.

An important debate concerns whether the different radio AGN classifications, in particular the division between WLRG/LEG and SLRG/HEG, are related to the {\it mode} rather than the {\it rate} of accretion. On the basis that
the powers of the jets of LEGs in the 3CR sample are consistent with fuelling via Bondi accretion of the hot ISM from the X-ray haloes of the objects, whereas HEGs are too powerful to be explained in this way, \citet{hardcastle07} proposed that LEGs are associated with hot mode accretion, whereas HEGs are associated with cold mode accretion which has the capability to fuel the AGN at the requisite higher accretion rates; in this model the division between FRI and FRII sources is explained in terms of environmental effects. \citet{buttiglione10} also favoured the mode of accretion as being the dominant factor, arguing that any hot gas being accreted will not be able to cool to form a thin accretion disk and will therefore be associated with WLRG/LEG nuclear activity in radio AGN. On the other  hand, cool gas in the process of being accreted will naturally form a standard accretion disk, leading to HEG-style nuclear activity.

Despite these arguments, it seems unlikely that all WLRG/LEG are fuelled by hot gas. Indeed, many have plentiful cold gas in their nuclear regions, as evidenced by nuclear dust lanes \citep{dekoff00}, at least one shows evidence at mid-IR wavelengths for a compact, warm dust structure close to the nucleus \citep[PKS0043-42:][]{ramos11b}, and the nuclear NLR of several WLRG/FRI sources --- albeit of low luminosity compared to SLRG --- appear compact and centred on nuclei in HST long-slit and imaging observations
\citep{verdoes99,capetti05}, thus providing evidence that there is at least some warm gas on a sub-100~pc scale in these objects, even if the gas masses are relatively low. Therefore, it seems more plausible that accretion rate is the dominant factor, and that some WLRG are fuelled by cold mode accretion at a relatively low rate, even if the majority are fuelled by hot mode accretion \citep[see also][]{best12}. 


Until recently, most studies of hot mode accretion assumed that the hot gas accretes as a spherical, unperturbed flow at the Bondi rate \citep{bondi52}. However, recent simulations that include realistic prescriptions for cooling, heating and turbulence, provide evidence that much of the hot gas in the nuclear regions on a scale of $\sim$0.1 --- 1~kpc can condense into filaments of cool gas that fall near-radially towards the nucleus, with cloud-cloud collisions helping to dissipate any angular momentum in the gas \citep[e.g.][]{gaspari13,gaspari15}. As a result, the accretion rates associated with such chaotic cold accretion (CCA) can be orders of magnitude higher than the Bondi rate --- sufficient to trigger a quasar-like AGN. The CCA simulations --- which clearly blur the distinction between hot and cold mode accretion --- also predict the presence of large masses of cold gas
and the formation of clumpy torus-like structures close to the AGN. Since these predictions are somewhat at odds with what we observe in the majority
of WLRG/FRI objects, the applicability of CCA models to such objects is currently uncertain. Furthermore, if the CCA mechanism is as effective
as the models suggest, we would expect luminous, quasar-like AGN to be common in rich clusters of galaxies, where the densities of hot ISM are relatively high; however, this appears inconsistent with the measured environments of most SLRG (see section 3.5).


On the basis of diagrams such as Figures \ref{ledlow_owen} and Figure \ref{oiii_mh} it might seem plausible that the radio morphological and optical spectroscopic classifications of radio AGN are linked through the nature of the accretion onto the black holes via the Eddington switch i.e. FRI objects are always WLRG and are associated with low rates accretion and a radiatively inefficient accretion flow, whereas FRII objects are always SLRG and are associated with thin, radiatively efficient accretion disks. 
However, the WLRG/FRII sources do not fit into this
picture. While the reason for this apparent discrepancy might be related to some aspect of the accretion flow onto the black hole that can generate powerful jets and an FRII radio morphology but not a SLRG nucleus, or perhaps to some particular combination of environmental and nuclear accretion factors, an alternative possibility is that it is due do the intermittency or even switching off the nuclear fuel supply. This possibility will be explored in the next subsection.

\subsection{Variability}
\label{sec:1.4}

All types of broad-line AGN are known to vary by up to a factor of a few at optical wavelengths on timescales of weeks to years \citep{matthews63,fitch67}. Radio AGN are no exception, and there are now detailed studies of bright, nearby BLRG that track the variability of both the broad lines and continuum, and use the time lags between the two types of emission to estimate the size of the broad-line region \citep[e.g.][]{dietrich12}. 

Apart from this ``normal'' AGN variability, which is likely to be due to relatively minor changes in the accretion flow that lead to a temporary increase or decrease in luminosity of the thermal accretion disk emission and hence
the flux of ionising photons in the BLR, it is also possible that there are longer-time-scale variations in the fuelling of the AGN that lead to larger amplitude (i.e. factor of 3 or more) variations. There are two situations to consider here:  high-amplitude intermittency  in the fuel supply {\it within} a particular cycle of SLRG activity; and the switch-off phase at the end of a SLRG cycle, in which a longer-term scarcity of fuel causes the central AGN to shut down completely or enter
a lower activity state associated with a low Eddington ratio.

\begin{figure}
\includegraphics[width=12.0cm]{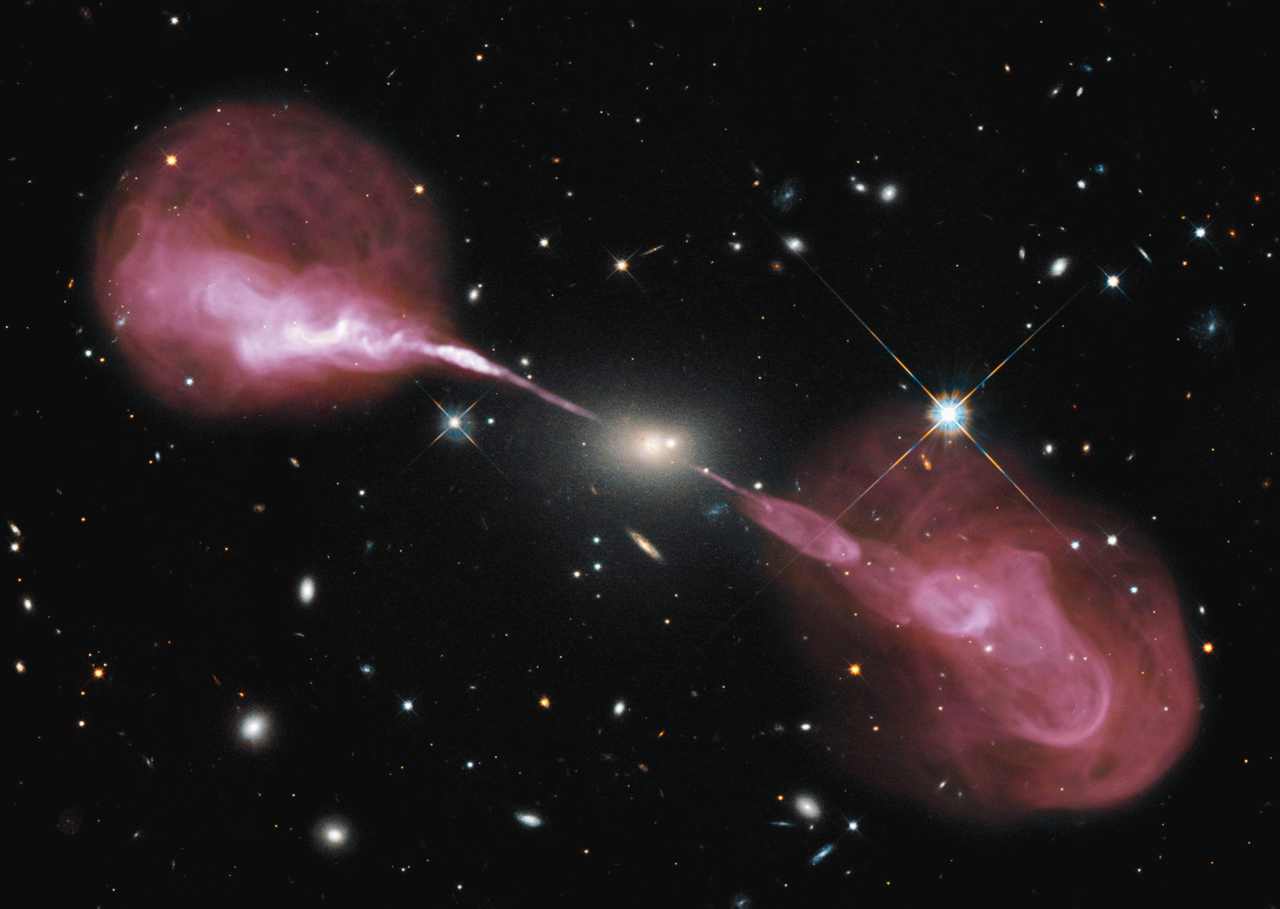}
\caption{Radio image  of the southern radio source Hercules~A (pink) superimposed
on an optical image of the same source. The radio image was made at frequencies between 4 and 9~GHz using a variety of VLA confirgurations, while the optical image represents a composite of images taken with the WFPC3 camera on the HST using
the F606W and F614W filters. Credit: NASA, ESA, S. Baum and C. O'Dea (RIT), R. Perley and W. Cotton (NRAO/AUI/NSF), and the Hubble Heritage Team (STScI/AURA).}
\label{hera}       
\end{figure}

Evidence for intermittency is provided by the radio structures of some radio sources. For example, the radio structure of the WLRG Hercules A (3C348: see Figure \ref{hera}),  shows hybrid FRI/FRII characteristics, including a series of bubble structures that may represent successive phases of high jet activity in the lifecycle of the source. Other objects show evidence for more than one phase of activity in the form of ``double-double'' radio structures \citep{schoenmakers00}, or high surface brightness, compact inner double structures combined with much larger-scale and lower-surface-brightness outer structures
\citep[e.g. PKS1345+12:][]{stanghellini05}.

The observational consequences of high-amplitude changes in the fuel supply depend on the timescales of the changes, and on the AGN components being considered. Observations show that the BLR, torus, NLR, and large-scale radio-emitting lobes of radio AGN have typical radial scales of $r_{\rm BLR} \sim 0.01 - 1$ light year,
$r_{\rm torus} \sim 0.1 - 100$~pc, $r_{\rm nlr} \sim 0.001 - 3$~kpc\footnote{This is the typical scale on which the narrow-line emitting gas accessed by the nuclear apertures used for the spectroscopic classification is emitted. Many radio galaxies show lower-surface-brightness extended emission line nebulosities on much larger scales \citep[up to $\sim$100~kpc:][]{baum88,tadhunter89a}, but these are not used for the optical spectral classifications of the sources.}, $r_{\rm lobe} \sim 0.05 - 1$~Mpc, corresponding to light crossing times of $\tau_{\rm BLR} \sim 0.01 - 1$~yr, 
$\tau_{\rm torus} \sim 0.3 - 300$~yr, $\tau_{\rm nlr} \sim 3 - 10,000$~yr and 
$\tau_{\rm lobe} \sim 150,000 - 3\times10^6$~yr respectively. 
Thus, when an AGN finally switches off due to the exhaustion of its fuel supply, this change will lead to a substantial decrease in the BLR and torus emission on a timescale of weeks to decades, in the NLR emission on a timescale of decades to thousands of years\footnote{Note that, for a typical cloud in the NLR, the time taken
for the [OIII] emission to fade following a substantial decline in the ionizing flux is negligible
compared with the light crossing time of the NLR \citep{capetti13}.}, and in the radio emission on a timescale of a hundred thousand to millions of years. It is notable that, due to their larger scales, the radio components access much longer time scales that the other components.

The different timescales of variability for the emission components on different scales in radio AGN could help to explain the apparent anomaly of the WLRG/FRII sources: perhaps these are objects in which the AGN has switched off or entered a low activity phase for a timescale of hundreds to thousands of years, such that there has been a substantial decrease in the nuclear NLR emission (hence the WLRG classification), but this information has yet to reach the hotspots in the radio lobes of the sources. This possibility has been discussed by \citet{buttiglione10} and  \citet{tadhunter12}; \citet{capetti11, capetti13} have also considered the possibility that  WLRG with unusually low excitation emission line spectra --- the so-called extreme low excitation radio galaxies (ELEG) --- represent objects caught in the act of the NLR switching off. 
The advantage of associating the WLRG/FRII sources with the switch-off phase is that it explains their ambiguous status in the unified schemes. Moreover, in the context of explanations of both radio morphological and optical spectroscopic classifications in terms of an Eddington switch, it allows the link between FRII and SLRG, and between FRI and WLRG to be preserved, without the need to explain the WLRG/FRII sources in terms of special accretion flow physics.

It is also possible that the radio AGN classified as  relaxed  or fat doubles \citep{owen89}, which lack  radio hotspots, but nonetheless have FRII-like radio lobes (e.g. 3C310, 3C314.1, 3C386 in the 3CR sample), represent a post-switch-off stage in the evolution of the radio sources, in which the hot spots have already responded to the decline in jet activity,  but the relic lobes have not yet faded below the
flux limoit of the sample. Significantly, $\sim$48\% of all the 25 WLRG/FRII sources, but only $\sim$11\% of all the 80 SLRG/FRII sources, in the 3CR and 2Jy samples of \citet{dicken09} and Buttiglione et al. (2009, 2011) could be classified as relaxed or fat doubles. 
It is also notable that spectral ageing studies of the lobes of at least two WLRG/FRII objects with relaxed radio lobes suggest that they represent 
relic radio sources in which the jets switched off $\sim$6 -- 20 Myr ago \citep{mazzotta04,harwood15}.

The switch-off phase {\it must} occur at the ends of the activity cycles of all powerful radio sources. However, it is also possible that the nuclear AGN/jet activity is intermittent, and switches off for timescales
$\tau_{\rm nlr} < t_{\rm off} < \tau_{\rm lobe}$, {\it within} a full activity cycle. In this case, the fraction of FRII that are WLRG sets an upper limit on the proportion of time that the radio sources are ``off'' for timescales longer than $\tau_{nlr}$ within the cycle:  $f_{\rm off} \le 0.25$ for the full 2Jy and 3CR samples\footnote{Note that WLRG with
hybrid FRII/FRI radio morphologies were counted as WLRG/FRII sources  when making this estimate.}, and $f_{off} \le 0.1$ for the 29 sources at the higher radio power end of the 2Jy sample ($P_{5GHz} > 10^{26}$ W Hz$^{-1}$). These estimates are upper limits on the in-cycle intermittency because some WLRG/FRII sources may represent objects observed in the switch-off phase, and there may be other explanations for the WLRG/FRII phenomenon. We deduce from this
that the fuel supply must be steady, and the AGN ``on'', for the overwhelming majority of activity cycle. This is remarkable given that the rate of infall of gas to the nuclear regions may be clumpy rather than smooth, and the feedback effect of the jets and winds associated with the AGN may disrupt the fuel supply.  

Note that, based on the existing spatially integrated spectra of radio galaxies, it is not possible to
rule out the existence high-amplitude intermittency on shorter timescales ($t_{\rm off} < \tau_{\rm nlr}$). Indeed, \citet{inskip07} have argued that the 2Jy BLRG PKS1932-46 represents a case in which the AGN has recently entered a low activity phase, but this is not yet reflected in its NLR properties, because its  [OIII] and mid-IR narrow emission line luminosities
are much higher than expected from its relatively low X-ray, near-IR and mid-IR continuum luminosities, and H$\alpha$ broad-emission-line luminosity. There are also several reports in the literature of ``changing state'' AGN in which broad lines near-disappeared, or appeared, on
timescale of years to decades \citep[e.g.][and references therein]{penston84,lamassa15,macleod15}. 

Before concluding that all WLRG/FRII sources represent powerful radio AGN in which the nuclear activity has switched off, either permanently of temporarily, it is important to consider some objections to this interpretation. First, it has been noted that the radio cores of WLRG/FRII sources  are stronger than expected in the case that the nuclear activity has switched off \citep{buttiglione10}. Second, some WLRG/FRII appear to be in richer environments and have more luminous host galaxies than their SLRG/FRII counterparts, suggesting that the gaseous environments of the radio sources may play a role in determining whether a source appears as a WLRG/FRII \citep{hardcastle07,ramos13,ineson15}. However, neither of these objections is insurmountable. For example, in the case of the radio cores, at least some of the WLRG/FRII sources  have radio
cores that {\it are} significantly weaker relative to their extended radio emission than in SLRG/FRII sources; examples from the 2Jy sample include PKS0043-42, PKS0347+05, PKS1648+05 and PKS2211-17, all of which have  core-to-extended radio flux ratios measured at 2.3~GHz of $R_{2.3GHz} < 10^{-3}$ 
compared with a median of $R_{2.3GHz} = 3\times10^{-3}$ for the NLRG/FRII sources
in the same sample \citep{morganti97a}. In addition, FRI radio galaxies have higher core/extended
radio flux ratios on average than FRII sources \citep{morganti97a}, so if a SLRG/FRII source were to drop down to a lower level of nuclear activity consistent with it eventually becoming a WLRG/FRI source, this change would not necessarily have a dramatic effect on the radio core flux. 

Considering the environments and host galaxies, it is notable that, even if WLRG/FRII sources are associated with 
richer environments and more luminous host galaxies on average than SLRG/FRII sources, there is a considerable range in, and overlap between, the environments and host galaxy properties of the two groups \citep{ramos13,ineson15}. It is also possible that the level of high amplitude intermittency depends on the environment and host galaxy properties (e.g. due to the nature of the fuel supply), such that FRII sources in rich environments and with massive host galaxies are more likely to be observed in a low activity (WLRG) phase. Alternatively, the confinement effect of the dense hot gaseous haloes in rich galaxy environments may lead to the relic radio sources remaining visible for longer than they would in lower density environments \citep[see discussion in][]{murgia11}, thus making in more likely that WLRG/FRII sources will be observed in rich environments.

\subsection{Relationship to lower luminosity radio AGN samples}
\label{sec:1.6}

So far, this review has concentrated on the results for samples of nearby radio sources selected in bright, flux-limited radio surveys such as the 3CR and the 2Jy. Because such surveys select relatively high luminosity radio sources, which are rare in the local universe, the sample sizes are modest. However, cross correlation of the SDSS optical and FIRST/NVSS radio catalogues has recently allowed the selection of much larger samples of nearby radio sources ($z < 0.2$) down to lower flux densities and radio luminosities \citep{best05,best12}, albeit lacking the detailed radio and optical morphological, X-ray and infrared spectral information available for the brighter samples. It is interesting to consider how the sources selected in the bright radio samples compare with those in the SDSS-selected samples.

\begin{figure}
\begin{center}
\includegraphics[width=10.0cm]{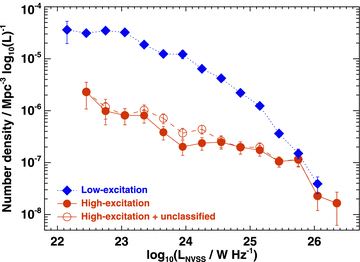}
\caption{The 1.4~GHz radio luminosity functions for LEGs (blue) and HEGs (red) derived from SDSS data by 
\citet{best12}. See \citet{best12} for details. Credits: this figure was originally
published as Figure 4 in \citet{best12}.
}
\label{lf}       
\end{center}
\end{figure}

The homogeneous spectral data available for these SDSS-selected samples has allowed the radio luminosity functions to be derived separately for the LEG and HEG populations. The results are shown in Figure \ref{lf}, which is taken from \citet{best12}. While the HEG population starts to dominate the radio source population at the highest radio luminsoities, and the LEG population dominates at lower radio luminosities, it is interesting that both types of objects are present in the SDSS sample at all radio luminosities. 

Although detailed radio morphological information is lacking for the SDSS-selected samples, based on the results for the 3CR and 2Jy samples, it seems highly likely that a large fraction of the LEG population at higher radio luminosities ($L_{1.4GHz} > 10^{24}$~W Hz$^{-1}$) are WLRG/FRI, with perhaps an admixture of WLRG/FRII sources, whereas the majority of the high radio power HEGs are SLRG/FRII sources. Moreover, the majority of these high radio luminosity sources are likely to be hosted by giant elliptical galaxies (see section 3.1 below). However,  considerable uncertainty surrounds the natures of the HEG and LEG radio sources and their host galaxies at lower radio luminosities: for a given optical spectral 
classification, do the lower power radio sources  have the same radio and optical morphologies as their higher radio power counterparts? Do the luminosity functions for the HEGs and LEGs each represent uniform populations of objects?

Concentrating first on the LEG objects, there is already evidence that the natures of the radio sources associated with the LEGs change dramatically with radio luminosity. For example, a recent radio morphological study of the low radio luminosity end of the LEG population ($L_{1.4GHz} < 10^{24}$~W Hz$^{-1}$) by \citet{baldi15} shows that most of the lower radio luminosity sources are compact and lack the prominent jets and diffuse lobes typical of more powerful LEG objects. \citet{baldi15} have labelled such sources FR0 radio galaxies, in recognition of the fact that they fall outside the usual FRI/FRII classification scheme \citep[see also][]{ghisellini11,sadler14}. However, the host galaxies of most of the FR0 objects appear to be massive, early-type galaxies, just like their more powerful LEG counterparts.

The situation for the HEG sources is less clear. While the high radio luminosity end of the HEG population is dominated by FRII radio sources that are hosted by giant elliptical galaxies, the lower power end of the luminosity function of the HEGs overlaps with the high power end of the radio luminosity function of Seyfert galaxies \citep{meurs84}, which are more likely to be hosted by spiral galaxies. This raises intriguing questions about the makeup of the HEG population: is there a gradual transition between spiral and elliptical host galaxies from low to high radio powers along the HEG sequence? How do the radio morphologies change along the HEG sequence?

\section{The host galaxies and environments of radio AGN}
\label{sec:2}

The ultimate aim of studies of the host galaxies and environments of radio AGN is to understand how their AGN are triggered as part of the evolution of the general (non-active) galaxy populations. An holistic approach is required. A limited study of, say, the host galaxy morphologies alone, or the star formation properties alone, would not be sufficient. Fortunately, this is another area that has profited from observations with the new generation of ground- and space-based telescope facilities, which have provided a far more complete picture of the host galaxies than was possible in the recent past.

\subsection{Overall morphologies, light profiles, kinematics and masses}
\label{sec:2.1}

It was already clear from the earliest photographic imaging studies that radio AGN are associated with early-type host galaxies.  \citet{matthews64} showed  that, above a certain radio power limit, the hosts of  AGN  are  elliptical, D, or cD galaxies, whereas at lower radio powers spiral morphologies are more prevalent.
However, there are exceptions. Table \ref{disks} shows a list of radio AGN for which extended disk or spiral components are clearly visible upon cursory inspection of optical images. Note that it is important to distinguish here between objects that show such morphologies, and those that merely harbour dust lanes that cross the nuclear regions of the otherwise elliptical host galaxies \citep{dekoff00}. 

There are several points to make about the radio AGN with clearly visible disk/spiral structures: (a) they are rare ($<$5\% of the objects in the 3CR and 2Jy samples); (b) they generally fall at the lower power end of the radio AGN population; (c) they are often associated with compact or double-double radio sources; and (d) their host galaxy stellar  and black hole masses
are comparable with those of the giant elliptical host galaxies of the majority of the radio AGN population
($M_{stellar} > 10^{11}$~M$_{\odot}$ and $M_{bh} > 10^{8}$~M$_{\odot}$). The latter point is particularly interesting because it suggests that it is only the most massive disk galaxies that are capable of hosting radio AGN; it also further reinforces the link between black hole mass and radio loudness in the AGN population (see below).


\begin{table}
\begin{center}
\begin{tabular}{llrllll}
Object &Redshift &$L_{1.4GHz}$ &Optical &Radio  &Ref \\
\hline \\
NGC612 &$0.0298$ &$2\times10^{25}$ &NLRG &FRII  &1 \\
0313-192 &$0.0671$ &$1\times10^{24}$ &NLRG  &FRI/DD?  &2,3 \\
J0832+0532 &$0.099$ &$1.5\times10^{24}$ &NLRG &FRII  &4 \\
3C223.1 &0.107 &$5.4\times10^{25}$ &NLRG  &FRII  &5,6 \\
J1159+5820 &$0.054$ &$2.3\times10^{24}$ &WLRG  &FRII/DD &7,4 \\
3C236 &$0.1007$   &$1.0\times10^{26}$   &NLRG    &FRII/DD   &8 \\
3C293 &$0.0450$ &$2\times10^{25}$ &WLRG &FRII?/DD &9,4 \\
3C305 &$0.0416$ &$1.2\times10^{25}$ &NLRG  &CSS  &10, 11 \\
Speca &$0.1378$ &$7\times10^{24}$ &WLRG  &FRII?/DD &12 \\
J1649+26 &$0.055$ &$1\times10^{24}$ &WLRG  &FRII  &14,4 \\
PKS1814-637 &$0.0641$ &$1.2\times10^{26}$ &NLRG  &CSS   &13 \\
J23345-0449 &$0.0755$ &$3\times10^{24}$ &WLRG  &FRII/DD &15\\
\end{tabular}
\caption{Radio AGN showing clear disk and/or spiral morphologies in optical
images. Reference key: 1. \citet{emonts08}; 2. \citet{ledlow01}; 3. \citet{keel06}; 4. \citet{singh15}; 5. \citet{dekoff00}; 6. \citet{madrid06}; 7. \citet{koziel12}; 8. \citet{odea01}; 9. \citet{vanbreugel84a}; 10. \citet{sandage66}; 11. \citet{heckman82}; 12. \citet{hota11}; 13. \citet{morganti11}; 14. \citet{mao15};
15. \citet{bagchi14}. The radio luminosities in column 3 are in units of W Hz$^{-1}$. Columns 4 and 5
give the optical spectroscopic and radio morphological classification respectively, with uncertain
classifications indicated by a question mark; a DD designation in column 5 indicates a double-double source.}
\end{center}
\label{disks}
\end{table}

With the development of sensitive, linear CCD detectors in the 1980s, it became possible to make quantitative studies of the radial light profiles of the hosts down to low surface brightness limits, and to compare them with those of non-active elliptical galaxies. Such studies have the potential to reveal disk-like components in the hosts of radio AGN that might not be immediately apparent upon cursory visual inspection of the images.  Many of these studies fit the 
radial fall off in surface brightness with  a Sersic profile:
\begin{equation}
\mu(r) =  a \left( \frac{r}{R_e} \right)^{1/n} + b
\end{equation}
where $r$ is the radius, $\mu(r)$ is the surface brightness in magnitudes arcsec$^{-2}$ at radius $r$, $n$ is the Sersic index,
$R_e$ is the effective radius that contains half the total light, and $a$ and $b$ are constants for a particular profile \citep{sersic63}. Note that $n=4$ represents an R$^{1/4}$  profile characteristic of elliptical galaxies \citep{devauc48} and $n=1$ an exponential profile characteristic of the disks of spiral galaxies. Some studies explicitly model a nuclear point source component --- especially important in the case of objects with unresolved broad-line nuclei --- and/or include more than one Sersic profile component.

The results of the profile fitting have been mixed. On the one hand, some studies have found that the surface brightness profiles of radio galaxies are similar to those of quiescent elliptical galaxies, with a preference for R$^{1/4}$ outer profiles in the majority of cases,  rather than the exponential profiles characteristic of the disks of the spiral galaxies \citep{heckman86,smith89a,smith89b,mclure99,dunlop03}. On the other hand, a number of studies have presented evidence for significant departures from pure R$^{1/4}$ profiles. Concentrating first on ground-based optical studies, \citet{colina95} found that the outer surface brightness profiles depart significantly from an R$^{1/4}$ law in 65\% of their sample of 44 nearby FRI radio galaxies, while \citet{govoni00} required an additional exponential component to fit the outer profiles of 29\% of their mixed sample of 72 nearby FRI and FRII radio sources. Further evidence for departures from a  single Sersic profile is provided by the H-band HST imaging results for a sample of 82 FRI and FRII 3CR sources presented by \citet{donzelli07},  who  require an additional exponential component to fit the outer surface brightness profiles in 45\% of their sample. 


The departures of the outer surface brightness profiles from  R$^{1/4}$ laws might be taken as evidence for major disk components in a significant fraction of the radio AGN population, but there are other explanations. In particular, given that radio AGN are hosted by massive elliptical galaxies that are sometimes at the centres of rich galaxy clusters, it is possible that the departures from a pure R$^{1/4}$ or Sersic law are due to the presence of cD-like outer envelopes, since the outer envelopes of cD galaxies often show significant excesses over and above the extrapolations of the inner Sersic or R$^{1/4}$ profiles \citep[e.g.][]{seigar07}. However,
\citet{colina95} have argued that the hosts of the FRI sources differ from cD galaxies in the sense that the departures from a pure R$^{1/4}$ profile set in at higher surface brightness levels in the FRI hosts than they do in cD galaxies. Alternatively, the departures might be explained in terms of the light of the accumulated debris of galaxies merging with the radio AGN hosts, with these mergers perhaps triggering the AGN activity \citep{colina95}.

It is also important to consider why some of the most recent HST-based studies provide apparently contradictory results. Most notably, \citet{mclure99} and \citet{dunlop03} were able to model the R-band light profiles {\it all} 20 intermediate redshift, radio AGN in their sample using single Sersic profiles with $n\sim4$ (range: $3.6 < n < 5.3$), whereas \citet{donzelli07} found evidence for departures from an R$^{1/4}$ law in the outer profiles of their larger sample of 3CR sources using H-band observations. At least some of the ambiguities in the results may be due to differences in samples/observations/analysis techniques. For example, it is notable that the detector used by \citet{dunlop03} has a much wider FOV than that used by \citet{donzelli07} for the objects in their sample in the overlapping redshift range, thus allowing more accurate sky subtraction; the \citet{dunlop03} observations are also more sensitive, reaching a magnitude fainter in surface brightness.


Although the current results on the outer surface brightness profiles may not be clear-cut, in other regards the hosts of the radio AGN show much in common with elliptical galaxies. For example, several studies have now demonstrated that radio galaxies follow the correlations between effective radius and the mean surface brightness at the effective radius (the so-called Kormendy relation) for non-active elliptical galaxies \citep[][]{smith89b,govoni00,dunlop03}. In addition, although data on the stellar kinematics are currently sparse for radio AGN, due to the observational difficulties in obtaining accurate results for all but the closest objects, the existing results provide evidence that radio galaxies fall on the fundamental plane that relates the kinematic and photometric properties of elliptical galaxies \citep{smith90b,bettoni01}.

The final important point to make about the hosts of radio AGN is that their stellar masses are relatively large:
optical and near-IR  results for low and intermediate
redshift samples yield estimates in the range $10^{11} < M_{stellar} < 2\times10^{12}$~M$_{\odot}$ \citep[][see Figure 18 below]{dunlop03,inskip10,tadhunter11}, corresponding to $\sim$0.7 --- 10$m_*$, where $m_*=1.4\times10^{11}$ is the characteristic mass of the galaxy mass function \citep{cole01}. This result appears to hold even in cases where the hosts show evidence for disk morphologies (see references in Table \ref{disks} above), or signs of recent star formation activity \citep{tadhunter11}. Given the correlations between black hole mass and host galaxy properties \citep{kormendy13}, the black hole masses of radio AGN are correspondingly large ($M_{bh} > 10^8$~M$_{\odot}$). Therefore it has been suggested that, along with black hole spin, the black hole mass is one of the key factors that determine whether an AGN is radio-loud \citep[e.g.][]{lacy01,dunlop03,sikora07,chiaberge11}.

%

\subsection{Morphological peculiarities}
\label{sec:2.2}

Despite their predominantly early-type character, it long been noted that the hosts of radio AGN show peculiarities in their {\it detailed} optical morphologies \citep{baade54,matthews64}. These peculiarities include: dust lanes, large-scale tidal tails, fans, shells and bridges; double nuclei; and, at a more subtle level, isophotal twists. The first truly systematic studies of these peculiarities were made by \citet{heckman86} and Smith \& Heckman (1989a,b), using CDD observations of a large sample of radio galaxies of all types, most selected from the 3CR catalogue. Based on isophotal analysis, these studies found that 50\% of all the radio AGN host galaxies show departures from elliptical symmetry. Further, it was noted that 50\% of the SLRG in the sample show  tidal tails, fans, shells and dust lanes at relatively high levels of surface brightness ($\mu_V < 25$ mag arcsec$^{-2}$), whereas only 7\% of WLRG show similar evidence. The rate of occurrence of tidal features in the SLRG was found to be the higher than the $\sim$10\% of
quiescent elliptical galaxies that show evidence for shell structures in photographic images \citep{malin83b}. Together, these results appeared consistent with the idea that the SLRG are triggered in galaxy mergers, with perhaps a different triggering mechanism for the WLRG.

\begin{figure}
\includegraphics[width=12.0cm]{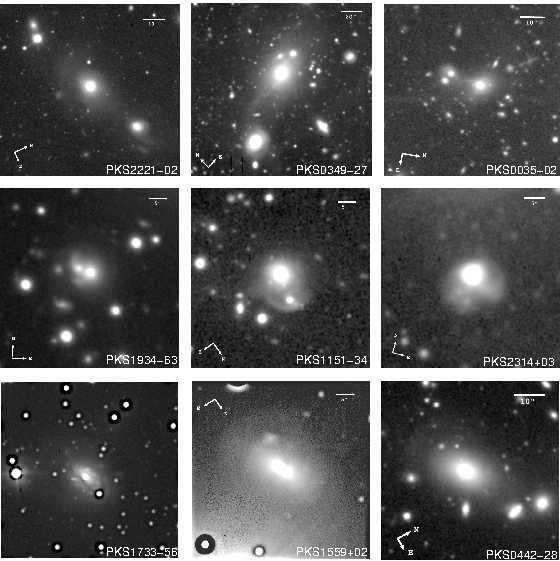}
\caption{Examples of optical r' Gemini-S images of 2Jy radio galaxies from the study of \citet{ramos11a}.
In each case, the radio source host galaxy is at the centre of the image. Note the diversity in the detailed structures: while PKS2221-02, PKS0349-27, PKS0035-02, PKS1934-63 and PKS1151-34 are pre-coalescence systems that show evidence for tidal interactions with neighbouring galaxies, 
PKS2314-03, PKS1733-56, PKS1559+02 and PKS0442-28 are post-coalescence systems that show evidence
for high surface brightness tidal features (PKS2314+03, PKS1733-56) or more subtle shell structures (PKS1559+02, PKS0442-28). See \citet{ramos11a,ramos12} for full details.}
\label{morphologies}       
\end{figure}

In apparent contradiction, some more recent HST-based imaging studies of nearby AGN found less evidence for tidal features. In particular, although some of the 20 radio AGN in the sample of \citet{dunlop03} show morphological peculiarities in their R-band HST images, the rate of such features was significantly lower than in the earlier Smith \& Heckman (1989a,b) study. Indeed, \citet{dunlop03} could find no clear difference between the rate of galaxy interactions in their luminous AGN sample and that of a matched sample of quiescent giant elliptical galaxies at the centres  Abell clusters of similar richness to those of the AGN hosts. However, despite their clearly superior spatial resolution, the surface brightness depths achieved in HST observations are sometimes inferior to those of the ground-based studies, because the HST is a relatively small telescope, and any diffuse tidal features are spread over many pixels of the HST detectors. This point is emphasised by the work of \citet{bennert08}, who repeated the HST imaging of 5 of the \citet{dunlop03} objects using longer exposure times (5 orbits rather than 1  orbit) and a more sensitive detector (the ACS rather than WFPC2), and found that 4/5 of the objects show tidal features that were not detected in the earlier study.

Recognising the importance of surface brightness depth in studies of this type, \citet{ramos11a,ramos12} undertook a deep r'- and i'-band imaging study of the complete intermediate redshift $0.05 < z < 0.7$ 2Jy sample using the Gemini South telescope, reaching a magnitude fainter in surface brightness depth compared with the earlier Smith \& Heckman (1989a,b) study, and with significantly better seeing ($0.55 < FWHM < 1.15$ arcsec compared with $1 < FWHM < 2$ arcsec). Crucially, they also compared their results with those for two comparison samples of quiescent early-type galaxies matched in absolute magnitude and imaged to similar surface brightness depths: the OBEY sample of nearby elliptical galaxies \citep{tal09} and a sample of intermediate redshift  early-type galaxy morphologically selected from deep Subaru images of the Extended Groth Strip (EGS).

Example Gemini images of some of the 2Jy sources are shown in Figure \ref{morphologies}. Strikingly, 94\% of the 35 SLRG in the 2Jy sample show tidal features\footnote{Dust lanes have not been counted as tidal features. Note that only two SLRG in the intermediate redshift 2Jy sample --- 3C105 (PKS0404+05) and PKS0252-71 --- show no evidence for tidal features or close double nuclei. It is significant that one of these objects (3C105) is affected by an unusually high level of dust extinction at optical wavelengths ($\sim$1 magnitude in the r'-band), which makes it harder to detect faint morphological features, while the other (PKS0252-71) is at higher redshifts and was observed in relatively poor seeing conditions. Therefore the true rate of occurrence of tidal features or close double nuclei in the 2Jy SLRG  could approach 100\%.} or close double nuclei ($<$10~kpc separation), whereas only 27\% of the 11 WLRG in the sample show similar features. This confirms the dichotomy in the incidence of morphological peculiarities between radio AGN with strong and weak emission lines originally noted by \citet{heckman86} and \citet{smith89b}, and is consistent with the idea that the AGN triggering mechanisms for the two groups are different (see discussion in section 3.6 below). 

Note that,  in terms of firmly establishing that radio AGN are triggered in galaxy mergers,  it is not sufficient simply to demonstrate a high incidence of tidal features. This is because a large proportion of the quiescent elliptical galaxy population show tidal features at faint surface brightness levels \citep{vandokkum05,tal09,duc15}. Therefore comparisons with control samples are important. Concentrating first on the incidence of morphological features with surface brightnesses $\mu_V < 26.2$ mag arsec$^{-2}$ --- this limit includes all the tidal features detected in the 2Jy radio galaxies --- \citet{ramos12} show that such features are present in 93\% and 95\% of the SLRG at redshifts $z < 0.2$ and $0.2 \le z < 0.7$ respectively, but in only 67\% of the 55 quiescent elliptical galaxies in $z < 0.01$ OBEY sample, and 55\% of the 109 early-type galaxies in the $0.2 < z < 0.7$ EGS sample. Moreover, 
the tidal features detected in the 2Jy radio galaxies have surface brightnesses that are 1 - 2 magnitudes brighter on average  than those detected in the comparison sample galaxies.

Overall, the deep optical studies provide compelling evidence that SLRG are triggered in galaxy  mergers. This is consistent with the recent results for radio-quiet AGN with quasar-like nuclei \citep[$L_{BOL} > 10^{38}$~W:][]{bessiere12,treister12} and  high redshift radio galaxies \citep[$z > 1$:][]{chiaberge15}, but in contrast with the results for samples of low-to-moderate luminosity AGN \citep[$L_{BOL} < 10^{37.5}$~W:][]{grogin05,cisternas11}. The dichotomy between high and low luminosity AGN suggests that, while luminous AGN favour galaxy mergers as a triggering mechanism, low/moderate luminosity AGN are triggered via secular processes. Interestingly, we see evidence for this dichotomy {\it within} the radio AGN population through the morphological differences between SLRG and WLRG.

The other striking feature of the deep imaging results is that radio galaxies are diverse in their detailed morphological properties: \citet{ramos11a} found that 37\% of the SLRG in the full 2Jy sample are pre-mergers in the sense that the host
galaxies are tidally interacting with companion galaxies (e.g. have bridges) or have close double nuclei, whereas 57\% show tidal features consistent with the hosts being observed at or after the time of coalescence of the merging nuclei. 


Despite the strong evidence they provide for the triggering of SLRG in galaxy mergers, the images alone provide only limited information about the {\it types} of mergers involved in the triggering events. Fortunately there are other facets of the triggering events that can provide further information, including the star formation properties (section 3.3), the cool gas contents (section 3.4), and the large-scale environments (section 3.5). For example, if the radio AGN were triggered at the peaks of major, gas-rich mergers we would expect them to appear similar to ultra luminous infrared galaxies \citep[ULIRGs:][]{sanders96}, with large masses of cool ISM, prodigious star formation activity, and relatively low density galaxy environments; detailed spectral synthesis modelling of the optical spectra also has the potential to provide information about the timing of the triggering of the AGN relative to the peak of the merger-induced starburst \citep[e.g.][]{canalizo01,tadhunter05,tadhunter11}. These aspects are considered in the following sections.


\subsection{Star formation}
\label{sec:2.4}

\begin{figure}
\includegraphics[width=12.0cm]{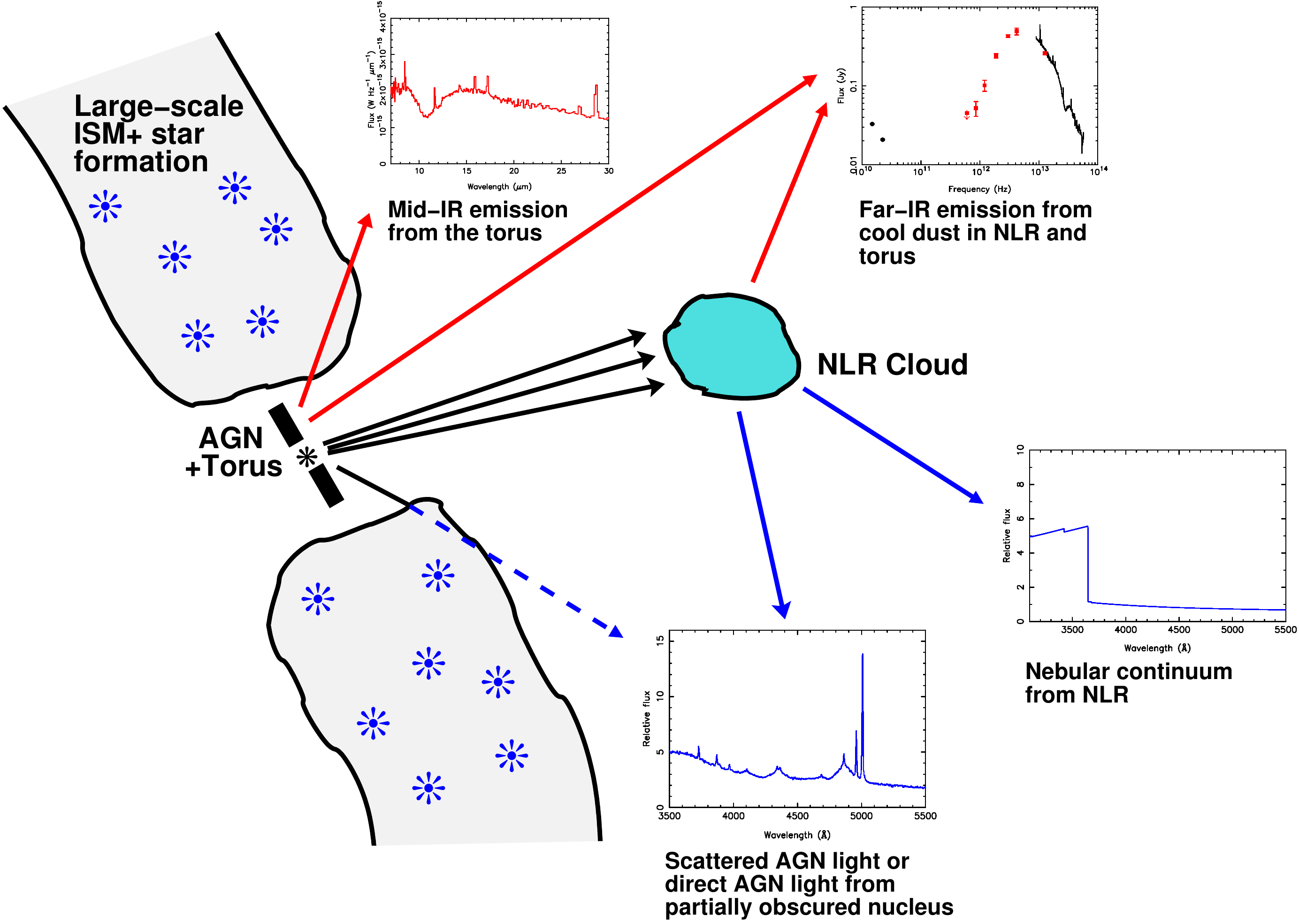}
\caption{Schematic showing various AGN-related continuum components that contribute at UV, optical, mid-IR and far-IR wavelengths. Blue lines indicate UV/optical components and red lines indicate mid- and far-IR components. By diluting the starlight (UV/optical wavelengths) and the emission of dust heated by recent star formation activity (mid- and far-IR wavelengths), these component make it challenging to 
determine the star formation properties of the host galaxies, especially for the objects with the
most luminous AGN. }
\label{contamination}       
\end{figure}

Measuring the level of star formation activity in the host galaxies of radio AGN is challenging. Figure \ref{contamination} illustrates the problem: as well as the direct starlight and emission from dust heated by recent star formation activity, several AGN-related components can contribute to the UV, optical and infrared continuum emission, even in cases where the broad-line AGN nucleus is  obscured. 
Apart from the components shown in Figure \ref{contamination}, non-thermal radiation from compact synchrotron-emitting components (e.g. cores, jets, and lobes) also has the potential to contribute at all wavelengths in radio AGN. 

Taking full account of the potential for AGN contamination, I now review the star formation properties of host galaxies of radio AGN at UV/optical, mid-IR, and far-IR wavelengths.

\subsubsection{UV/optical diagnostics}

The earliest studies of star formation in radio AGN were based on the optical or optical/near-IR colours provided by deep photometric observations. For example, \citet{smith89b} found that a subset of  their sample of $z < 0.4$ 3CR radio galaxies 
show blue (B-V) colours relative to quiescent elliptical galaxies. Interestingly, the bluer colours are associated with the SLRG in their sample rather than the WLRG, most of which tend to show red, elliptical-like colours. These results are broadly consistent with those obtained using spectroscopic measurements of the 4000\AA\, break (D4000)\footnote{The 4000\AA\, break measures the ratio of fluxes measured in wavelength bins above and below the metal line blanketing break for old stars at 4000\AA. The version of D4000 defined by \citet{tadhunter02} uses continuum bins that avoid strong emission lines ($3750 < \lambda < 3850$\AA\, and $4150 < \lambda < 4250$\AA\,).} by \citet{tadhunter02}
and \citet{herbert10}, and HST photometric measurements of the UV/optical colours by \citet{allen02} and \citet{baldi08}. However, it is important to emphasise that the presence of blue or UV excesses does not necessarily imply high levels of star formation. For example, in their detailed study of the UV emission in 3C236 --- one of the nearby radio galaxies with the brightest off-nuclear UV structures ---  \citet{odea01} show that the star formation rates in the UV knots are relatively modest: typically a few solar masses per year.

The higher incidence of blue/UV excesses in SLRG than in WLRG might be taken as evidence for distinct star formation histories for the two classes, perhaps related to differeces in the triggering mechanisms \citep[e.g.][]{smith89b,allen02,baldi08,herbert10}.  However, at UV/optical wavelengths there is a strong potential for contamination by AGN-related continuum components. These include: scattered AGN light \citep[e.g.][]{tadhunter90,tadhunter92} and nebular continuum \citep{dickson95}\footnote{The nebular continuum comprises recombination, 2-photon and free-free emission from the warm ionised gas in the NLR: see Dickson et al. (1995) for details.} from the NLR; and direct AGN emission from weak or partially obscured broad-line AGN \citep{shaw95}. Since the luminosities of the nebular continuum, scattered AGN continuum, and narrow emission lines all depend strongly on the intrinsic AGN luminosity and the covering factor of the NLR, the absolute level of the AGN-related continuum components is expected to strongly correlate with the emission line luminosity, with the highest degrees AGN contamination relative to the level of the stellar continuum found in objects with the highest equivalent width emission lines. Therefore, at least some of the correlation between the optical spectroscopic class and blue/UV excess could be due to AGN contamination. 

Recognising the potential for AGN contamination, \citet{tadhunter02} used a combination of spectroscopy and polarimetry observations to directly quantify the contribution of various AGN continuum components to the optical and near-UV emission of the integrated light of a complete sample of 22 $0.15 < z < 0.7$ 2Jy radio galaxies. Their polarimetry measurements show that scattered AGN light makes a significant contribution (up to 10 --- 30\% in some cases) to the UV  continuum below 3600\AA\, in 32\% of the full sample (50\% of the 11 NLRG and 20\% of the 9 BLRG). Moreover, based on measurements of the Balmer emission lines, they show that nebular continuum is important in all the SLRG objects in the sample, contributing as much as 30\% of the UV continuum. In a further 40\% of objects, emission from  the direct light of weak or partially obscured broad-line AGN also contributes at UV wavelengths. 

Taking full account of the AGN-related components, \citet{tadhunter02} found that the spectra of 85\% of their complete sample could be adequately modelled using a combination of quiescent elliptical template  and a power law (PL), with the latter included to represent scattered or direct AGN light. However, in 3 objects (14\% of the full sample) an additional young stellar population (YSP) with age $t_{YSP} < 2$~Gyr was required to model the continuum in the region of the Balmer break; all three of these objects also show higher order Balmer lines  in absorption, providing direct confirmation of the presence of a YSP. Subsequent deeper spectroscopy observations reported in \citet{holt07} have detected high order Balmer lines in one further object in the sample. Less directly, in a further 7 objects the PL contributions to the UV continuum are larger than expected on the basis of the polarimetry results if all the PL represents scattered light; since a very young YSP can mimic a PL, potentially this might indicate a YSP contribution to the UV continuum in these objects. Overall, between 18\% (objects with detected Balmer absorption lines) and 55\% (including objects with less direct evidence for YSP) of the $0.15 < z < 0.7$ 2Jy sample show evidence for recent star formation activity in their UV/optical spectra. This rate of detection of YSP components is similar to the $\sim$30 --- 40\% deduced by \citet{aretxaga01} and \citet{wills02} based on spectral synthesis modelling of low redshift 3CR sources. Note, however, that a large proportion of radio AGN may have lower levels of star formation activity that have not so far been detected due to the strength of light of the old stellar populations and/or AGN-related continuum components. 

The detection of significant YSP in some radio AGN opens the prospect of deducing their detailed properties (masses,  ages, reddening)  in order to further investigate possible triggering mechanisms. This is challenging work because of the difficulty of adequately accounting for the AGN-related continuum components, as well as the degeneracies inherent in modelling multiple stellar components of different ages, some of which may be heavily reddened.  Only the small proportion of radio AGN host galaxies that are already known to have significant YSP are suitable for detailed spectral synthesis modelling studies. The results of such studies reveal a substantial diversity in the YSP properties. On the one hand, in some nearby radio AGN the YSP have intermediate ages ($0.1 < t_{ysp} < 2$~Gyr) and relatively large masses, suggestive of post-starburst populations  \citep{tadhunter05,emonts06,holt07,tadhunter11}. In these cases, the ages of  YSP are older than the typical lifetimes of the radio sources, consistent with the idea that the radio AGN have been triggered {\it after} the peaks of the merger-induced starbursts. On the other hand, in other radio AGN the YSP are much younger ($t_{ysp} < 0.1$~Gyr), suggesting that the starbursts and AGN are triggered quasi-simultaneously. Interestingly, younger YSP ages tend to be associated with radio AGN 
with quasar-like AGN luminosities, whereas older YSP ages are more frequently found in objects
with lower AGN luminosities.  In the case that the AGN with detected YSP are triggered in mergers, this difference is consistent with the idea that the lower luminosity radio AGN  are triggered later in the merger sequence  \citep{tadhunter11}.

\subsubsection{Mid-IR diagnostics}

A major advantage of the mid-IR wavelength region ($5 < \lambda < 30$$\mu$m) is that it is much less affected by dust extinction than the UV/optical. Potentially, this property allows the detection of regions of star formation that
suffer high levels of dust extinction. However,
contamination by the AGN is particularly pronounced in the mid-IR, where the continuum is dominated by 
emission of the warm dust in the circum-nuclear torus (mainly SLRG), or by the synchrotron emission of the inner jets (most WLRG). Only a few radio AGN with particularly strong starbursts show evidence that the emission from dust heated by regions of recent star formation makes a significant contribution to their continuum emission at 24$\mu$m (Dicken et al. 2012). Otherwise, the main diagnostics of star formation activity at these wavelengths are the broad polyaromatic hydrocarbon (PAH) features --- emitted by large, sheet-like organic molecules --- which are strong in starbursts, but might be destroyed by the hard radiation fields of AGN \citep{siebenmorgen04}. 

The Spitzer Observatory made a major impact this field, because its high sensitivity allowed for the first time deep mid-IR spectroscopy observations to be made of large samples of radio AGN \citep[e.g.][]{ogle06,cleary07,leipski09}. \citet{dicken12} exploited these capabilities to  make a Spitzer/IRS survey of the complete $0.05 < z < 0.7$ 2Jy sample, combining the results with those for a complete sample of 3CRR FRII sources with redshifts $z < 0.11$ (see Figure \ref{mir_spectra} for example spectra).

The principal mid-IR PAH features are emitted at 6.6. 7.7 and 11.3~$\mu$m. Unfortunately, the strongest PAH blend  at 7.7~$\mu$m is particularly broad, is potentially contaminated by narrow fine-structure emission lines, and falls at the blue edge of the 10~$\mu$m silicate absorption feature. Therefore, the 11.3~$\mu$m PAH feature, which is considerably narrower than the 7.7~$\mu$m feature, is preferred when attempting to establish the occurrence rate of PAH features, and hence the incidence of recent star formation activity, in the radio AGN population. \citet{dicken12} detected this feature in only 30\% of their combined 2Jy and 3CRR sample. This immediately suggests that the hosts of most radio AGN do not harbour major starbursts, although the dilution effect of the strong AGN continuum at mid-IR wavelengths would make it difficult to detect low levels of star formation activity based on the PAH features alone.


Another technique involves using the mid- to far-IR (MFIR) colours (e.g. $L_{70 \mu m}/L_{24 \mu m}$), since an increase in the contribution from cool dust heated by young stars will tend to boost the far-IR relative to the mid-IR emission. Indeed, a number of studies have shown that the MFIR colours correlate well with a range of other diagnostics of the relative contributions of AGN and star formation activity \citep[e.g.][]{veilleux09,dicken09}. In the case of the combined 2Jy and 3CRR sample studied by \citet{dicken10,dicken12}, 24\% of the 55 objects with the requisite photometric data were found to have red MFIR colours $L_{70 \mu m}/L_{24 \mu m} \ge 5$, consistent with a significant contribution at far-IR wavelengths from dust heated by young stars.

\subsubsection{Far-IR diagnostics}

The strength of far-IR ($30 < \lambda < 500$$\mu$m) continuum is often considered to provide a clean diagnostic of star formation activity. This is because the dust heated by starbursts is relatively cool, such that its thermal emission peaks in the far-IR, whereas the AGN-heated dust in the compact, circum-nuclear torus is much warmer and radiates predominantly at shorter, mid-IR wavelengths. However, this division is not as clear-cut as it might at first seem. Mirroring the controversy surrounding the nature of the UV excess in radio AGN host galaxies (see section 3.3.1 above), the main issue here is the extent to which AGN-related continuum components contaminate the far-IR continuum: although the warmer dust components in the torus will radiate mostly at mid-IR wavelengths, clumpy torus models and those with  large outer torus radii allow the possibility of significant cool dust components in the torus that radiate in the far-IR; AGN illumination of the larger-scale NLR may also power cool dust that radiates at far-IR wavelengths \citep{tadhunter07}; and, for radio AGN, contamination by synchrotron-emitting jets and lobes may also be an issue in some sources \citep[e.g.][]{cleary07,dicken08,leipski09,vanderwolk10}. Most attempts to investigate the degree of AGN contamination of the far-IR continuum have adopted the approach of fitting the mid- to far-IR continuum spectral energy distributions using a combination of starburst and AGN templates \citep[e.g.][]{netzer07,mullaney11}.

\begin{figure}
\begin{center}
\includegraphics[width=12.0cm]{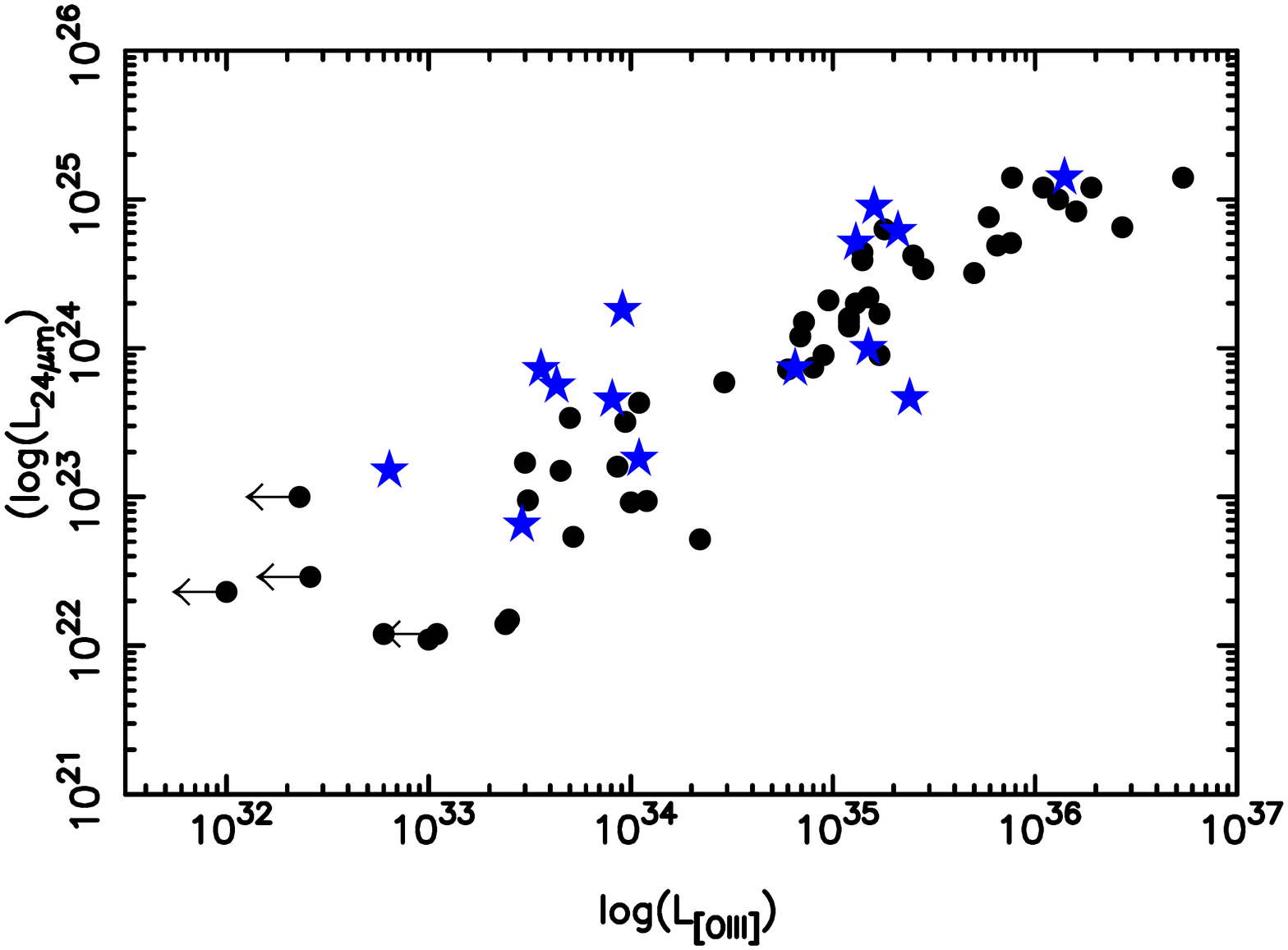}
\includegraphics[width=12.0cm]{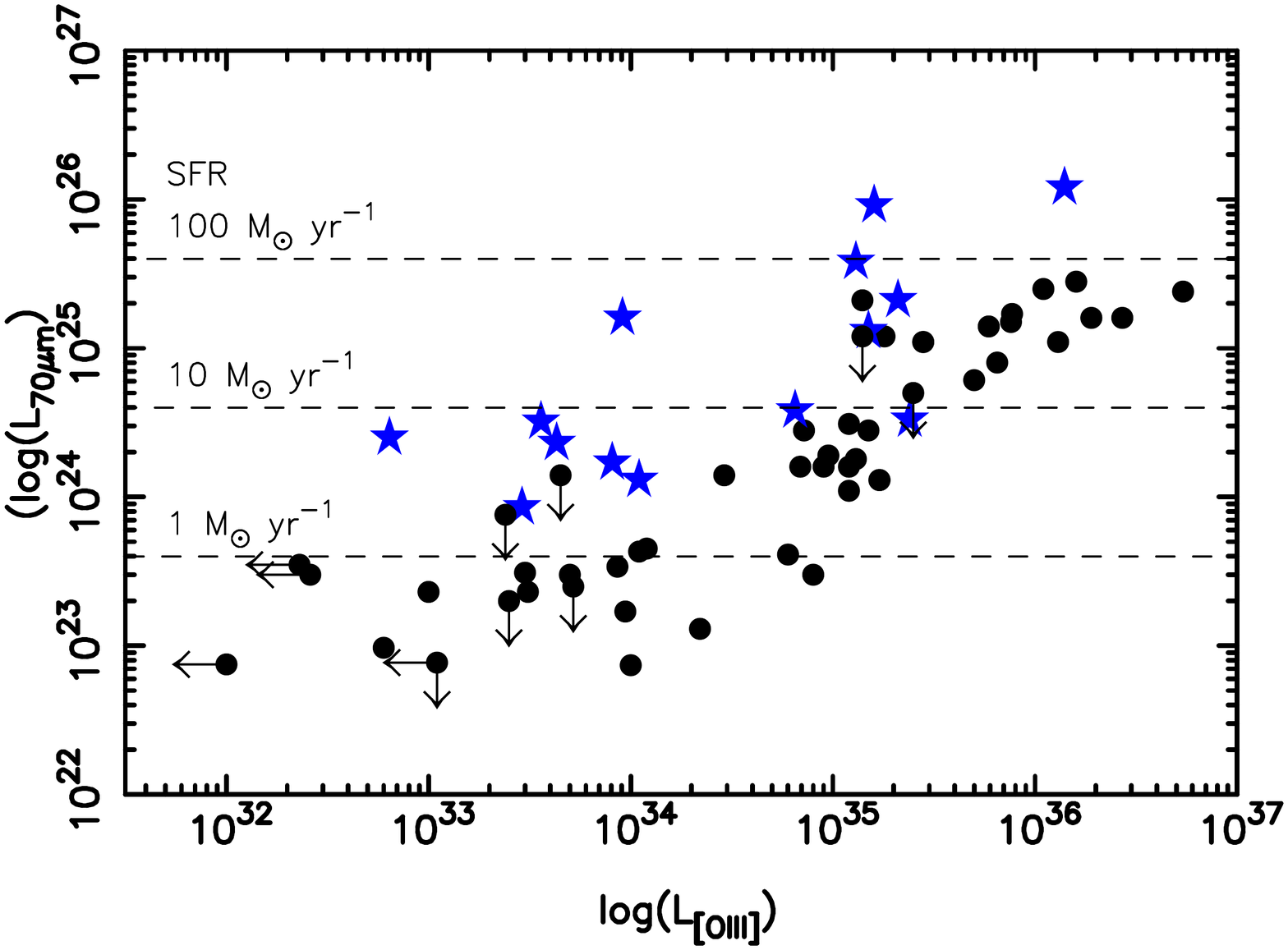}
\caption{Correlation plots for  $L_{24 \mu m}$ vs $L_{[OIII]}$ (top) and  
$L_{70 \mu m}$ vs $L_{[OIII]}$ (bottom) for the
combined $0.05 < z < 0.7$ 2Jy and $z < 0.11$ 3CRR FRII sample of \citep{dicken10}. The 24$\mu$m and 70$\mu$m   monochromatic continuum luminosities have units of W Hz$^{-1}$, while the [OIII] luminosities have units of W. Objects showing evidence for recent star formation based on optical spectroscopy and/or PAH detection are indicated by blue stars, whereas those that show
no such evidence are indicated by black, filled circles. Note the larger scatter in the $L_{70 \mu m}$ vs $L_{[OIII]}$ correlation. In the lower plot, the 
horizonal dashed lines show the equivalent star formation rates, based on the calibration of \citet{calzetti10} and assuming that all the far-IR emission is due to dust heated by young stars.}
\label{mfir_corr_sf}       
\end{center}
\end{figure}

The sensitive mid- and far-IR continuum observations of radio AGN made possible by Spitzer and Herschel have helped to illuminate this debate. First, analysis of the mid-IR to radio SEDs has shown that contamination by synchrotron emitting components is unlikely to be a serious issue at mid- to far-IR wavelengths for most SLRG, but may be important for WLRG \citep{cleary07,dicken08,leipski09,vanderwolk10,dicken16}. Second, Dicken et al. (2009,2010) have found strong correlations  between the optical [OIII] emission line luminosity --- often taken as a proxy of the overall level of AGN activity --- and both the mid-IR 24$\mu$m and the far-IR 70$\mu$m continuum luminosities for their combined sample of 2Jy and 3CRR radio galaxies (see Figure \ref{mfir_corr_sf}), although the scatter is larger for the
$L_{70 \mu m}$ vs $L_{[OIII]}$ correlation. A partial rank correlation analysis shows that these correlations are not simply the result of the well-known correlations between redshift, radio power and emission line luminosity for flux limited samples \citep{dicken09,dicken10}. 

The strength $L_{24 \mu m}$ vs $L_{[OIII]}$ correlation is not surprising given that the 24$\mu$m and [OIII] emission are both produced by AGN illumination of the circum-nuclear structures: the dusty torus and/or the NLR in the case of the 24$\mu$m emission, and the NLR in the case of the [OIII] emission; as the bolometric luminosity of the AGN increases, the level of illumination of both the torus and the NLR increases, hence the correlation. The ratio of the mid-IR to the [OIII] luminosity depends on the relative covering factors of the torus and the NLR, and \citet{dicken09} have shown on energetic grounds that the observed correlation is consistent with the expected torus/NLR covering factor ratios.

There are two interpretations of the $L_{70 \mu m}$ vs $L_{[OIII]}$ correlation. First, the $L_{70 \mu m}$ far-IR continuum luminosity measures the rate of star formation, while the [OIII] luminosity measures intrinsic power of the AGN, and the correlation then arises because the star formation rate and AGN power are correlated \citep{schweitzer06,netzer07}. Second, the far-IR continuum is produced by AGN illumination of the extended dust in the NLR or circumnuclear torus, and is therefore strongly correlated with the [OIII] luminosity, which is  produced by AGN illumination of the NLR \citep{tadhunter07,dicken09}. Support for the latter interpretation is provided by the fact that the slopes of the $L_{24 \mu m}$ vs $L_{[OIII]}$ and $L_{70 \mu m}$ vs $L_{[OIII]}$ are similar. Moreover, \citet{dicken09} have shown that luminosity of the far-IR continuum in most SLRG can be reproduced if the covering factors of the dust structures emitting the far-IR continuum are similar to those of the NLR. An important caveat on the latter argument is that, in order for the dust in the NLR
to radiate significantly at far-IR wavelengths, it must be distributed on sufficiently large scales to avoid being heated too highly by the AGN. Whether the latter condition is fulfilled is currently uncertain, but it is notable that some SLRG show NLR that are extended on radial scales of kpc.

If the far-IR and mid-IR continuum are both produced by AGN illumination, why is the scatter larger for the  
$L_{70 \mu m}$ vs $L_{[OIII]}$ correlation? Clues are provided by Figure \ref{mfir_corr_sf}, where the objects showing independent evidence for recent star formation activity 
are highlighted in the correlations. It is clear that all of the objects that form the upper envelope of the $L_{70 \mu m}$ vs $L_{[OIII]}$  correlation show independent evidence for star formation activity, whereas the main correlation is made
up of the majority of radio AGN that show no such evidence. Therefore much of the enhanced scatter in the 
$L_{70 \mu m}$ vs $L_{[OIII]}$ correlation can be explained in terms of the boosting of the far-IR emission in a minority of the sources by starburst-heated dust, with the degree of boosting reaching a factor of 10 or more in some cases. It is also plausible that the far-IR continuum of most of the 
SLRG on the main correlation is dominated by the emission of cool, AGN-illuminated dust in the NLR.  
In this case, the far-IR results provide evidence for a wide range of star formation properties in radio AGN, with star formation rates varying by more than an order of magnitude for a given intrinsic AGN luminosity.

Uncertainties about the degree of contamination by AGN heated dust make it difficult to use the far-IR observations to derive accurate star formation rates (SFR) for individual objects; however, upper limits on the SFR can be obtained by assuming that {\it all} the far-IR continuum is associated with star formation activity. The horizontal dashed lines in Figure \ref{mfir_corr_sf} show the star formation rates corresponding to different 70$\mu$m luminosities, as determined using the calibration of \citet{calzetti10}. The two most far-IR luminous radio AGN plotted in Figure \ref{mfir_corr_sf} --- 3C459 and PKS2135-20 --- have IR luminosities that qualify them as ULIRGs with star formation rates  $SFR > $100~M$_{\odot}$ yr$^{-1}$. However, for the majority of SLRG on the main correlation in Figure \ref{mfir_corr_sf} the upper limiting star formation rates are much lower, and many have  
$SFR <$10~M$_{\odot}$ yr$^{-1}$ (see also Figure \ref{main_sequence} below).

\subsubsection{Star formation summary}

\begin{figure}
\includegraphics[width=12.0cm]{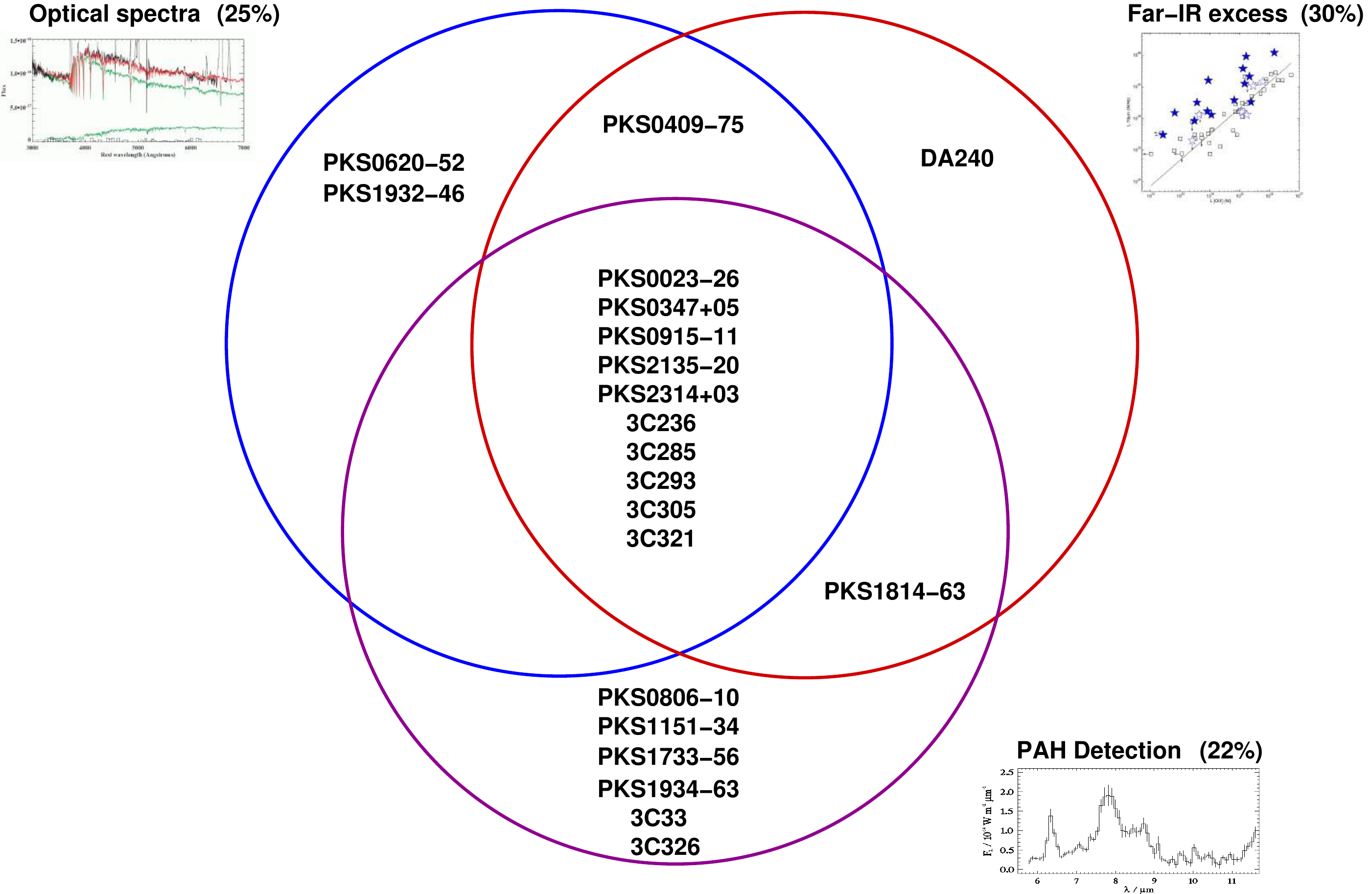}
\caption{Venn diagram comparing  the detection of recent star formation activity in individual
objects in the combined 2Jy and 3CRR FRII sample of \citet{dicken12} using  three different techniques: far-IR excess, UV/optical spectroscopy and PAH detection. In the case of the far-IR excess, objects lying more than 0.5 dex (i.e. a factor of 3) above the main correlation shown in 
Figure \ref{mfir_corr_sf} are considered to show a far-IR excess that indicates recent star formation
activity. For each technique, the detection rate of recent star formation activity using that technique across the full sample is shown in brackets. }
\label{venn}       
\end{figure}

Although the UV/optical, mid-IR and far-IR results on the star formation properties have been presented separately, there is a strong correlation between the results from the different diagnostics: objects showing clear signs of star formation activity based on their UV/optical spectra generally also show PAH features at mid-IR wavelengths, as well as a far-IR continuum excesses \citet{dicken12}. This is demonstrated by the Venn diagram shown in Figure \ref{venn}. Individually, each of the techniques gives a relatively low incidence of radio AGN with clear evidence for recent star formation activity ($\sim$20 - 30\%), and it is notable that the proportion of objects in the combined 2Jy and 3CRR sample of \citet{dicken12} that show {\it any} evidence for recent star formation activity is only 33\%. Clearly the star formation properties of radio AGN are diverse. A major implication of the low detection rate of star formation activity is that a large fraction of radio AGN in the local universe cannot have been triggered at the peaks of major, gas-rich mergers, given that such mergers lead to prodigious star formation activity around the time of coalescence of the merging nuclei which would be detectable using at least one, but probably all, of the techniques considered above. 

\begin{figure}
\includegraphics[width=14.0cm]{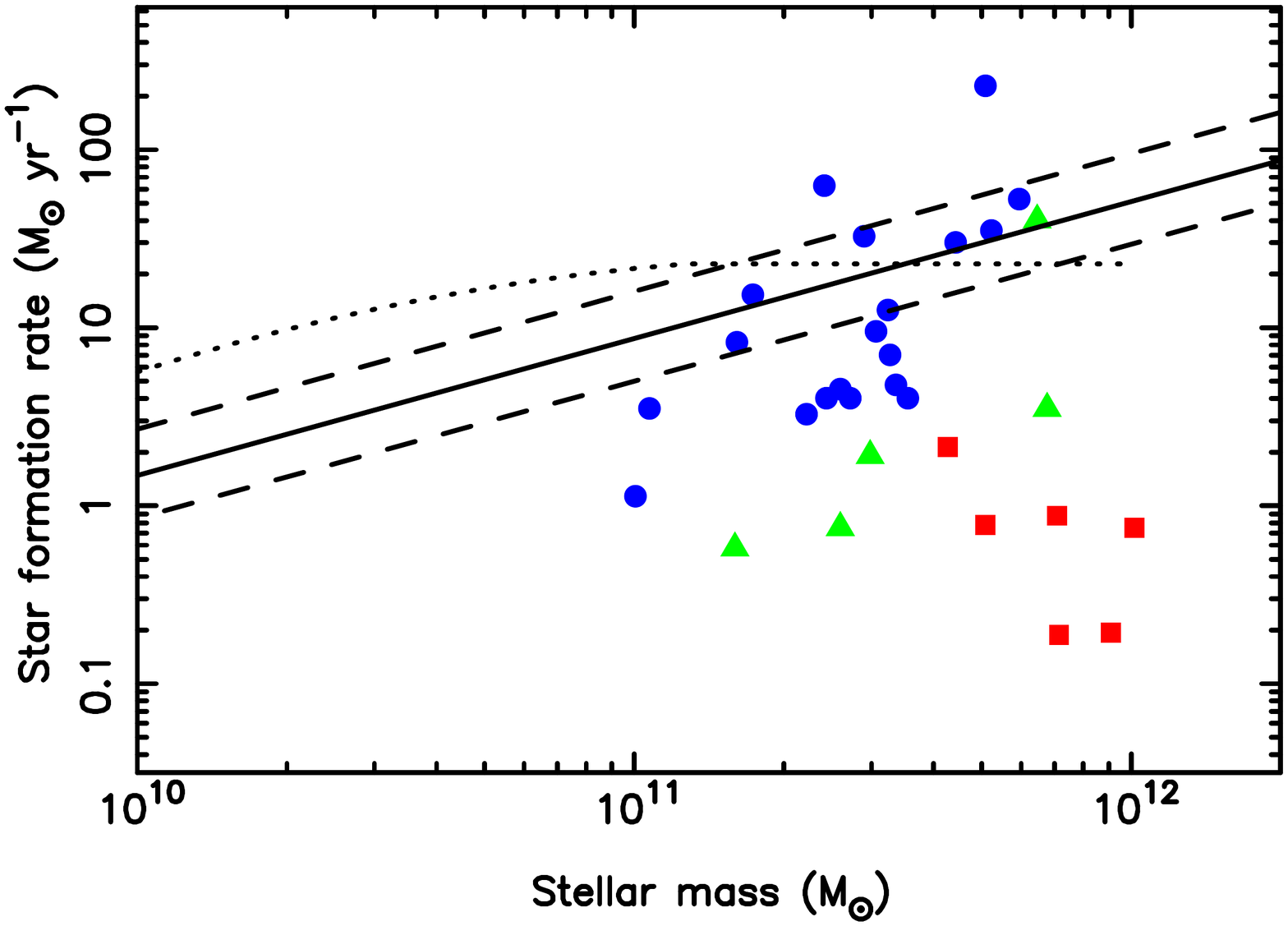}
\caption{Star formation rate vs total stellar mass for a complete subset of 30 sources from the \citet{dicken09} 2Jy sample (see Table 1) with redshifts in the range $0.05 < z < 0.5$, excluding quasars and BLRG whose K-band light has a major contribution from AGN emission. Blue circles, green triangles and red squares represent SLRG/FRII, WLRG/FRII and WLRG/FRI sources respectively.
The solid line shows the main sequence for nearby star forming galaxies ($z < 0.2$) derived by \citet{elbaz07} 
from SDSS data, with the dashed lines representing the 1$\sigma$ scatter about this relationship, while the dotted line
shows a more recent determination of the main sequence for $z = 0.5$ star forming galaxies taken from \citet{schreiber15}. The stellar masses were estimated using  K-band photometry under the assuption
of a constant  mass-to-light ratio: $(L_K/M_*)_{\odot} = 0.9$. The
64~kpc aperture K-band magnitudes of \citet{inskip10} were used for the majority of objects, but K-band magnitudes from 2MASS were used for the four sources not included in \citet{inskip10} study. The star formation rates were estimated from the Spitzer 70$\mu$m luminosities of \citet{dicken09} by using
the calibration of \citet{calzetti10}. Note that the star formation rates derived for the radio AGN are likely to represent upper limits because they have
not been corrected for contributions to the 70$\mu$m luminosities by AGN heated dust and synchrotron emission.}
\label{main_sequence}       
\end{figure}

This lack of evidence for major star formation activity in the majority of radio AGN is further demonstrated by Figure  \ref{main_sequence}, which plots the maximum star formation rates derived from the far-IR photometry against the total stellar masses for a complete sample of nearby 2Jy sources with redshifts in the range $0.05 < z < 0.5$. Apart from  
one clear starburst object, the majority of sources fall on or well below the ``main sequence'' of star formation derived for nearby star forming galaxies, and towards the part of the diagram normally occupied by ``red and dead'' galaxies. It is also notable that, while all but one of the WLRG fall well below the main sequence, a significant
proportion of the SLRG ($\sim$50\%) fall close to or above the main sequence. However, this difference may in part reflect an enhanced level of contamination of far-IR light by AGN heated dust for the SLRG (see section 3.3.3), given their generally higher levels of AGN activity.

Although the majority of the local radio AGN population lacks evidence for the moderate-to-high levels of star formation that would be readily detected using existing techniques, this does not necessarily imply that their star formation rates are zero. Indeed, based on the typical cool ISM masses estimated from the Herschel results ($10^8 < M_{gas} < 2\times10^9$~M$_{\odot}$: see section 3.4 below), and the relationships between total gas mass and SFR deduced for the star forming galaxies at both low and high redshifts \citep{daddi10}, we would expect typical radio AGN to form stars at a rate of $\sim$0.5 -- 30~M$_{\odot}$ yr$^{-1}$ if the radio AGN hosts have the type of efficient star formation that is characteristic of
starbursts. However, the star formation rates would be at least a factor of ten lower than this if the star formation efficiency were more typical of the disks of spiral galaxies. Such low levels of star formation would be difficult to detect because of AGN-related components that effectively ``hide'' low levels of star formation activity, particularly in the objects with the most luminous AGN. Dilution by the strong continuum emission of the old stellar populations in the giant elliptical galaxy host galaxies  also disfavours the detection of low levels of star formation activity at optical wavelengths. 

\subsection{Dust and gas contents}
\label{sec:2.5}

Most triggering mechanisms for the AGN in SLRG involve the accretion of cool gas in a particular event such as a galaxy merger. Large reservoirs of cool gas are required in the triggering event to fuel the AGN and simultaneously grow the bulge of the host galaxy: a cool ISM reservoir of at least $\sim$10$^9$~M$_{\odot}$ is required for an AGN at the lower end of the quasar luminosity range if the quasar has a lifetime of $10^7$~yr \citep{tadhunter14}. Therefore, measurements  of the cool gas contents of radio AGN have the potential to provide key information about the nature of the triggering events.

Until recently, most studies of the cool gas in radio AGN involved observations of the mm-wavelength CO lines \citep{evans05,smolcic09,ocana10}. Unfortunately, even using long integration times, the sensitivity of such observations was limited. Moreover, the CO observations  tended to involve relatively small, heterogeneous samples of radio AGN, or samples of objects that are unusually bright at far-IR wavelengths. Overall, the CO detection rates are low ($\sim$20 -- 60\%), and the upper limits on the gas masses derived for the undetected sources are often large --- well above the predicted minimum mass of the total reservoir required to trigger a quasar event.

Fortunately, the high sensitivity of the Herschel Observatory at the longer far-IR wavelengths ($\ge 100$~$\mu$m) has provided an alternative way of estimating  the cool ISM masses using the emission of the associated dust, under the assumption that the dust radiates as a modified black body, and that the gas-to-dust ratio  is constant. 
The deep Herschel observations of the complete $0.05 < z < 0.7$ 2Jy sample reported in \citet{dicken16} detect 100\% of the sources at 100~$\mu$m and 85\% at 160~$\mu$m. Considering first the 35 SLRG objects in the sample, the derived dust and cool ISM masses cover a wide range: $7\times10^5 < M_{dust} < 3\times10^8$~M$_{\odot}$ and $10^8 < M_{gas} < 4\times10^{10}$~M$_{\odot}$ \citep[see][for details]{tadhunter14}. For reference, the median cool ISM mass for the SLRG  in the 2Jy sample ($1.2\times10^{9}$~M$_{\odot}$) is a factor $\sim$4$\times$ lower than the total cool ISM mass of the Milky Way, and remarkably close to the predicted minimum total mass of the gas reservoir required for a quasar triggering event. 

\begin{figure}
\begin{center}
\includegraphics[width=12.0cm]{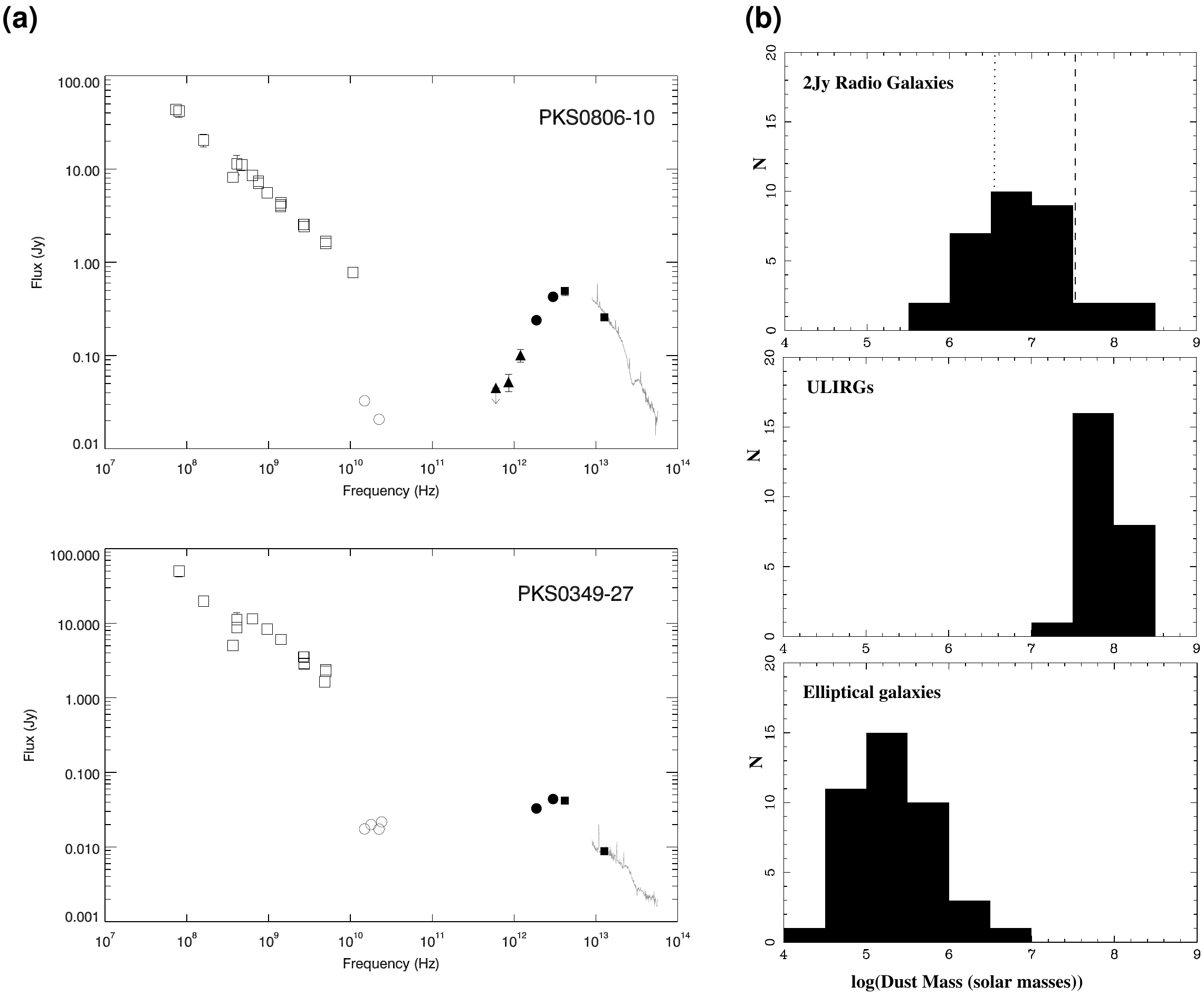}
\caption{Dust emission as a probe of the ISM contents of radio AGN host galaxies: (a)
example radio to mid-IR SEDs of the 2Jy NLRG/FRII objects PKS0806-10 and 
PKS0349-27; (b) dust masses derived for the SLRG in the $0.05 < z < 0.7$ 2Jy sample of \citep{dicken12} (top)
compared with those for nearby ULRGS (middle) and elliptical galaxies (bottom). The vertical dashed and dotted lines indicate the dust mass of the Milky Way and LMC respectively. See \citet{tadhunter14} for details.}
\label{dust_masses}       
\end{center}
\end{figure}


Figure \ref{dust_masses} compares the distribution of dust masses for the 2Jy SLRG with those measured for samples of local ULIRGs and quiescent elliptical galaxies using similar techniques. The dust masses of the SLRG are typically a factor of 10 higher than those of quiescent elliptical galaxies, but a factor of 10 lower than those of ULIRGs; however, the high mass tail of the elliptical galaxy distribution overlaps with lower end of the SLRG distribution, and some SLRG have dust masses comparable with those of ULIRGs. It is notable that many of the latter group belong to the rare subset of radio galaxies that show evidence for recent star formation activity (see section 3.3 above).

Analysis of the long-wavelength SEDs demonstrates that the far-IR emission of 5/6 of the WLRG/FRI sources in the 2Jy sample is likely to be dominated by non-thermal synchrotron emission \citep[see][]{dicken16}. However, even if all the far-IR emission were entirely due to dust emission in these WLRG/FRI sources, their dust dust masses would fall at or below the lower end of the dust mass distribution of the SLRG.   In contrast, all 5 WLRG/FRII sources in 2Jy sample show evidence for cool dust emission, with a dust mass distribution similar to that of the SLRG. This apparent difference between the dust properties of WLRG/FRI and WLRG/FRII sources has important implications for our understanding of the nature of the WLRG/FRII sources, potentially supporting the idea that the SLRG/FRII represent objects in which the AGN has recently switched off (see section 2.5). However, observations of a larger sample will be required to put this result on a firmer footing.


\subsection{Large-scale environments}
\label{sec:2.6}

A further important property that might provide clues to the triggering mechanism is the large-scale environment of the host galaxy. For example, a high density cluster environment might favour fuelling of the radio AGN via the accretion of the hot ISM that is expected to be relatively dense in such environments, whereas a lower density group environment is more likely to be consistent with galaxy mergers as the main triggering mechanism, since major mergers are most common in such environments.

Several studies have sought to measure the environments of radio AGN using number counts derived from optical images, with much of the analysis based on the spatial clustering amplitude ($B_{gg}$). The most recent study by \citet{ramos13} for the $0.05 < z < 0.7$ 2Jy sample found clear differences between the environments of WLRG and SLRG: while the 82\% of the 11 WLRG in the 2Jy sample  are in relatively rich cluster environments
($400 < B_{gg} < 1600$, corresponding to Abell class 0, 1, 2 and 3 clusters), only 31\% of the 35 SLRG in the same sample are in similarly dense environments; however, there is a significant overlap between the environmental properties of the two groups. 
Perhaps not surprisingly given the correlations between optical and radio classifications, similar results are obtained when comparing the environments of FRI and FRII galaxies, in the sense that the FRI sources inhabit significantly denser environments on average than FRII sources.  All of these results for the 2Jy sample agree with those obtained for  other samples of nearby radio AGN using number count techniques \citep{longair79,prestage88,zirbel97}.

An alternative method involves using X-ray observations to measure the luminosities of the extended hot gas surrounding the host galaxies. Such luminosities can be used as a proxy for the large scale galaxy environments, since richer galaxy environments tend to be associated with denser, more massive hot gaseous halos. \citet{ineson15} used this technique to make an extensive study of the environments of 2Jy and 3CRR sources at $z < 0.2$, finding a clear dichotomy between the extended X-ray luminosities of SLRG and WLRG, in the sense that WLRG show significantly higher extended X-ray luminosities for a given radio power; notwithstanding the uncertainties inherent in both techniques, they also found reasonable concordance with the  number count analysis of \citet{ramos13} for 2Jy objects in the overlapping redshift range. 

To summarize, there is now clear evidence from various studies at low and intermediate redshifts that WLRG  occupy higher density environments than SLRG on average, but there is clearly some overlap between the two groups in terms of their environmental properties.  

\subsection{The triggering of radio AGN}

Although the majority of radio AGN hold in common the feature that they are hosted by giant elliptical galaxies, the host galaxies are surprisingly diverse in their detailed properties. 
Together, the detailed information on the optical morphologies, star formation properties, gas contents and environments allows us to develop a coherent picture of the dominant triggering mechanisms for the different sub-types of radio AGN. Table \ref{summary} summarises the typical properties of the main SLRG/FRII and WLRG/FRI classes.

\begin{table}
\begin{center}
\begin{tabular}{lll}
{\bf Property} &{\bf SLRG/FRII} &{\bf WLRG/FRI} \\
\hline \\
{\bf Host galaxy} &Giant ellipticals  &Giant ellipticals \\
{\bf Tidal features} &Common &Rare \\
{\bf Environment} &Group or weak cluster &Modest or rich cluster \\
{\bf Star formation rate} &Low or modest &Low \\
{\bf Cool gas content} &Modest &Low \\
{\bf AGN accretion mode} &Geometrically thin, radiatively &Radiatively inefficient \\
                       &thick accretion disk                        &accretion flow \\
{\bf Triggering mechanism} &Modest merger &Hot ISM/ICM accretion \\
\hline \\
\end{tabular}
\caption{Summary of the typical host galaxy properties, accretion modes and triggering mechanisms of the two main classes of radio AGN. The host galaxy properties are quantified in sections 3.1 to 3.5. It is important to emphasise that these are the typical properties; however, there are important exceptions which are discussed in the main text. }
\label{summary}
\end{center}
\end{table}

The hosts of SLRG/FRII sources frequently show optical peculiarities such a tidal tails, fans, shells and bridges that have higher surface brightnesses than the tidal features detected in comparison samples of quiescent elliptical galaxies matched in absolute magnitude. They also show a tendency to be associated with modest, but not rich, galaxy environments. These two features are fully consistent with the idea that their AGN have been triggered in galaxy mergers. On the other hand, most SLRG/FRII hosts lack evidence for high levels of recent star formation activity, and contain relatively small amounts of cool ISM compared with ULIRGs, even if the cool ISM masses are larger than those of typical quiescent elliptical galaxies. Therefore, for the majority of SLRG/FRII sources, the triggering mergers are likely to have been relatively modest: a merger between a giant elliptical  and a galaxy with a cold gas mass twice that of the Large Magellanic Cloud (LMC) would suffice; in these objects we are likely to be witnessing the late-time re-triggering of radio-AGN activity in mature elliptical galaxies with existing super-massive black holes. 

The majority of SLRG/FRII sources  could therefore represent a fleeting active phase in the evolution of a subset of general population of early-type galaxies with high stellar masses $M_{star} > 10^{11}$~M$_{\odot}$, which formed the bulk of its stellar
mass at $z > 1$ and has not shown major evolution in stellar mass or co-moving number density at more recent epochs \citep[e.g.][]{pozzetti10,mcdermid15}. Interestingly, there 
is evidence that minor mergers have contributed most of the growth in stellar mass of this population since $z = 1$ \citep{kaviraj11}. 

However, a minority of SLRG/FRII objects ($<$20\% of the full population)  have cool gas masses and levels of star formation activity that are more comparable with those of ULIRGs. In these objects, major, gas-rich mergers are likely to have been involved in the triggering of the activity, and we may be witnessing the main event in terms of the formation of both the host galaxy bulges and the supermassive black holes.

Of course, these conclusions for SLRG/FRII objects refer to the redshift and radio power ranges covered by the bright radio AGN samples in the local universe ($z < 0.7$), and it is possible that the situation changes at higher redshifts. Indeed, a recent Herschel study  of a sample of 3CR radio galaxies at high redshifts ($1 < z < 2.5$) found evidence for higher rates of star formation and larger dust masses than those of the local samples \citep{podigachoski15}, thus suggesting that major, gas-rich mergers may become a more common triggering mechanism for radio AGN at higher redshifts and/or radio powers.

It is important to draw the distinction here between modest or minor mergers\footnote{Minor mergers
are generally defined to be those in which one object involved in the merger has
a mass of 25\% or less of that of the other.}  and major, gas-rich mergers. The debate about the triggering mechanism for luminous  AGN (whether radio-loud or radio-quiet) has often been framed in terms of major, gas-rich mergers, with quasars perhaps representing a late, post-coalescence phase in the evolution of  ULIRGs \citep{sanders88a,sanders88b}. Therefore, the hypothesis that luminous AGN are triggered in mergers has sometimes been rejected on the grounds that the host galaxies do not appear as major, ULIRG-like mergers. However, the results for the nearby radio AGN clearly demonstrate that luminous AGN activity can be triggered in more modest mergers, and also that the triggering occurs at a variety of merger stages, including pre-coalescence. Since the  relatively subtle 
morphological, star formation and gas content signs of modest mergers may be difficult to detect in surveys of high redshift AGN, clearly some caution is required when interpreting the results from such surveys in terms of  triggering mechanisms.

In stark contrast to the SLRG/FRII objects, the WLRG/FRI sources tend to favour more massive host galaxies, inhabit richer large-scale environments, show less evidence for recent star formation activity and tidal features, and have lower masses of cool ISM. These properties are consistent with the idea that, rather than being {\it triggered} by a sharp accretion event such as merger, the WLRG/FRI sources are {\it fuelled} by the more gradual accretion of the hot ISM from the host galaxy or cluster; certainly analysis of the LEGs in the SDSS sample suggests a high duty cycle for this type of radio AGN activity \citep{best05}. However, it is always dangerous to generalise, and a subset of the WLRG/FRI objects have host galaxy and/or environmental properties that are similar to those of SLRG/FRII sources. An obvious example of this is the closest radio AGN, the WLRG/FRI Centaurus A, which contains massive reservoir of cool gas in its prominent dust lane \citep[$\sim3\times10^9$~M$_{\odot}$:][]{parkin12}, is located in a group environment \citep[e.g.][]{cote97}, and shows large-scale tidal features in deep imaging observations \citep{malin83a}. In such objects is possible  that the original galaxy merger delivered a large reservoir of cool gas to the system and perhaps triggered a luminous, SLRG/FRII event at the merger peak, but subsequently the gas reservoir has settled into a more stable dynamical configuration, reducing the rate of accretion of cool gas into the nuclear regions and leading to lower-level WLRG/FRI activity. This would be consistent with the evidence for a delay between the merger-induced starburst and the triggering of the current phase of AGN activity based on analysis of the stellar populations in lower-luminosity radio AGN with YSP \citep{tadhunter05,emonts06,tadhunter11}.

One further possible fuel supply, which may be particularly relevant to the WLRG objects, is gas cooling from the  hot X-ray haloes of the host galaxies and clusters and falling into the nuclear regions: the so-called ``cooling flows''.  Indeed, it has been argued that the kinematics of the extended emission line gas in some radio AGN, particuarly WLRG in cluster environments, are consistent with a cooling flow origin for the warm gas \citep{tadhunter89a,baum92}. 

Considering more luminous AGN, cooling flows could be particularly effective under the conditions of chaotic cold accretion 
(Gaspari et al. 2013: see discussion in section 2.4 above) in the centres of rich clusters of galaxies. In this context, it is notable that a cooling flow has been implicated in the triggering of the luminous, quasar-like AGN and starburst at the centre of the Phoenix cluster \citep[$z = 0.596$:][]{mcdonald12}. However, the Phoenix cluster is extreme in its X-ray properties --- it is one of the most X-ray luminous galaxy clusters with one of the highest cooling rates. Therefore, it is not clear whether this mechanism is significant for the majority of SLRG, which are in relatively low density galaxy environments.

This discussion of triggering/fuelling mechanisms feeds into the question of whether it is the rate or the mode of accretion that is most important for determining the optical spectral properties of radio AGN (e.g. whether WLRG/LEG or SLRG/HEG: see section 2.4). Broadly, the results on the host galaxies and environments of the radio AGN are consistent with cold mode accretion via galaxy mergers for the SLRG objects, and hot mode accretion of gas from the gaseous haloes of the host galaxies/clusters for the WLRG. However, the modes are not independent of the rates. For example,  because of the rarefied nature of the hot ISM at the centres of galaxies and clusters, hot mode will generally {\it tend} be associated with low rates to accretion. Moreover, it is likely that the types of events associated with cold mode accretion (e.g. galaxy mergers) {\it tend} to lead to higher rates of accretion and Eddington ratios above that required to produce a
SLRG. However, the link between rate and mode is not one-to-one, and cases like Centaurus A suggest that cold accretion at low rates may be possible following a merger, once the gas has settled to a stable configuration. 

Finally, it is interesting that the WLRG/FRII objects, which can be considered misfits in terms of their radio/optical classifications, also appear to show hybrid host/environment properties: while on average they tend to inhabit relatively rich environments that are similar to those WLRG/FRI sources \citep{ramos13,ineson15}, evidence is emerging that their cool gas contents are more similar to those of SLRG/FRII sources \citep{dicken16}; and  one of the WLRG/FRII objects in the 2Jy sample  (PKS0347+05) appears to be involved in a major, gas-rich galaxy merger \citep{tadhunter11}. 
Although intermittency/switch off  may help to explain some of their properties (see section 2.5), the WLRG/FRII sources clearly present a challenge to our current understanding of radio AGN phenomenology. 

\section{Conclusions}
\label{sec:2}

The population of radio AGN shows considerable diversity in  both its AGN and its host galaxy properties. By investigating the causes of this diversity using a multi-wavelength approach, considerable recent progress has been made in understanding the underlying physical mechanisms.
 
\begin{itemize}

\item[-] {\bf Anisotropy and orientation.} A broad range of observations now lend strong support to the  unified schemes that explain the relationship between broad- and narrow-line radio AGN within the SLRG/FRII category in terms of anisotropy and orientation effects.  
The observations also demonstrate that two of the key indicators of the bolometric luminosities of the AGN --- the [OIII] emission line luminosity (L$_{[OIII]}$) and the 24$\mu$m continuum luminosity (L$_{24}$) --- suffer mild (factor $\sim$2 -- 3) attenuation in NLRG, due to the extinction effects of the circum-nuclear dust.

\item[-] {\bf Accretion rates.} Differences in  accretion rates onto the central supermassive black holes can further help to explain the diversity of the optical spectra and radio morphologies of the radio AGN population: the properties of SLRG/FRII sources are consistent with high Eddington accretion ratios,  and those of WLRG/FRI sources with lower Eddington ratios, with the break between the two occurring at $(L_{bol}+Q_{jet})/L_{edd} \sim 10^{-2}$.

\item[-] {\bf Variability.} Long-term, high-amplitude variability of the AGN --- either within a cycle, or in the switch-off phase at the end of a cycle --- can help to explain the properties of the WLRG/FRII objects, whose relationship with the other classes of radio AGN is otherwise difficult to explain in terms of the effects of anisotropy/orientation and/or different accretion physics.

\item[-] {\bf The triggering of SLRG.} The detailed morphological, star formation, cool ISM and environmental properties of SLRG are consistent with them being triggered in galaxy mergers. However, the mergers are likely to be relatively minor in most cases: equivalent to the accretion of twice the gas mass of the LMC; only a minority of the hosts of radio AGN in the local universe ($<$20\%) have the high star formation rates and large cool ISM masses typical of major, gas-rich mergers that would lead to substantial growth of the supermassive black holes and stellar masses. Simulations of galaxy mergers indicate that timescale of this type of quasar-like radio AGN activity is likely to be relatively short \citep[$<$100~Myr:][]{dimatteo05}.

\item[-] {\bf The fuelling of WLRG.} In contrast to the SLRG, the detailed properties of the WLRG are more consistent with their supermassive black holes
being fuelled at low rates: by the accretion of the hot ISM in the host galaxies/clusters, the cool ISM from cooling flows, or cool ISM in the near-nuclear regions that has reached a relatively stable dynamical configuration. This type of radio AGN activity is likely to have a high duty cycle.

\end{itemize}

\section{Outlook}

Promising directions for future research in this field include the following.

\begin{itemize}

\item[*] {\bf Understanding the WLRG/FRII sources.} It is important to use observations of larger samples of WLRG/FRII sources to determine whether they represent a truly uniform population in terms of their host galaxy masses, detailed optical morphologies, large-scale environments, star formation properties and gas contents. By comparison with the properties of SLRG/FRII objects, this will help to establish whether all, or a only a subset, of the WLRG/FRII sources represent objects in which the AGN has recently switched off. In addition,  further detailed spectral ageing studies at radio wavelengths will be key to establishing whether the radio sources of some WLRG/FRII objects represent relics, consistent with the switch-off idea.

\item[*] {\bf What types of mergers trigger luminous radio AGN activity?} The star formation properties and gas contents of the majority of SLRG/FRII sources suggest that they have been triggered in galaxy mergers that are relatively modest, at least in terms of their cool gas masses. Such mergers have the potential to deliver total reservoirs of cool gas that are sufficiently massive to sustain quasar activity on the requisite timescale. However, further numerical simulations are required to establish what types of modest mergers can trigger luminous AGN activity, considering  a range of galaxy properties (e.g. total stellar masses), not just the cool gas contents. On the observational side, it will also be important to use deep high resolution imaging and/or integral field observations of the tidal features to establish the stellar masses of the galaxies merging with the 
giant elliptical hosts of the radio AGN.

\item[*] {\bf The nature of the lower luminosity radio AGN populations.} This review has concentrated the high-luminosity end of the radio source population ($L_{1.4GHz} > 10^{24}$~W Hz$^{-1}$). However, the majority of the radio AGN detected in large-area surveys such as NVSS and FIRST have lower luminosities. Although the SDSS has been used to establish the optical spectral classes of such sources, we know relatively little about their detailed radio and optical morphologies. Therefore further deep, high resolution imaging observations of samples of lower-luminosity radio sources at both radio and optical wavelengths are required to establish their relationship with the high-luminosity radio AGN population, and hence investigate the natures of their AGN and how they are triggered. 


\item[*] {\bf High spatial resolution studies of the cool gas kinematics in the near-nuclear regions.} The new generation of mm-wavelength interferometers such as ALMA has the capability to study the kinematics of the cool gas in the central regions of the nearest radio AGN on scales of $\sim$0.1 --- 1~kpc via observations of the CO emission lines. On such scales, the dynamical timescale of the gas ($\le 10^7$~yr) is comparable with, or less than, the typical lifetimes of the AGN. In this case, detailed mapping of the gas kinematics has the potential to reveal how the gas is transported from kpc-scale dust lanes to sub-100pc-scales --- one of the key outstanding problems in AGN studies. The observations will be challenging because the cool gas masses are modest in most cases (see section 3.4). However, it will be particularly interesting to compare the kinematics of the cool gas in nearby SLRG/FRII and WLRG/FRII objects, to determine whether the relatively dim AGN in the WLRG/FRII (perhaps related to the switch-off phase) are linked to a particular dynamical state of the near-nuclear gas.

\end{itemize}

\begin{acknowledgements}
I am grateful to all my collaborators on the 2Jy project over the last 25 years for their valuable contributions, in particular Raffaella Morganti, Dan Dicken, Cristina Ramos Almeida, Martin Shaw, Bob Dickson, Sperello di Serego Alighieri, Katherine Inskip, Bob Fosbury, Andy Robinson, Joanna Holt, Martin Hardcastle, Beatriz Mingo and Montse Villar-Mart\'in. I also thank Raffaella Morganti and Francesco Palla for suggestions that have improved the manuscript. I acknowledge the NASA Astrophysics Data System, which has greatly assisted me in constructing the bibliography for this review. 
\end{acknowledgements}


\begin{thebibliography}{aa}

\bibitem[\protect\citeauthoryear{Adams} {1977}]{adams77}
Adams, T.F., 1977, ApJS, 33, 19


\bibitem[\protect\citeauthoryear{Allen et al.} {2002}]{allen02}
Allen, M.G., Sparks, W.B., Koekemoer, A., Martel, A.R., O'Dea, C.P., Baum, S.A.,
Chiaberge, M., Macchetto, F.D., Miley, G.K., ApJSS, 139, 411


\bibitem[\protect\citeauthoryear{Antonucci} {1982}]{antonucci82}
Antonucci, R.R.J., 1982, Nat, 299, 605

\bibitem[\protect\citeauthoryear{Antonucci} {1984}]{antonucci84}
Antonucci, R.R.J., 1984, ApJ, 278, 499

\bibitem[\protect\citeauthoryear{Antonucci \& Ulvestad} {1985a}]{antonucci85a}
Antonucci, R.R.J., Ulvestad, J.S., 1985a, ApJ, 294, 158

\bibitem[\protect\citeauthoryear{Antonucci \& Miller} {1985b}]{antonucci85b}
Antonucci, R.R.J., Miller, J.S., 1985b, ApJ, 297, 621


\bibitem[\protect\citeauthoryear{Appleton et al.} {2004}]{appleton04}
Appleton, P.N., et al., 2006, ApJS, 639, L54

\bibitem[\protect\citeauthoryear{Aretxaga et al.} {2001}]{aretxaga01}
Aretxaga, I., Terlevich, E., Terlevich, R., Cotter, G., Diaz, A.I., 2011, MNRAS, 325, 636

\bibitem[\protect\citeauthoryear{Baade \& Minkowski} {1954}]{baade54}
Baade, W., Minkowski, R., 1954, ApJ, 119, 215

\bibitem[\protect\citeauthoryear{Bagchi et al.} {2014}]{bagchi14}
Bagchi, J., Vivek, M., Vikram, V., Hota, A., Biju, K., Sirothia, S.K., Srianand, R., 
Gopal-Krishna, Jacob, J., 2014, ApJ, 788, 174

\bibitem[\protect\citeauthoryear{Baldi \& Capetti} {2008}]{baldi08}
Baldi, R.D., Capetti, A., 2008, A\&A, 489, 989

\bibitem[\protect\citeauthoryear{Baldi et al.} {2010}]{baldi10}
Baldi, R., Chiaberge, M., Capetti, A., Sparks, W., Machetto, F.D., O\'Dea, C.P.,
Axon, D.J., Baum, S.A., Quillen, A.C., 2010, ApJ, 725, 2426

\bibitem[\protect\citeauthoryear{Baldi et al.} {2013}]{baldi13}
Baldi, R.D., Chiaberge, M., Capetti, A., Rodr\'iguez-Zaur\'in, J., 
Deustua, S., Sparks, W.B., 2013, ApJ, 762, 30

\bibitem[\protect\citeauthoryear{Baldi et al.} {2015}]{baldi15}
Baldi, R.D., Capetti, A., Giovannini, G., 2015, A\&A, 576, 38

\bibitem[\protect\citeauthoryear{Balmaverde, Capetti \& Grandi} {2006}]{balmaverde06}
Balmaverde, B., Capetti, A., Grandi, P., 2006, A\&A, 451, 35

\bibitem[\protect\citeauthoryear{Barthel} {1989}]{barthel89}
Barthel, P.D., 1989, ApJ, 336, 606

\bibitem[\protect\citeauthoryear{Barthel \& Arnaud} {1996}]{barthel96}
Barthel, P.D., Arnaud, K.A., 1996, MNRAS, 283, L45


\bibitem[\protect\citeauthoryear{Baum et al.} {1988}]{baum88}
Baum, S.A., Heckman, T.M., Bridle, A., van Breugel, W.J.M., Miley, G.K., 1988, ApJS, 68, 643

\bibitem[\protect\citeauthoryear{Baum, Heckman \& van Breugel} {1992}]{baum92}
Baum, S.A., Heckman, T.M., van Breugel, W., 1992, ApJ, 389, 208

\bibitem[\protect\citeauthoryear{Baum et al.} {2010}]{baum10}
Baum, S.A., Gallimore, J.F., O\'Dea, C., Buchanan, C.L., Noel-Storr, J., Axon, D.J., 
Robinson, A., Elitzer, M., Dorn, M., Staudaher, S., 2010, ApJ, 710, 289

\bibitem[\protect\citeauthoryear{Bennert et al.} {2008}]{bennert08}
Bennert, N., Canalizo, G., Jungwiert, B., Stockton, A., Schweizer, F., Peng, C.Y., Lacy, M., 2008, ApJ, 677, 846

\bibitem[\protect\citeauthoryear{Bennett} {1962a}]{bennett62a}
Bennett, A.S., 1962, MNRAS, 125, 75

\bibitem[\protect\citeauthoryear{Bennett} {1962b}]{bennett62b}
Bennett, A.S., 1962, MmRAS, 68, 163

\bibitem[\protect\citeauthoryear{Benson et a.} {2003}]{benson03}
Benson, A.J., Bower, R.G., Frenk, C.S., Lacey, C.G., Baugh, C.M.,
Cole, S., 2003, ApJ, 599, 38

\bibitem[\protect\citeauthoryear{Bessiere et al.} {2012}]{bessiere12}
Bessiere, P.S., Tadhunter, C.N., Ramos Almeida, C., Villar-Mart\'in, M., 2012, MNRAS, 426, 276

\bibitem[\protect\citeauthoryear{Best et al.} {2005}]{best05}
Best, P.N., Kauffmann, G., Heckman, T.M., Brinchmann, J., Charlot, S., Ivezi\'c, \u{Z}., White, S.D.M., 2005,
MNRAS, 362, 25


\bibitem[\protect\citeauthoryear{Best \& Heckman} {2012}]{best12}
Best, P.N., Heckman, T.M., 2012, MNRAS, 421, 1569



\bibitem[\protect\citeauthoryear{Bettoni et al.} {2001}]{bettoni01}
Bettoni, D., Falomo, R., Fasano, G., Govoni, F., Salvo, M., Scarpa, R., 2001, A\&A, 380, 471

\bibitem[\protect\citeauthoryear{Black et al.} {1992}]{black92}
Black, A.R.S., Baum, S.A., Leahey, J.P., Riley, J.M., Scheuer, P.A.G., 1992, MNRAS, 256, 186

\bibitem[\protect\citeauthoryear{Bolton, Stanley \& Slee} {1949}]{bolton49}
Bolton, J.G., Stanley, G.J., Slee, O.B., 1949, Nat, 164, 101

\bibitem[\protect\citeauthoryear{Bolton, Gardner \& Mackey} {1963}]{bolton63}
Bolton, J.G., Gardner, F.F., Mackey, M.B., 1963, Nat, 199, 681

\bibitem[\protect\citeauthoryear{Bondi} {1952}]{bondi52}
Bondi, H., 1952, MNRAS, 112, 195

\bibitem[\protect\citeauthoryear{Buttiglione et al.} {2009}]{buttiglione09}
Buttiglione, S., Capetti, A., Celotti, A., Axon, D.J., Chaiberge, M., Machetto, F.D., Sparks, W.B., 2009,
A\&A, 495, 1033

\bibitem[\protect\citeauthoryear{Buttiglione et al.} {2010}]{buttiglione10}
Buttiglione, S., Capetti, A., Celotti, A., Axon, D. J., Chiaberge, M., Macchetto, F. D., Sparks, W. B., 
2010, A\&A, 509, 6

\bibitem[\protect\citeauthoryear{Buttiglione et al} {2011}]{buttiglione11}
Buttiglione, S., Capetti, A., Celotti, A., Axon, D.J., Chaiberge, M., Machetto, F.D., Sparks, W.B., 2011, 
A\&A, 525, 28

\bibitem[\protect\citeauthoryear{Calzetti et al.} {2010}]{calzetti10}
Calzetti, D., et al., 2010, ApJ, 714, 1256

\bibitem[\protect\citeauthoryear{Canalizo \& Stockton} {2001}]{canalizo01}
Canalizo, G., Stockton, A., 2001, ApJ, 555, 719

\bibitem[\protect\citeauthoryear{Cao \& Rawlings} {2004}]{cao04}
Cao, X., Rawlings, S., 2004, MNRAS, 349, 1419

\bibitem[\protect\citeauthoryear{Capetti et al.} {2002}]{capetti02}
Capetti, A., Celotti, A., Chiaberge, M., de Ruiter, H.R., Fanti, R., Morganti, R., Parma, P., 2002, A\&A, 383, 104

\bibitem[\protect\citeauthoryear{Capetti, Verdoes Keijn \& Chiaberge} {2005}]{capetti05}
Capetti, A., Verdoes Kleijn, G., Chiaberge, M., 2005, A\&A, 439, 935



\bibitem[\protect\citeauthoryear{Capetti et al.} {2007}]{capetti07}
Capetti, A., Axon, D.J., Chiaberge, M., Sparks, W.B., Macchetto, F.D., Cracraft, M., Celotti, A., 2007, A\&A, 471, 137

\bibitem[\protect\citeauthoryear{Capetti et al.} {2011}]{capetti11}
Capetti, A., Buttiglione, S., Axon, D.J., Robinson, A., Celotti, A., Baldi, R.D., Chiaberge, M., Macchetto, F.D., Sparks, W.B., 2011, A\&A, 527, L2

\bibitem[\protect\citeauthoryear{Capetti et al.} {2013}]{capetti13}
Capetti, A., Robinson, A., Baldi, R.D., Buttiglione, S., Axon, D.J., Celotti, A., Chiaberge, M., 2013, A\&A,
551, 55

\bibitem[\protect\citeauthoryear{Carilli \& Barthel} {1996}]{carilli96}
Carilli, C.L., Barthel, P.D., 1996, A\&ARv, 7, 1



\bibitem[\protect\citeauthoryear{Chiaberge, Capetti \& Celotti} {1999}]{chiaberge99}
Chiaberge, M. Capetti, A., Celotti, A., 1999, A\&A, 349, 77

\bibitem[\protect\citeauthoryear{Chiaberge, Capetti \& Celotti} {2002}]{chiaberge02}
Chiaberge, M., Capetti, A., Celotti, A., 2002, A\&A, 394, 791

\bibitem[\protect\citeauthoryear{Chiaberge \& Marconi} {2011}]{chiaberge11}
Chiaberge, M., Marconi, A., 2011, MNRAS, 416, 917

\bibitem[\protect\citeauthoryear{Chiaberge et al.} {2015}]{chiaberge15}
Chiaberge, M., Gilli, R., Lotz, J.M., Norman, C., 2015, ApJ, 806, 147

\bibitem[\protect\citeauthoryear{Cisternas et al.} {2011}]{cisternas11}
Cisternas, M., et al., 2011, ApJ, 726, 57

\bibitem[\protect\citeauthoryear{Clavel \& Wamsteker} {1987}]{clavel87}
Clavel, J., Wamsteker, W., 1987, ApJ, 320, L9

\bibitem[\protect\citeauthoryear{Cleary et al.} {2007}]{cleary07}
Cleary, K., Lawrence, C.R., Marshall, J.A., Hao, L., Meier, D., 2007, ApJ, 660, 117

\bibitem[\protect\citeauthoryear{Cohen et al.} {1977}]{cohen77}
Cohen, M.H., Linfield, R.P., Moffet, A.T., Seielstad, G.A., Kellermann, K.I., Shaffer, D.B., Pauliny-Toth, I.I.K., 
Preuss, E., Witzel, A., Romney, J.D., 1977, Nat, 268, 405

\bibitem[\protect\citeauthoryear{Cohen et al.} {1999}]{cohen99}
Cohen, M.H., Ogle, P.M., Tran, H.D., Goodrich, R.W., Miller, J.S., 1999, AJ, 118, 1963

\bibitem[\protect\citeauthoryear{Cole et al.} {2001}]{cole01}
Cole, S., et al., 2001, MNRAS, 326, 255

\bibitem[\protect\citeauthoryear{Colina \& De Juan} {1995}]{colina95}
Colina, L., de Juan, L., 1995, ApJ, 448, 548

\bibitem[\protect\citeauthoryear{Condon} {1989}]{condon89}
Condon, J.J., 1989, ApJ, 338, 13

\bibitem[\protect\citeauthoryear{Corbett et al.} {1996}]{corbett96}
Corbett, E.A., Robinson, A., Axon, D.J., Hough, J.H., Jeffries, R.D., Thurston, M.R., 
Young, S., 1996, MNRAS, 281, 737

\bibitem[\protect\citeauthoryear{Costero \& Osterbrock} {1977}]{costero77}
Costero, R., Osterbrock, D.E., 1977, ApJ, 211, 675

\bibitem[\protect\citeauthoryear{C\^ot\'e et al.} {1997}]{cote97}
C\^ot\'e, S., Freeman, K.C., Carignan, C., Quinn, P.J., 1997, ApJS, 103, 81

\bibitem[\protect\citeauthoryear{Daddi et al.} {2010}]{daddi10}
Daddi, E., et al., 2010, ApJ, 714, L118


\bibitem[\protect\citeauthoryear{de Koff et al.} {2000}]{dekoff00}
de Koff, S., Best, P., Baum, S.A., Sparks, W., R\"ottgering, H., Miley, G., Golombek, D., 
Macchetto, F., Martel, A., 2000, ApJS, 129, 33

\bibitem[\protect\citeauthoryear{de Vaucouleurs} {1948}]{devauc48}
de Vaucouleurs, G., 1948, AnAp, 11, 247

\bibitem[\protect\citeauthoryear{di Matteo, Springel \& Hernquist} {2005}]{dimatteo05}
di Matteo, T., Springel, V., Hernquist, L., 2005, Nat, 433, 604

\bibitem[\protect\citeauthoryear{Dicken et al.} {2008}]{dicken08}
Dicken, D., Tadhunter, C., Morganti, R., Buchanan, C., Oosterloo, T., Axon, D.,2008, ApJ, 678, 712

\bibitem[\protect\citeauthoryear{Dicken et al.} {2009}]{dicken09}
Dicken, D., Tadhunter, C., Morganti, R.,Axon, D., Morganti, R., Inskip, K. J., Holt, J., Gonz\'alez Delgado, R., 
Groves, B., 2009,
ApJ, 694, 268

\bibitem[\protect\citeauthoryear{Dicken et al.} {2010}]{dicken10}
Dicken, D., Tadhunter, C., Axon, D., Robinson, A., Morganti, R., Kharb, P., 2010, ApJ, 722, 1333

\bibitem[\protect\citeauthoryear{Dicken et al.} {2012}]{dicken12}
Dicken, D., et al., 2012, ApJ, 745, 172

\bibitem[\protect\citeauthoryear{Dicken et al.} {2014}]{dicken14}
Dicken, D., et al., 2014, ApJ, 788, 98

\bibitem[\protect\citeauthoryear{Dicken et al.} {2016}]{dicken16}
Dicken, D., et al., 2016, MNRAS, in preparation

\bibitem[\protect\citeauthoryear{Dickson et al.} {1995}]{dickson95}
Dickson, R., Tadhunter, C., Shaw, M., Clark, N., Morganti, R., 1995, MNRAS, 273, L29

\bibitem[\protect\citeauthoryear{Dietrich et al.} {2012}]{dietrich12}
Dietrich, M., Peterson, B.M., Grier, C.J., Bentz, M.C., Eastman, J., Frank, S., Gonzalez, R., Marshall, J.L., DePoy, D.L., 
Prieto, J.L., 2012, ApJ, 757, 53

\bibitem[\protect\citeauthoryear{di Serego Alighieri et al.} {1994}]{diserego94}
di Serego-Alighieri, S., Danziger, I.J., Morganti, R., Tadhunter, C.N., 1994, MNRAS, 269, 998


\bibitem[\protect\citeauthoryear{Djorgovski et al.} {1991}]{djorgovski91}
Djorgovski, S., Weir, N., Matthews, K., Graham, J.R., 1991, ApJ, 372, L67

\bibitem[\protect\citeauthoryear{Donzelli et al.} {2007}]{donzelli07}
Donzelli, C.J., Chiaberge, M., Machetto, F.D., Madrid, J.P., Capetti, A., Marchesini, D., 2007, ApJ, 667, 780

\bibitem[\protect\citeauthoryear{Duc et al.} {2015}]{duc15}
Duc, P-A., et al., 2015, MNRAS, 446, 120

\bibitem[\protect\citeauthoryear{Dunlop et al.} {2003}]{dunlop03}
Dunlop, J.S., McLure, R.J., Kukula, M.J., Baum, S.A., O'Dea, C.P.O., Hughes, D.H., 2003, MNRAS, 340, 1095

\bibitem[\protect\citeauthoryear{Elbaz et al.} {2007}]{elbaz07}
Elbaz, D., et al., 2007, A\&A, 468, 33

\bibitem[\protect\citeauthoryear{Elvis} {2012}]{elvis12}
Elvis, M., 2013, JPhCS, 372, 2032

\bibitem[\protect\citeauthoryear{Emonts et al.} {2006}]{emonts06}
Emonts, B.H.C., Morganti, R., Tadhunter, C.N., Holt, J., Oosterloo, T.A., van der Hulst, J.M., Wills, K.A., 2006,
A\&A, 454, 125

\bibitem[\protect\citeauthoryear{Emonts et al.} {2008}]{emonts08}
Emonts, B.H.C., Morganti, R., van Gorkham, J.H., Oosterloo, T.A., Brogt, E., Tadhunter, C.N., 2008, A\&A, 488, 519

\bibitem[\protect\citeauthoryear{Evans et al.} {2005}]{evans05}
Evans, A.S., Mazzarella, J.M., Surace, J.A., Frayer, D.T., Iwasawa, K., Sanders, D.B., 2005, ApJSS, 159, 197

\bibitem[\protect\citeauthoryear{Fabian} {2012}]{fabian12}
Fabian, A.C., 2012, ARA\&A, 50, 455

\bibitem[\protect\citeauthoryear{Falcke, K\"ording \& Markoff} {2004}]{falcke04}
Falcke, H., K\"ording, E., Markoff, S., 2004, A\&A, 414, 895

\bibitem[\protect\citeauthoryear{Fanaroff \& Riley} {1974}]{fanaroff74}
Fanaroff, B.L., Riley, J.M., 1974, MNRAS, 167, P31

\bibitem[\protect\citeauthoryear{Fanti et al.} {1990}]{fanti90}
Fanti, R., Fanti, C., Schilizzi, R.T., Spencer, R.E., Nam Rendong, Parma, P,
van Breugel, W.J.M., Venturi, T., 1990, A\&A, 231, 333

\bibitem[\protect\citeauthoryear{Fanti et al.} {1995}]{fanti95}
Fanti, C., Fanti, R., Dallacasa, D., Schillizi, R.T., Spencer, R.E., Stanghellini, C.,
1995, A\&A, 302, 317

\bibitem[\protect\citeauthoryear{Fitch et al.} {1967}]{fitch67}
Fitch, W.S., Pacholczyk, A.G., Weymann, R.J., ApJ, 150, L67

\bibitem[\protect\citeauthoryear{Garrington et al.} {1988}]{garrington88}
Garrington, S.T., Leahey, J.P., Conway, R.G., Laing, R.A., 1988, Nat, 331, 147

\bibitem[\protect\citeauthoryear{Gaspari, Peng \& Oh} {2013}]{gaspari13}
Gaspari, M., Ruszkowski, M., Oh, S.P., MNRAS, 432, 3401

\bibitem[\protect\citeauthoryear{Gaspari, Brighenti \& Temi} {2015}]{gaspari15}
Gaspari, M., Brighenti, F., Temi, P., 2015, A\&A, 579, 62


\bibitem[\protect\citeauthoryear{Ghisellini \& Celotti} {2001}]{ghisellini01}
Ghisellini, G., Celotti, A., 2001, A\&A, 379, L1

\bibitem[\protect\citeauthoryear{Gopal-Krishna, Kulkarni \& Wiita} {1996}]{gopal96}
Gopal-Krishna, Kulkarni, V.K., Wiita, P.J., 1996, ApJ, 463, L1

\bibitem[\protect\citeauthoryear{Govoni et al.} {2000}]{govoni00}
Govoni, E., Falomo, G., Fasano, G., Scarpa, R., A\&A, 353, 507


\bibitem[\protect\citeauthoryear{Grogin et al.} {2005}]{grogin05}
Grogin, N.A., et al., 2005, ApJ, 627, 97

\bibitem[\protect\citeauthoryear{Haas et al.} {2004}]{haas04}
Haas, M., et al., 2004, A\&A, 424, 531

\bibitem[\protect\citeauthoryear{Haas et al.} {2005}]{haas05}
Haas, M., Siebenmorgen, R., Schulz, B., Krugel, E., Chini, R., 2005, A\&A, 442, L39

\bibitem[\protect\citeauthoryear{Hardcastle et al.} {1997}]{hardcastle97}
Hardcastle, M.J., Alexander, P., Pooley, G.G., Riley, J.M., 1997, MNRAS, 288, 859


\bibitem[\protect\citeauthoryear{Hardcastle et al.} {2006}]{hardcastle06}
Hardcastle, M., Evans, D., Croston, J., 2006, MNRAS, 370, 1893

\bibitem[\protect\citeauthoryear{Hardcastle, Evans \& Croston} {2007}]{hardcastle07}
Hardcastle, M., Evans, D.A., Croston, J.H., 2007, MNRAS, 376, 1849

\bibitem[\protect\citeauthoryear{Hardcastle et al.} {2009}]{hardcastle09}
Hardcastle, M.J., Evans, D.A., Croston, J.H., 2009, MNRAS, 396, 1929

\bibitem[\protect\citeauthoryear{Harwood, Hardcastle \& Croston} {2015}]{harwood15}
Harwood, J.J., Hardcastle, M.J., Croston, J.H., 2015, MNRAS, in press (arXiv:1509.06757v1)

\bibitem[\protect\citeauthoryear{Heckman} {1980}]{heckman80}
Heckman, T.M., 1980, A\&A, 87, 152

\bibitem[\protect\citeauthoryear{Heckman et al.} {1982}]{heckman82}
Heckman, T., Miley, G., van Breugel, W., Butcher, H., 1982, ApJ, 262, 529

\bibitem[\protect\citeauthoryear{Heckman et al.} {1986}]{heckman86}
Heckman, T.M., Smith, E.P., Baum, S.A., van Breugel, W.J.M., Miley, G.K., Illingworth, G.D., Bothun, G.D., 
1986, ApJ, 311, 526

\bibitem[\protect\citeauthoryear{Heckman et al.} {2005}]{heckman05}
Heckman, T.M., Ptak, A., Hornschemeier, A., Kaukkmann, G., 2005, ApJ, 634, 161

\bibitem[\protect\citeauthoryear{Heckman \& Best} {2014}]{heckman14}
Heckman, T.M., Best, P.N., 2014, ARA\&A, 52, 589

\bibitem[\protect\citeauthoryear{Herbert et al.} {2010}]{herbert10}
Herbert, P.D., Jarvis, M.J., Willott, C.J., McLure, R.J., Mitchell, E., Rawlings, S., Hill, G.J., Dunlop, J.S., 2010,
MNRAS, 406, 1841

\bibitem[\protect\citeauthoryear{Hes, Barthel \& Fosbury} {1993}]{hes93}
Hes, R., Barthel, P.D., Fosbury, R.A.E., 1993, Nat, 362, 326

\bibitem[\protect\citeauthoryear{Hill et al.} {1996}]{hill96}
Hill, G.J., Goodrich, R.W., Depoy, D.L., 1996, ApJ, 462, 163

\bibitem[\protect\citeauthoryear{Hine \& Longair} {1979}]{hine79}
Hine, R.G., Longair, M.S., 1979, MNRAS, 188, 111

\bibitem[\protect\citeauthoryear{Holt, Tadhunter \& Morganti} {2003}]{holt03}
Holt, J., Tadhunter, C.N., Morganti, R., 2003, MNRAS, 342, 227

\bibitem[\protect\citeauthoryear{Holt, Tadhunter \& Morganti} {2008}]{holt08}
Holt, J., Tadhunter, C.N., Morganti, R., 2008, MNRAS, 387, 639


\bibitem[\protect\citeauthoryear{Holt et al.} {2007}]{holt07}
Holt, J., Tadhunter, C.N., Gonz\'alez Delgado, R.M., Inskip, K.J., Rodr\'iguez Zaur\'in, J., Emonts, B.H.C., 
Morganti, R., Wills, K.A., 2007, MNRAS, 381, 611


\bibitem[\protect\citeauthoryear{Hota et al.} {2011}]{hota11}
Hota, A., Sirothia, S.K., Ohyama, Y., Konar, C., Kim, S., Rey, S-C, Saikia, D., Croston, J.H., 
Matsushita, S., 2011, MNRAS, 417, L36

\bibitem[\protect\citeauthoryear{Ghisellini} {2011}]{ghisellini11}
Ghisellini, G., 2011, American Institute of Physics Conference Series, 1381, 180

\bibitem[\protect\citeauthoryear{Ibar et al.} {2008}]{ibar08}
Ibar, E., et al., 2008, MNRAS, 386, 953

\bibitem[\protect\citeauthoryear{Ineson et al.} {2015}]{ineson15}
Ineson, J., Croston, J.H., Hardcastle, M.J., Kraft, R.P., Evans, D.A., Jarvis, M., 2015, MNRAS, 453, 268

\bibitem[\protect\citeauthoryear{Inskip et al.} {2007}]{inskip07}
Inskip, K.J., Tadhunter, C.N., Dicken, D., Holt, J., Villar-Mart\'in, M., Morganti, R., 2007, MNRAS, 382, 95

\bibitem[\protect\citeauthoryear{Inskip et al.} {2010}]{inskip10}
Inskip, K., Tadhunter, C.N., Morganti, R., Holt, J., Ramos Almeida, C., Dicken, D., 2010, MNRAS, 407, 1739

\bibitem[\protect\citeauthoryear{Israel} {1998}]{israel98}
Israel, F.P., 1998, A\&ARv, 8, 237

\bibitem[\protect\citeauthoryear{Jackson \& Browne} {1990}]{jackson90}
Jackson, N., Browne, I.W.A., 1990, Nat, 343, 43

\bibitem[\protect\citeauthoryear{Jackson \& Rawlings} {1997}]{jackson97}
Jackson, N., Rawlings, S., 1997, MNRAS, 286, 241

\bibitem[\protect\citeauthoryear{Jackson, Tadhunter \& Sparks} {1998}]{jackson98}
Jackson, N., Tadhunter, C., Sparks, W.B., 1998, MNRAS, 301, 131

\bibitem[\protect\citeauthoryear{Kapahi et al.} {1995}]{kapahi95}
Kapahi, V.K., Athreya, R.M., Subrahmanya, C.R., Hunstead, R.W., Baker, J.C., McCarthy, P.J., 
van Breugel, W., Hunstead, R.W., Baker, J.C., McCarthy, P.J., van Breugel, W., 1995, ApAS, 16, 125

\bibitem[\protect\citeauthoryear{Kaviraj et al.} {2011}]{kaviraj11}
Kaviraj, S., Tan, K.-M., Ellis, R.S., Silk, J., 2011, MNRAS, 411, 2148

\bibitem[\protect\citeauthoryear{Keel et al.} {2006}]{keel06}
Keel, W.C., While, R.E., Owen, F.N., Ledlow, M.J., 2006, AJ, 132

\bibitem[\protect\citeauthoryear{Kellermann} {1966}]{kellermann66}
Kellermann, K.I., 1966, AuJPh, 19, 577

\bibitem[\protect\citeauthoryear{Kellermann et al.} {1989}]{kellermann89}
Kellermann, K.I., Sramek, R., Schmidt, M., Shaffer, D.B., Green, R., 1989, AJ, 98, 1195

\bibitem[\protect\citeauthoryear{Kewley et al.} {2006}]{kewley06}
Kewley, L.J., Groves, B., Kauffmann, G., Heckman, T., 2006, MNRAS, 372, 961

\bibitem[\protect\citeauthoryear{Khachikian \& Weedman} {1971}]{khachikian71}
Khachikian, E. Ye., Weedman, D.W., 1971, Astrofizika, 7, 389


\bibitem[\protect\citeauthoryear{K\"ording, Jester \& Fender} {2006}]{kording06}
K\"ording, E.G., Jester, S., Fender, R., 2006, MNRAS, 372, 1366

\bibitem[\protect\citeauthoryear{Kormendy \& Ho} {2013}]{kormendy13}
Kormendy, J. Ho, L.C., 2013, ARA\&A, 2013, 51, 511

\bibitem[\protect\citeauthoryear{Kozie\l-Wierzbowska et al.} {2012}]{koziel12}
Kozie\l-Wierzbowska, D., Jamrozy, M., Zola, S., Stachowski, G., Ku\'zmicz, A., 2012, MNRAS, 422, 1546

\bibitem[\protect\citeauthoryear{Lacy et al.} {2001}]{lacy01}
Lacy, M., Laurent-Muehleisen, S.A., Ridgway, S.E., Becker, R.H., White, R.L., 2001, ApJ, 551, L17

\bibitem[\protect\citeauthoryear{Laing, Riley \& Longair} {1983}]{laing83}
Laing, R.A., Riley, J.M., Longair, M.S., 1983, MNRAS, 204, 151

\bibitem[\protect\citeauthoryear{Laing} {1988}]{laing88}
Laing, R.A., 1988, Nat, 331, 149

\bibitem[\protect\citeauthoryear{Laing et al.} {1994}]{laing94}
Laing, R.A., Jenkins, C.R., Wall, J.V., Unger, S.W., 1994, In: The First Stromlo Symposium: The Physics of Active Galaxies, Bicknell, G.V., Dopita,M.A., Quinn, P.J., Eds., PASP Conference Series, 54, 201

\bibitem[\protect\citeauthoryear{Laing \& Bridle} {2014}]{laing14}
Laing, R.A., Bridle, a.H., 2014, MNRAS, 437, 340

\bibitem[\protect\citeauthoryear{LaMassa et al.} {2010}]{lamassa10}
LaMassa, S.M., Heckman, T.M., Ptak, A., Martins, L., Wild, V., Sonnentrucker, P., 2010, ApJ, 720, 786

\bibitem[\protect\citeauthoryear{LaMassa et al.} {2015}]{lamassa15}
LaMassa, S.M., Cales, S., Moran, E.C., Myers, A.D., Richards, G.T., Eracleous, M., Heckman, T.M., Gallo, L.G.,
Urry, C.M., 2015, ApJ, 800, 144


\bibitem[\protect\citeauthoryear{Lawrence} {1991}]{lawrence91}
Lawrence, A., 1991, MNRAS, 252, 586

\bibitem[\protect\citeauthoryear{Lawrence \& Elvis} {2010}]{lawrence10}
Lawrence, A., Elvis, M., 2010, ApJ, 714, 561

\bibitem[\protect\citeauthoryear{Leahy et al.} {1997}]{leahy97}
Leahy, J.P., Black, A.R.S., Dennett-Thorpe, J., Hardcastle, M.J., Komissarov, S.,
Perley, R.A., Riley, J.M., Scheuer, P.A.G., 1997, MNRAS, 291, 20

\bibitem[\protect\citeauthoryear{Leahy} {1993}]{leahy93}
Leahy, J.P., 1993, LNP, 421, 1

\bibitem[\protect\citeauthoryear{Ledlow \& Owen} {1996}]{ledlow96}
Ledlow, M.J., Owen, F. N., 1996, AJ, 112, 9

\bibitem[\protect\citeauthoryear{Ledlow et al.} {2001}]{ledlow01}
Ledlow, M.J., Owen, F.N., Yun, M.S., Hill, J.M., 2001, ApJ, 552, 120

\bibitem[\protect\citeauthoryear{Leipski et al.} {2009}]{leipski09}
Leipski, C., Antonucci, R., Ogle, P., Whysong, D., 2009, ApJ, 701, 891

\bibitem[\protect\citeauthoryear{Lilly \& Longair} {1984}]{lilly84}
Lilly, S.J., Longair, M.S., 1984, MNRAS, 211, 833


\bibitem[\protect\citeauthoryear{Longair \& Seldner} {1979}]{longair79}
Longair, M.S., Seldner, M., 1979, MNRAS, 189, 433

\bibitem[\protect\citeauthoryear{Lopez-Rodriguez et al.} {2014}]{lopez14}
Lopez-Rodriguez, E., et al., 2014, ApJ,  793, 81

\bibitem[\protect\citeauthoryear{Macarrone} {2003}]{maccarone03}
Maccarone, T.J., 2003, A\&A, 409, 697

\bibitem[\protect\citeauthoryear{MacLeod et al.} {2015}]{macleod15}
MacLeod, C.L., et al., 2015, MNRAS, in press (arXiv:1509.08393v1)

\bibitem[\protect\citeauthoryear{Madrid et al.} {2006}]{madrid06}
Madrid, J.P., et al., 2006, ApJS, 164, 307

\bibitem[\protect\citeauthoryear{Maiolino et al.} {2001}]{maiolino01}
Maiolino, R., Marconi, A., Salvati, M., Risaliti, G., Severgini, P.,
Oliva, E., La Franca, F., Vanzi, L., 2001, A\&A, 365, 28

\bibitem[\protect\citeauthoryear{Mao et al.} {2015}]{mao15}
Mao, M.Y., et al., 2015, MNRAS, 446, 4176

\bibitem[\protect\citeauthoryear{Malin, Quinn \& Graham} {1983}]{malin83a}
Malin, D.F., Quinn, P.J., Graham, J.A., 1983, ApJ, 272, L5

\bibitem[\protect\citeauthoryear{Malin \& Carter} {1983}]{malin83b}
Malin, D.F., Carter, D., 1983, ApJ, 274, 534


\bibitem[\protect\citeauthoryear{Massaro et al.} {2010}]{massaro10}
Massaro, F., et al., 2010, ApJ, 714,589

\bibitem[\protect\citeauthoryear{Massaro et al.} {2012}]{massaro12}
Massaro, F., et al., 2012, ApJS, 203, 31

\bibitem[\protect\citeauthoryear{Massaro et al.} {2015}]{massaro15}
Massaro, F., et al., 2015, ApJS, 220, 5

\bibitem[\protect\citeauthoryear{Matthews \& Sandage} {1963}]{matthews63}
Matthews, T.A., Sandage, A.R., 1963, AJ, 68, 77

\bibitem[\protect\citeauthoryear{Matthews, Morgan \& Schmidt} {1964}]{matthews64}
Matthews, T.A., Morgan, W.W., Schmidt, M., 1964, ApJ, 140, 35

\bibitem[\protect\citeauthoryear{Mazzotta et al.} {2004}]{mazzotta04}
Mazzotta, P., Brunetti, G., Giancintucci, S., Venturi, T., Bardelli, S., 2004, JKAS, 37, 381

\bibitem[\protect\citeauthoryear{McCarthy} {1993}]{mccarthy93}
McCarthy, P.J., 1993, ARA\&A, 31, 639

\bibitem[\protect\citeauthoryear{McDermid et al.} {2015}]{mcdermid15}
McDermid, R.M., et al., 2015, MNRAS, 448, 3484

\bibitem[\protect\citeauthoryear{McDonald et al.} {2012}]{mcdonald12}
McDonald, M., et al., 2012, Nat, 488, 349

\bibitem[\protect\citeauthoryear{McLure et al.} {1999}]{mclure99}
McLure, R.J., Kukula, M.J., Dunlop, J.S., Baum, S.A., O'Dea, C.P., Hughes, D.H., 1999, MNRAS, 308, 377

\bibitem[\protect\citeauthoryear{McNamara \& Nulsen} {2007}]{mcnamara07}
McNamara, B.J., Nulsen, P.E.J., 2007, ARA\&A, 45, 117

\bibitem[\protect\citeauthoryear{Merloni, Heinz \& di Matteo} {2003}]{merloni03}
Merloni, A., Heinz, S., di Matteo, T., MNRAS, 345, 1057

\bibitem[\protect\citeauthoryear{Meurs \& Wilson} {1984}]{meurs84}
Meurs, E.J.A., Wilson, A.S., 1984, A\&A, 136, 206

\bibitem[\protect\citeauthoryear{Miley} {1980}]{miley80}
Miley, G., 1980, ARA\&A, 18, 165

\bibitem[\protect\citeauthoryear{Miley \& de Breuck} {2008}]{miley08}
Miley, G., de Breuck, C., 2008, A\&ARv, 15, 67

\bibitem[\protect\citeauthoryear{Mingo et al.} {2014}]{mingo14}
Mingo, B., Hardcastle, M.J., Croston, J.H., Dicken, D., Evans, D. A., Morganti, R., Tadhunter, C., 2014, 
MNRAS, 440, 269

\bibitem[\protect\citeauthoryear{Morganti et al.} {1993}]{morganti93}
Morganti, R., KIlleen, N.E.B., Tadhunter, C.N., 1993, MNRAS, 263, 1023

\bibitem[\protect\citeauthoryear{Morganti et al.} {1997a}]{morganti97a}
Morganti, R., Oosterloo, T.A., Reynolds, J.E., Tadhunter, C.N., Migenes, V., 1997a, MNRAS, 284, 541

\bibitem[\protect\citeauthoryear{Morganti et al.} {1997b}]{morganti97b}
Morganti, R., Tadhunter, C.N., Dickson, R., Shaw, M., 1997b, A\&A, 326, 130

\bibitem[\protect\citeauthoryear{Morganti et al.} {1999}]{morganti99}
Morganti, R., Oosterloo, T, Tadhunter, C.N., Aiudi, R., Jones, P., Villar-Mart\'in, M., 1999, A\&ASS, 140, 355

\bibitem[\protect\citeauthoryear{Morganti et al.} {2005}]{morganti05}
Morganti, R., Tadhunter, C.N., Oosterloo, T.A., 2005, A\&A, 444, L9

\bibitem[\protect\citeauthoryear{Morganti et al.} {2011}]{morganti11}
Morganti, R., Holt, J., Tadhunter, C., Ramos Almeida, C., Dicken, D., Inskip, K., Oosterloo, T., Tzioumis, T., 2011,
A\&A, 535, 97

\bibitem[\protect\citeauthoryear{Morganti et al.} {2013}]{morganti13}
Morganti, R., Fogasy, J., Paragi, Z., Oosterloo, T., 2013, Sci, 341, 1082

\bibitem[\protect\citeauthoryear{Mullaney et al.} {2011}]{mullaney11}
Mullaney, J.R., Alexander, D.M., Goulding, A.D., Hickox, R.C., 2011, MNRAS, 414, 1082



\bibitem[\protect\citeauthoryear{Murgia et al.} {2011}]{murgia11}
Murgia, M., Parma, P., Mack, K.-H., de Ruiter, H.R., Fanti, R., Govoni, F.,
Tarchi, A., Giancintucci, S., Markevitch, M., 2011, A\&A, 526, 148



\bibitem[\protect\citeauthoryear{Netzer et al} {2007}]{netzer07}
Netzer, H., et al., 2007, ApJ, 666, 806

\bibitem[\protect\citeauthoryear{Nipoti, Blundell \& Binney} {2005}]{nipoti05}
Nipoti, C., Blundell, K.M., Binney, J., 2005, MNRAS, 361, 633

\bibitem[\protect\citeauthoryear{Oca\~na Flaquer} {2010}]{ocana10}
Oca\~na Flaquer, B., Leon, S., Combes, F., Lim, J., 2010, A\&A, 518, 9

\bibitem[\protect\citeauthoryear{O'Dea, Baum \& Stanghellini} {1991}]{odea91}
O'Dea, C. P., Baum, S.A., Stanghellini, C., 1991, ApJ, 380, 66


\bibitem[\protect\citeauthoryear{O'Dea} {1998}]{odea98}
O'Dea, C.P., 1998, PASP, 192, 1603

\bibitem[\protect\citeauthoryear{O'Dea et al.} {2001}]{odea01}
O'Dea, C., Koekemoer, A.M., Baum, S.A., Sparks, W.B., Martel, A.R., Allen, M.G., Macchetto, F.D., 2001, AJ, 121


\bibitem[\protect\citeauthoryear{Ogle et al.} {1997}]{ogle97}
Ogle, P.M., Cohen, M.H., Miller, J.S., Tran, H.D., Fosbury, R.A.E., Goodrich, R.W., 1997, ApJ, 482, L37 

\bibitem[\protect\citeauthoryear{Ogle et al.} {2006}]{ogle06}
Ogle, P., Whysong, D., Antonucci, R., 2006, ApJ, 647, 161

\bibitem[\protect\citeauthoryear{Osterbrock \& Miller} {1975}]{osterbrock75}
Osterbrock, D.E., Miller, J.S., 1975, ApJ, 197, 535

\bibitem[\protect\citeauthoryear{Osterbrock, Koski \& Philips} {1976}]{osterbrock76}
Osterbrock, D.E., Koski, A.T., Phillips, M.M., 1976, ApJ, 206, 898

\bibitem[\protect\citeauthoryear{Owen \& Laing} {1989}]{owen89}
Owen, F.N., Laing, R.A., 1989, MNRAS, 238, 357

\bibitem[\protect\citeauthoryear{Owen} {1993}]{owen93}
Owen, F.N., 1993, Lecture notes in Physics, 421, 273

\bibitem[\protect\citeauthoryear{Owsianik, Conway \& Polatidis} {1998}]{owsianik98}
Owsianik, I., Conway, J.E., Polatidis, A.G., 1998, A\&A, 336, L37

\bibitem[\protect\citeauthoryear{Parkin et al.} {2012}]{parkin12}
Parkin, T.J., et al., 2012, MNRAS, 422, 2291

\bibitem[\protect\citeauthoryear{Pedlar et al.} {1990}]{pedlar90}
Pedlar, A., Ghataure, H.S., Davies, R.D., Harrison, B.A., Perley, R., Crane, P.C., Unger, S.W.,
1990, MNRAS, 246, 477

\bibitem[\protect\citeauthoryear{Penston \& Cannon} {1971}]{penston71}
Penston, M.V., Cannon, R.D., 1970, R. Obs. Bull., No. 159, 85

\bibitem[\protect\citeauthoryear{Penston \& P\'erez} {1984}]{penston84}
Penston, M.V., P\'erez, E., 1984, MNRAS, 211, P33

\bibitem[\protect\citeauthoryear{Podigachoski et al.} {2015}]{podigachoski15}
Podigachoski, O., et al., 2015, A\&A, 575, 80

\bibitem[\protect\citeauthoryear{Pogge} {1988}]{pogge88}
Pogge, R.W., 1988, ApJ, 328, 519

\bibitem[\protect\citeauthoryear{Polatidis \& Conway} {2003}]{polatidis03}
Polatidis, A.G., Conway, J.E., 2003, PASA, 2003, 20, 69

\bibitem[\protect\citeauthoryear{Pozzetti et al.} {2010}]{pozzetti10}
Pozzetti, L., et al., 2010, A\&A, 523, 13

\bibitem[\protect\citeauthoryear{Prestage \& Peacock} {1988}]{prestage88}
Prestage, R.M., Peacock, J.A., 1988, MNRAS, 230, 131

\bibitem[\protect\citeauthoryear{Ram\'irez et al.} {2009}]{ramirez09}
Ram\'irez, E.A., Tadhunter, C.N., Axon, D., Batcheldor, D., Young, S., Packham, C., Sparks, W.B.,
2009, AJ, 138, 991

\bibitem[\protect\citeauthoryear{Ram\'irez et al.} {2014a}]{ramirez14a}
Ram\'irez, E.A., Tadhunter, C.N., Dicken, D., Rose, M., Axon, D., Sparks, W., Packham, C., 2014a, MNRAS, 439, 1270

\bibitem[\protect\citeauthoryear{Ram\'irez et al.} {2014b}]{ramirez14b}
Ram\'irez, E.A., Tadhunter, C.N., Axon, D., Batcheldor, D., Packham, C., Lopez-Rodriguez, E.
Sparks, W., Young, S., 2014b, MNRAS, 444, 466

\bibitem[\protect\citeauthoryear{Ramos Almeida et al.} {2011a}]{ramos11a}
Ramos Almeida, C., Ramos Almeida, C., Tadhunter, C. N., Inskip, K. J., Morganti, R., Holt, J., Dicken, D. 2011, MNRAS, 410, 1550

\bibitem[\protect\citeauthoryear{Ramos Almeida et al.} {2011b}]{ramos11b}
Ramos Almeida, C., Dicken, D., Tadhunter, C., Asensio Ramos, A., Inskip, K.J., Hardcastle, M.J., Mingo, B.,
2011, MNRAS, 413, 2358

\bibitem[\protect\citeauthoryear{Ramos Almeida et al.} {2012}]{ramos12}
Ramos Almeida, C., Bessiere, P. S., Tadhunter, C. N., P\'erez-Gonz\'alez, P. G., Barro, G., Inskip, K. J., Morganti, R., Holt, J., Dicken, D., 2012, MNRAS, 419, 687

\bibitem[\protect\citeauthoryear{Ramos Almeida et al.} {2013}]{ramos13}
Ramos Almeida, C., Bessiere, P.S., Tadhunter, C.N., Inskip, K.J., Morganti, R., Dicken, D., Gonz\'alez-Serrano, J.I.,
Holy, J., 2013, MNRAS, 436, 997

\bibitem[\protect\citeauthoryear{Rawlings \& Saunders} {1991}]{rawlings91}
Rawlings, S., Saunders, R., 1991, Nat, 349, 138

\bibitem[\protect\citeauthoryear{Sadler et al.} {2002}]{sadler02}
Sadler, E.M., et al., 2002, MNRAS, 329, 227

\bibitem[\protect\citeauthoryear{Sadler et al.} {2014}]{sadler14}
Sadler, E.M., Ekers, R.D., Mahoney, E.K., Mauch, T., Murphy, T., 2014, MNRAS, 438, 796

\bibitem[\protect\citeauthoryear{Sandage} {1966}]{sandage66}
Sandage, A., 1966, ApJ, 145, 1

\bibitem[\protect\citeauthoryear{Sanders \& Mirabel} {1996}]{sanders96}
Sanders, D.B., Mirabel, I.F., 1996, ARA\&A, 34, 749

\bibitem[\protect\citeauthoryear{Sanders et al.} {1988a}]{sanders88a}
Sanders, D.B., Soifer, B.T., Elias, J.H., Madore, B.F., Matthews, K., 
Neugebauer, G., Scoville, N.Z., 1988a, ApJ, 325, 74

\bibitem[\protect\citeauthoryear{Sanders et al.} {1988b}]{sanders88b}
Sanders, D.B., Saifer, B.T., Elias, J.H., Neugebauer, G., Matthews, K., 1988b, ApJ, 328, L35


\bibitem[\protect\citeauthoryear{Schmidt} {1963}]{schmidt63} 
Schmidt, M., 1963, ApJ, 137, 758

\bibitem[\protect\citeauthoryear{Schoenmakers et al.} {2000}]{schoenmakers00}
Schoenmakers, A.P., de Bruyn, A.G., Rottgering, H.J.A., van der Laan, H., Kaiser, C.R., 2000, MNRAS, 315, 371

\bibitem[\protect\citeauthoryear{Schreiber et al.} {2015}]{schreiber15}
Schreiber, C., et al., 2015, A\&A, 575, 74

\bibitem[\protect\citeauthoryear{Schweitzer et al.} {2006}]{schweitzer06}
Schweitzer, M., et al., 2006, ApJ, 649, 79

\bibitem[\protect\citeauthoryear{Seigar, Graham \& Jergen} {2007}]{seigar07}
Seigar, M.S., Graham, A.W., Jergen, H., 2007, MNRAS, 378, 1575


\bibitem[\protect\citeauthoryear{Sersic} {1963}]{sersic63}
Sersic, J.L., 1963, BAAA, 6, 41

\bibitem[\protect\citeauthoryear{Shaw et al.}{1995}]{shaw95}
Shaw, M., Tadhunter, C., Dickson, R., Morganti, R., 1995, MNRAS, 275, 703

\bibitem[\protect\citeauthoryear{Siebenmorgen, Kr\"{u}gel \& Spoon} {2004}]{siebenmorgen04}
Siebenmorgen, R., Kr\"{u}gel, E., Spoon, H.W.W., 2004, A\&A, 414, 123

\bibitem[\protect\citeauthoryear{Siebert et al.} {1996}]{siebert96}
Siebert, J., Brinkmann, W., Morganti, R., Tadhunter, C.N.,
Danziger, I.J., Fosbury, R.A.E., di Serego Alighieri, S., 1996, MNRAS, 279, 1331

\bibitem[\protect\citeauthoryear{Sikora, Stawarz \& Lasota} {2007}]{sikora07}
Sikora, M., Stawarz, \L., Lasota, J-P., 2007, ApJ, 658, 815

\bibitem[\protect\citeauthoryear{Simpson, Ward \& Wilson} {1995}]{simpson95}
Simpson, C., Ward, M.J., Wilson, A.S., 1995, ApJ, 454, 683

\bibitem[\protect\citeauthoryear{Singal et al.} {1993}]{singal93}
Singal, A.K., 1993, MNRAS, 262, L27

\bibitem[\protect\citeauthoryear{Singh et al.} {2015}]{singh15}
Singh, V., Ishwara-Chandra, C.H., Sievers, J., Wadadekar, Y., Hilton, M., Beelen, A., 2015, MNRAS, in press 
(arXiv:1509.01559)

\bibitem[\protect\citeauthoryear{Smith \& Heckman} {1989a}]{smith89a}
Smith, E.P., Heckman, T.M., 1989a, ApJSS, 69, 365

\bibitem[\protect\citeauthoryear{Smith \& Heckman} {1989b}]{smith89b}
Smith, E.P., Heckman, T.M., 1989b, ApJ, 341, 658


\bibitem[\protect\citeauthoryear{Smith, Heckman \& Illingworth} {1990b}]{smith90b}
Smith, E.P., Heckman, T.M., Illingworth, G.D., 1990b, ApJ, 356, 399

\bibitem[\protect\citeauthoryear{Smol\u{c}i\'c \& Riechers} {2009}]{smolcic09}
Smol\u{c}i\'c, V., Riechers, D.A., 2011, ApJ, 730, 64

\bibitem[\protect\citeauthoryear{Spinrad et al.} {1985}]{spinrad85}
Spinrad, H., Marr, J., Aguilar, L., Djorgovski, S., 1995, PASP, 97, 932

\bibitem[\protect\citeauthoryear{Stanghellini et al.} {2005}]{stanghellini05}
Stanghellini, C., O'Dea, C., Dallacasa, D., Cassaro, P., Baum, S., Fanti, R., Fanti, C., 2005, A\&A, 443, 443, 891


\bibitem[\protect\citeauthoryear{Tadhunter, Fosbury \& Quinn} {1989a}]{tadhunter89a}
Tadhunter, C.N., Fosbury, R.A.E., Quinn, P.J., 1989, MNRAS, 240, 225

\bibitem[\protect\citeauthoryear{Tadhunter \& Tsvetanov} {1989b}]{tadhunter89b}
Tadhunter, C., Tsvetanov, Z., 1989, Nat, 341, 422

\bibitem[\protect\citeauthoryear{Tadhunter, Scarrott \& Rolph} {1990}]{tadhunter90}
Tadhunter, C., Scarrott, S.M., Rolph, C.D., 1990, MNRAS, 246, 163

\bibitem[\protect\citeauthoryear{Tadhunter et al.} {1992}]{tadhunter92}
Tadhunter, C.N., Scarrott, S.M., Draper, P., Rolph, C., 1992, MNRAS, 256, P53

\bibitem[\protect\citeauthoryear{Tadhunter et al.} {1993}]{tadhunter93}
Tadhunter, C., Morganti, R., di Serego-Alighieri, S., Fosbury, R. A. E., Danziger, I. J., 1993, MNRAS, 263, 999 


\bibitem[\protect\citeauthoryear{Tadhunter et al.} {1998}]{tadhunter98}
Tadhunter, C., Morganti, R., Robinson, A., Dickson, R., Villar-Mart\'in, M., Fosbury, R. A. E., 1998, MNRAS, 298, 1035

\bibitem[\protect\citeauthoryear{Tadhunter et al.} {1999}]{tadhunter99}
Tadhunter, C.N., Packham, C., Axon, D.J., Jackson, N.J., Hough, J.H., Robinson, A., Young, S., Sparks, W., 1999, ApJ, 512, L94

\bibitem[\protect\citeauthoryear{Tadhunter et al.} {2000}]{tadhunter00}
Tadhunter, C.N., Sparks, W., Axon, D.J., Bergeron, L., Jackson, N.J., Packham, C., Hough, J.H., Robinson, A., Young, S., 2000, MNRAS, 313, L52

\bibitem[\protect\citeauthoryear{Tadhunter et al.} {2002}]{tadhunter02}
Tadhunter, C., Dickson, R., Morganti, R., Robinson, T. G., Wills, K., Villar-Mart\'in, M., Hughes, M., 2002, MNRAS, 330, 997

\bibitem[\protect\citeauthoryear{Tadhunter et al.} {2005}]{tadhunter05}
Tadhunter, C., Robinson, T.G., Gonz\'alez Delgado, R.M., Wills, K., Morganti, R., 2005, MNRAS, 356, 480

\bibitem[\protect\citeauthoryear{Tadhunter et al.} {2007}]{tadhunter07}
Tadhunter, C., Dicken, D., Holt, J., Inskip, K., Morganti, R.,
Axon, D., Buchanan, C., Gonz\'alez Delgado, R., Barthel, P., van Bemmel, I., 2007, ApJ, 661, L13

\bibitem[\protect\citeauthoryear{Tadhunter et al.} {2011}]{tadhunter11}
Tadhunter, C.,Gonz\'alez Delgado, R., Rodr\'iguez Zaur\'in, J., Villar-Mart\'in, M., Morganti, R., Emonts, B., Ramos 
Almeida, C., Inskip, K., 2011, MNRAS, 412, 960

\bibitem[\protect\citeauthoryear{Tadhunter et al.} {2012}]{tadhunter12}
Tadhunter, C.N., Ramos Almeida, C., Morganti, R., Holt, J., Rose, M., Dicken, D., Inskip, K., 2012, MNRAS, 427, 1603

\bibitem[\protect\citeauthoryear{Tadhunter et al.} {2014}]{tadhunter14}
Tadhunter, C., Dicken, D., Morganti, R., Konyves, V., Ysard, N., Nesvadba, N., Ramos Almeida, C., 2014, MNRAS, 445, L51


\bibitem[\protect\citeauthoryear{Tal et al.} {2009}]{tal09}
Tal, T., van Dokkum, P.G., Nelan, J., Bezanson, R., 2009, AJ, 138, 1417

\bibitem[\protect\citeauthoryear{Treister et al.} {2012}]{treister12}
Treister, E., Schawinski, K., Urry, C.M., Simmons, B.D., 2012, ApJ, 758, L39

\bibitem[\protect\citeauthoryear{Tzioumis et al.} {2002}]{tzioumis02}
Tzioumis, A., et al., 2002, A\&A, 392, 841

\bibitem[\protect\citeauthoryear{Ulrich} {1981}]{ulrich81}
Ulrich, M.H., 1981, A\&A, 103, L1

\bibitem[\protect\citeauthoryear{Urry, Padovani \& Stickel} {1991}]{urry91}
Urry, C.M., Padovani, P., Stickel, M., 1991, ApJ, 382, 501

\bibitem[\protect\citeauthoryear{Urry \& Padovani} {1995}]{urry95}
Urry, C., Padovani, P., 1995, PASP, 107, 803

\bibitem[\protect\citeauthoryear{van Breugel et al.} {1984}]{vanbreugel84a}
van Breugel, W., Heckman, T., Butcher, H., Miley, G., 1984, ApJ, 277, 82



\bibitem[\protect\citeauthoryear{van der Wolk et al.} {2010}]{vanderwolk10}
van der Wolk, G., Barthel, P.D., Peletier, R.F., Pel, J.W., 2010, A\&A, 511, 64

\bibitem[\protect\citeauthoryear{van Dokkum} {2005}]{vandokkum05}
van Dokkum, P.G., 2005, AJ, 130, 2647

\bibitem[\protect\citeauthoryear{Veilleux et al.} {2009}]{veilleux09}
Veilleux, S., et al., 2009, ApJS, 182, 628

\bibitem[\protect\citeauthoryear{Verdoes Kleijn et al.} {1999}]{verdoes99}
Verdoes Klein, G.A., Baum, S., de Zeeuw, P., O'Dea, C.P., 1999, AJ, 118, 2592

\bibitem[\protect\citeauthoryear{Wall \& Peacock} {1985}]{wall85}
Wall, J.V., Peacock, J.A., 1985, MNRAS, 216, 173


\bibitem[\protect\citeauthoryear{Wills et al.} {2002}]{wills02}
Wills, K.A., Tadhunter, C.N., Robinson, T.G., Morganti, R., 2002, MNRAS, 333, 211

\bibitem[\protect\citeauthoryear{Wills et al.} {2004}]{wills04}
Wills, K.A., Morganti, R., Tadhunter, C.N., Robinson, T.G., Villar-Mart\'in, M., 2004, MNRAS, 347, 771

\bibitem[\protect\citeauthoryear{Wold, Lacy \& Armus} {2007}]{wold07}
Wold, M., Lacy, M., Armus, L., 2007, A\&A, 470, 531

\bibitem[\protect\citeauthoryear{Worrall} {2009}]{worrall09}
Worrall, D.M., 2009, A\&ARv, 17, 1

\bibitem[\protect\citeauthoryear{Zheng et al.} {1995}]{zheng95}
Zheng, W., P\'erez, E., Grandi, S.A., Penston, M.V., 1995, AJ, 109, 2355

\bibitem[\protect\citeauthoryear{Zirbel} {1997}]{zirbel97}
Zirbel, E.L., 1997, ApJ, 478, 489

\bibitem[\protect\citeauthoryear{ } { }]{ }




%
%
\end{thebibliography}


\end{document}